\documentclass[AMS,STIX1COL]{WileyNJDv5}
\articletype{Research article}%
\received{Date Month Year}
\revised{Date Month Year}
\accepted{Date Month Year}
\journal{Journal}
\volume{00}
\copyyear{2025}
\startpage{1}
\usepackage{bbm}
\usepackage{mathrsfs}
\usepackage{tikz}
\usepackage{tikz-cd}
\usepackage{enumitem}
\usepackage{empheq}
\usepackage{cases}
\usepackage{hyperref}
\usetikzlibrary{arrows.meta,positioning,calc}
\newcommand{\Int }{\displaystyle \int}

\newcommand{\vect}{\mathbf{b}}
\newcommand{\flow}{\mathbf{X}}
\newcommand{\pos}{\mathbf{x}}
\newcommand{\SP}{{\cal S}p{\cal S}t}
\newcommand{\be}{\begin{equation}}
\newcommand{\de}{\end{equation}}

\newcommand{\R}{R}
\newcommand{\alg}{\mathscr{D}_{pow}}
\newcommand{\blg}{\mathscr{H}_{pow}}
\newcommand{\V}{V_0}
\newcommand{\LLT}{L^2(\mathbb{T}^2)}
\newcommand{\LSPO}{OLSP}

\makeatletter
\newcommand{\customtoc}{\@starttoc{toc}}
\makeatother

\newtheorem*{propositionav}{Proposition 5.5 \cite{AV24}}
\newtheorem*{lemma5_6}{Lemma 5.6}

\newtheorem*{theoremav}{Theorem 1.1 \cite{AV24}}
\newtheorem*{propositionavCORE}{Proposition 5.2 \cite{AV24}}
\title{Spontaneous stochasticity in the Armstrong-Vicol passive scalar}
\author[1]{Wandrille Ruffenach}
\author[2,3]{Eric Simonnet}
\author[3]{Nicolas Valade}
\authormark{Ruffenach \textsc{et al.}}
\titlemark{Spontaneous stochasticity in the Armstrong-Vicol passive scalar}
\address[1]{\orgdiv{LPENSL}, \orgname{ENS de Lyon, CNRS}, \orgaddress{UMR5672, 69342 \state{Lyon cedex 07}, \country{France}}}
\address[2]{\orgdiv{Institut de Physique de Nice}, \orgname{Université C\^ote d'Azur et CNRS}, \orgaddress{\state{State Name}, \country{France}}}
\address[3]{\orgdiv{Department Name}, \orgname{INRIA Sophia-Antipolis}, \orgaddress{\state{State Name}, \country{France}}}
\corres{Eric Simonnet. \email{eric.simonnet@univ-cotedazur.fr}}

\abstract[Abstract]{
Spontaneous stochasticity refers to the emergence of intrinsic randomness in deterministic systems under singular limits, a phenomenon conjectured to be fundamental in turbulence. Armstrong and Vicol~\cite{AV24} recently constructed a deterministic, divergence-free multiscale vector field arbitrarily close to a weak Euler solution, proving that a passive scalar transported by this field exhibits anomalous dissipation and lacks a selection principle in the vanishing–diffusivity limit.

We show that this advection-diffusion PDE also selects a non-Dirac measure in the space of weak solutions in the inviscid limit, thereby exhibiting Eulerian spontaneous stochasticity. We further provide numerical evidence of Lagrangian spontaneous stochasticity, together with a numerical illustration of the Obukhov-Corrsin conjecture for this system.

We formulate a general framework for spontaneous stochasticity in arbitrary finite-dimensional systems under arbitrary regularizations, distinguishing two regimes: \emph{weak}, where different probability measures may arise along subsequences of inviscid limits, and \emph{strong}, where the limit measure is unique. The advection–diffusion system of~\cite{AV24} lies in the strong regime.

We prove that the set of selected measures is compact and equals the closed convex hull of Dirac measures. Moreover, for any non-Dirac measure supported on the set of nonunique solutions of the inviscid system, there exists a regularization that produces strong spontaneous stochasticity.

Finally, we relate this framework to renormalization-group methods \`a la 
Feigenbaum and examine how the underlying dynamical system influences the inviscid limit. The discussion is complemented by elementary finite-dimensional examples illustrating a variety of cases.

}

\date{\today}

\begin{document}
\maketitle
%\tableofcontents
%\customtoc
\section{Introduction}\label{mainintro}
This project can be understood as a very modest attempt to reconcile two perspectives -- those of mathematicians and physicists -- on the nature of spontaneous stochasticity.

This work is motivated in large part by a recent mathematical proof establishing the existence of a dissipative anomaly in a passive-scalar transport system \cite{AV23,AV24}, 
as well as the absence of a selection principle. Similar results have been observed in recent studies of advection-diffusion partial differential equations (PDEs) \cite{Drivas_Elgindi,colombo,Titi2023,burczak2023anomalous}.

The main message of this paper is twofold. On the one hand, the absence of a selection principle, as seen by mathematicians, is often perceived negatively as a mathematical pathology of the model in the inviscid limit.
The consequence is that such a pathology further distances the scientific community from a deeper understanding of turbulence, adding yet another layer of mystery to the connection between the Navier-Stokes equations and the Euler equations.

On the other hand, the notion of spontaneous stochasticity is not so well defined among physicists: it is often identified with Lagrangian spontaneous stochasticity as introduced by \cite{Chaves2003}.
Alternatively, it might also be confused sometimes with a notion of classical chaos or with a problem of representing complexity -- for instance, the need to represent complexity in a probabilistic manner.
As a result, spontaneous stochasticity may be wrongly perceived as merely a methodological or representational issue rather than as a genuine physical phenomenon in its own right.

The goal here is to reconcile and clarify these perspectives.
For mathematicians, the absence of a selection principle should not lead to the rejection of the model and is certainly not a mathematical pathology. Its very existence is of a physical nature. Whether it is more or less realistic is a question of a different kind.
In the system of  \cite{AV23,AV24}, as well as by \cite{burczak2023anomalous}, 
it appears in a much more realistic manner than ever seen before.
Remarkably, it can be analyzed mathematically, largely thanks to the development of convex integration over the past decade
\cite{hprinciple,Isett,Admi_Onsager,BuckVic}.

For physicists, the fact that transport equations for passive scalars -- despite being linear PDEs -- can exhibit Eulerian spontaneous stochasticity should be regarded as a significant result.

In fact, the role that these recent models of passive scalar transport play in understanding Eulerian spontaneous stochasticity is reminiscent of the role that the Kraichnan model \cite{Kraichnan} has played in the turbulence community in understanding the dissipative anomaly and Lagrangian spontaneous stochasticity \cite{Chaves2003,Drivas17}.

\subsection{Historical review}
The modern notion of {\it spontaneous stochasticity} has been quite elusive since the first time it
was coined as such by \cite{Chaves2003}. It emerged from a long and intricate history mostly
related to turbulence and the Navier-Stokes equations. 
Spontaneous stochasticity is the property that a deterministic system, in some singular limit, can exhibit an intrinsic source of randomness.
It has often been wrongly identified
with the phenomenon of chaos in nonlinear systems: it is not, not even close... 
The confusion indeed arises simply because in turbulence: chaos and spontaneous stochasticity conspire together.

There are in fact two different parallel stories: a Lagrangian tale
and an Eulerian one. It seems we are still far from reconciliating the two.
\subsubsection{Lagrangian spontaneous stochasticity}\label{subsec_lss}
One can probably root the concept 
with the work of Richardson on turbulence diffusion \cite{Richardson1926}.
In essence, the separation ${\bf R}_{12}(t) = {\bf R}_1(t)-{\bf R}_2(t)$ between two close fluid particles in a given flow would satisfy $\frac{d}{dt} {\bf R}_{12} = {\bf v}(t,{\bf R}_1)
-{\bf v}(t,{\bf R}_2)$ or taking the norm $\rho := ||{\bf R}_{12}||, \frac{d}{dt} \rho^2 =
2 \rho \cdot \Delta {\bf v}$. If the fluid velocity is smooth 
$||{\bf v}(t,{\bf R}_1)
-{\bf v}(t,{\bf R}_2)|| \propto ||{\bf R_1-R_2}||$ (Lipschitz), then one would obtain
$\frac{d}{dt} \rho^2 \propto \rho^2$. This yields an exponential growth.
In a turbulent fluid however, velocities are not smooth. It is known from Kolmogorov
\cite{Kolmogorov1941}
that, neglecting intermittent effects, a reasonable approximation is the K41 velocity
having $\frac13$ H\"older regularity: $||{\bf v}(t,{\bf R}_1)
-{\bf v}(t,{\bf R}_2)|| \propto ||{\bf R_1-R_2}||^{\frac{1}{3}}$. It implies that 
one has $\frac{d}{dt} \rho^2 \propto (\rho^2)^\frac23$. This yields a totally different behavior
where now, $\rho$ grows as a power of time, even more important this is true no matter how close are the two particles initially. This is the famous Richardson's law:
$\langle \rho^2 \rangle \propto t^3$ observed numerically in a consistent way
, see e.g. \cite{Falkovich2001,salazar2009,BEC} and references therein.
It is much less clear at the experimental level  \cite{Jullien2000}. 

Meanwhile, one of the fundamental aspect of turbulence, already suggested
by Taylor \cite{Taylor}, is the  fact that kinetic energy can be dissipated 
for fluid of "infinitesimal viscosity". It was indeed phrased as a conjecture by
Onsager \cite{Onsager49} for 3D Euler, that we would like to quote: 
"{\it It is of some interest to note that in principle, turbulent dissipation as
described could take place just as readily without the final assistance by
viscosity. In the absence of viscosity, the standard proof of the conservation
of energy does not apply, because the velocity field does not remain differentiable! In fact it is possible to show that the velocity field in such "ideal"
turbulence cannot obey any LIPSCHITZ condition of the form $|{\bf v(r'+r)-v(r')}| < {\rm Const.}r^n$,
for any order $n$ greater than 1/3; otherwise the energy is conserved.}"

This remarkable phenomenon of "dissipative anomaly" (sometimes referred to as "the zeroth law of turbulence") also appears in simpler models and in particular
in passive scalar models: linear advection-diffusion equations for a scalar field
$\theta=\theta(t,{\bf x})$ taking the simple form $\partial_t \theta + {\bf v} \cdot \nabla \theta = 
\kappa \Delta \theta$, and where the velocity ${\bf v}(t,{\bf x})$ and initial condition $\theta(0,{\bf x})$ are prescribed. The dissipative anomaly then is expressed mathematically simply as
$\limsup_{\kappa \to 0} \kappa ||\nabla \theta||_{L^2_{x,t}}^2 > 0$.
A synthetic model of turbulence which has played a key role in turbulence studies is
the so-called Kraichnan model \cite{Kraichnan}. It is again an advection-diffusion system where the
velocity mimics some key aspects of velocity fields solving 3-D Navier-Stokes at high Reynolds number. It is a random white-in-time, correlated in space, Gaussian field having two free parameters
controlling its spatial regularity over a given range of scales.
In a remarkable series of works \cite{Gawedzki98,Chaves2003,Cardy2008}, Gaw\c{e}dzky and colleagues were able to establish
using the Kraichnan ensemble that indeed the classical determinism of Lagrangian trajectories breaks down
in the limit of small diffusion. In particular, they showed that dissipative anomaly was
the sole consequence of the phenomenon of spontaneous stochasticity: the Lagrangian particles have nonunique trajectories even when starting from the same initial condition and fixed
realization of the velocity, and this was essentially a consequence of the spatial
roughness of the velocity. This work was a breakthrough in the understanding on two seemingly different phenomena: spontaneous stochasticity and dissipative anomaly. The same phenomenon was also observed and studied in \cite{EVDE},\cite{E2003} under the name intrinsic stochasticity using the notion of generalized flows.  Some decades later, the equivalence between spontaneous stochasticity and dissipative flows was established in more general cases by the work of \cite{Drivas17} as a Lagrangian fluctuation-dissipation theorem. In simple words:
spontaneous stochasticity and anomalous diffusion are equivalent, at least for passive scalars.

Of importance is to recall that either in the work of Richardson or later, 
the notion of spontaneous stochasticity must be called in fact {\it Lagrangian spontaneous stochasticity}. "Lagrangian" was later on becoming quite implicit and dropped.
The story of course does not end there. Even more fascinating aspects emerged starting from
the year 1969.
\subsubsection{Eulerian spontaneous stochasticity}
\label{subsec_ess}
In the year 1969, E.N. Lorenz came with a surprising statement that we quote:
"{\it It is proposed that certain formally deterministic fluid systems which possess many
scales of motion are observationally indistinguishable from indeterministic systems;
specifically, that two states of the system differing initially by a small “observational
error” will evolve into two states differing as greatly as randomly chosen states of the
system within a finite time interval, which cannot be lengthened by reducing the
amplitude of the initial error}" \cite{Lorenz69}. Reading twice, this has exactly the same flavor that
Richardson particle pair separation, but it is now the full (Navier-Stokes) Eulerian field which is involved! The existence of such a {\it finite-time} predictability barrier is of course deeply connected to 3-D turbulence itself, in particular it requires the typical Kolmogorov $k^{-5/3}$ spectrum.
This work has been barely noticed for decades, probably shadowed by the famous 1963 paper \cite{Lorenz63} on unpredictability (sensitive dependence to initial conditions) of low-order chaos.
As explained in \cite{Rotunno_Snyder08,Palmer14}, in the 70s, people also believed at that time that the atmospheric spectrum was essentially 2-D with a $k^{-3}$ spectrum and a predictability barrier more consistent with classical chaos. Nowadays, the situation has changed dramatically: computers routinely
solve atmospheric scales much below 100 km where a clear $k^{-5/3}$ spectrum emerges. The intuition
of Lorenz becomes highly relevant. 
A name has been given to this sensitivity in 3-D multiscale flows: Palmer \cite{Palmer14} called it 
\emph{the real butterfly effect} which, incidentally, was originally a seagull. We refer to it as Eulerian spontaneous stochasticity.
Interestingly, the term Eulerian spontaneous stochasticity was not coined until much later -- first introduced by G. Eyink at the APS 2016 conference and only formally discussed in \cite{Eyink_Bandak20}.

When this phenomenon occurs, some Lyapunov exponents -- possibly finite-time ones -- are expected to diverge as the Reynolds number increases. This behavior contrasts with classical chaos, where Lyapunov exponents may be positive but remain bounded.

For all these reasons, people started to take a closer look at Lorenz's idea.
The first immediate systems were the 2-D surface quasi-geostrophic equations 
which possesses some direct $k^{-5/3}$ cascade \cite{Rotunno_Snyder08},\cite{Palmer14}
and more recently \cite{Nicolasetal24}, as well as the Kelvin-Helmholtz instability in 2-D Navier-Stokes
\cite{Simon_Jeremie_AM20} and Rayleigh-Taylor 3-D turbulence \cite{biferale2018rayleigh} (see also \cite{crisanti1993intermittency} for shell models). Indeed, these results confirm the intuition of Lorenz to various extents. A more detailed review of known results can be found in \cite{Eyink_Bandak20}.

Until now, we do not have clearly stated what this concept is. In general, ideal fluids or
wave dynamics are ill-posed. Typical examples are the inviscid Burgers equations and the incompressible Euler equations. It is of common practice to regularize such systems by adding other terms like (hyper) viscosity and/or small noise terms, e.g.  Navier-Stokes
or viscous Burgers equations. The main question is then: how such systems behave in the limit of vanishing regularization? This is commonly referred to as the {\it inviscid limit}. Whenever a system becomes stochastic in the inviscid limit, we say it is spontaneously stochastic. A difficult and important question is then to understand how the statistics, in the inviscid limit, depend on the way the system has been regularized.
One of the hope in the case of Euler/Navier-Stokes is that for a wide class of regularizations, those statistics are independent of the set of regularizations considered, therefore exhibiting a form of {\it universality}. 
These questions have been tackled thoroughly by A. Mailybaev and colleagues for the last 10 years 
using the concept of {\it renormalization group} (RG) \cite{Maily2012,Maily2016,AM_Ra23,AM_Rb23,AM_24} (see also \cite{Eyink_Bandak20}). The RG approach is the one akin to the dynamical system theory approach for the universality of the Feigenbaum period-doubling cascade \cite{Feigenbaum1976}.
These studies mainly focus on simpler shell models of turbulence, like Sabra or dyadic systems which possess many scales (of motion). 
Physicists and mathematicians widely recognize these shell models for sharing many desirable properties with the Euler, Navier-Stokes, or Burgers equations such as power laws, multiscaling \cite{Gilson1998,Mailybaev2012}, making them much easier alternatives given the considerable difficulty of directly tackling the formidable 3-D incompressible Euler equations. The reason to consider a RG approach is that ideal fluids have many symmetries, associated with spatio-temporal self-similar (fractal) structures. Therefore, it is possible to define a particular self-consistent notion of (canonical) regularizations \cite{AM_24}. The RG is then interpreted as a genuine dynamical system in the space of flow maps, where the successive group iterates reflect the level of regularization of the inviscid system. The lesson  is that, provided there exists some RG attractor (it can be a fixed point), one can infer some universal statistical properties
of the inviscid limit provided a given set of regularizations belong to the basin of attraction. We would like to insist on another misunderstanding that one must not look at
a fixed regularization, say Navier-Stokes at fixed Reynolds number $Re$ but rather at the inviscid limit $Re \to \infty$. However, signatures of spontaneous stochasticity are visible well before
the limit is actually taken as pointed out in \cite{Eyink_Bandak20}.

In order to exhibit spontaneous stochasticity, the inviscid system must necessarily be ill-posed. Well-posedness is
existence, unicity and continuous dependence on the initial conditions.
In this case, this is rather the nonuniqueness of the inviscid system which is responsible
for the ill-posedness and as a consequence loss of continuity w.r.t. initial conditions. 
It is easy to find infinite-dimensional systems that satisfy existence and uniqueness but are not well-posed. To fix the idea, let us write an inviscid system, say Euler equations as 
$\dot x = f(x), x(0)=x_0$. For smooth initial data, it is possible
to establish some (local) well-posedness until some time $T^\star > 0$. One of the most important
open question in mathematical fluid mechanics is whether $T^\star < +\infty$ or not:
{\it does three-dimensional incompressible Euler flow with smooth initial conditions develop a singularity with infinite vorticity after a finite time?} \cite{FrischBec1}. When
such thing happens, the solution blows up in finite-time: nonuniqueness emerges as a pure manifestation
of the solution hitting a (non-Lipschitz) singularity in finite-time. It turns out
that this is precisely what is observed in simpler models of Euler equations, e.g. in log-lattice Sabra
\cite{Ciro} or for Rayleigh-Taylor instability \cite{mailybaev2017toward}. The deep connection between spontaneous stochasticity and the presence of singularities in the inviscid system has been rigorously investigated by \cite{Drivas21} and
\cite{Drivas24} in simpler finite-dimensional dynamical systems. This is also the approach adopted in Sections \ref{Defs},\ref{allprob},\ref{RGSP},\ref{CNforSP} but in a more general framework.

Assuming the solution reaches some singularity in finite-time,
its post blow-up life might face nonuniqueness. In such a case, which solutions 
of the inviscid system among the infinity of them is chosen in the inviscid limit?
This selection must a priori depend on the way the system is regularized.
This {\it selection principle} is one of the fundamental questions underlying the concept of spontaneous stochasticity. In this work, we convey the idea that one must not look for a "good/physical" regularization but rather ask for the universality and/or robustness of families of regularizations.
 This is a difficult question, but again it is best addressed using  RG-like approaches.

A not so well-known example where spontaneous stochasticity is strongly believed to occur is the one-dimensional Kuramoto--Sivashinsky (KS) equation,
$\partial_t \phi + \phi \partial_x \phi + \partial_{xx} \phi + \partial_{xxxx} \phi = 0$, on a periodic domain of size $L \gg 1$. In the early 1980s, Yakhot~\cite{Yakhot1981Kuramoto} conjectured that \emph{in the hydrodynamic limit, the probability distribution of the KS equation may be mimicked by that of the noise-driven (inviscid) Burgers equation}~\cite{ChowHwa1995}. See also \cite{pomeau1984intrinsic} for early observations and its link with turbulence, where they explicitly used the term \emph{intrinsic stochasticity}. Thus, the KS equation can be regarded as an alternative regularization of Burgers, with a destabilizing negative-viscosity term and a stabilizing hyperviscous (bi-harmonic) dissipation. In the inviscid limit, its statistics fall within the Kardar--Parisi--Zhang (KPZ) universality class~\cite{RoyKS}, in contrast to the classical viscous regularization of Burgers, which yields a unique entropy solution.

\subsubsection{The convex integration revolution}\label{subsec_convex}
Until now, we have only described the physicist viewpoint. What do mathematicians say about this subject? 
Strictly speaking the notion of spontaneous stochasticity has not been addressed
in those terms. Rather, there has been a strong focus on the Millennium prize regarding global well-posedness of
the 3D incompressible Navier-Stokes. In particular, central is the understanding of weak solutions
not only for Navier-Stokes, (non)uniqueness of  Leray weak solutions \cite{Leray}, but for the Euler 
weak solutions as well. For the last case, a great achievement of the recent years is the mathematical proof of the Onsager conjecture (see Onsager quote in Section \ref{subsec_lss}). 

The story begins with the Nash-Kuiper $C^1$ isometric embedding theorems for the torus \cite{Nash,Kuiper}. A detailed review can be found in \cite{hprinciple}. Roughly speaking, isometric embedding of the torus involves taking a flat piece of paper and bending it into the shape of a doughnut. However, anyone attempting this would fight against Gauss's Theorema Egregium, which states that Gaussian curvature must be preserved unless the paper is stretched. In simpler terms, this "rigidity" makes it impossible to transform a flat sheet into a sphere or a doughnut without bending --something most people have experienced once.

Crucially, such transformations typically require $C^2$ regularity. This led to an intriguing question posed to J. Nash: could it still be possible to achieve this using only $C^1$ transformations? Surprisingly, the answer was a resounding yes. Striking images can be found nowadays showing how the doughnut looks like \cite{Borelli}: it involves successive embeddings
yielding $C^1$ fractals. It was indeed the birth of convex integration.

What is the link with Euler equations? It took some time before people started to realize there was a strong analogy with Onsager's conjecture. This analogy is a suitable variant of the $h$-principle -- $h$ for homotopy -- found by Gromov \cite{Gromov}, and developed by De Lellis and Sz\'ekelyhidi 
\cite{hprinciple}. The Egregium Gauss theorem is now played by the conservation of energy for solutions
having enough regularity above the critical 1/3 H\"older exponent. This rigid part
(the analogy is the $C^2$ folding of a piece of paper) is the easiest to prove \cite{CET94}. 
Proving the nonuniqueness of dissipative weak solutions in 3-D Euler equations turns out to be much more 
difficult: this is the flexible part of the Onsager's conjecture. It took almost a decade to climb the ladder of weak dissipative solutions regularity from $C^0$ up to $C^{1/3}$ by improving the convex integration schemes until the
conjecture was finally proven by Isett \cite{Isett} and Buckmaster et al. \cite{Admi_Onsager} for the admissible case.

Convex integration schemes have recently attained some high level of sophistication (intermittent schemes) so that it is now possible to handle weak solutions of the 3-D Navier-Stokes as well in the groundbreaking work \cite{BuckVic}. In particular, the viscous Laplacian term is adding much difficulties for controlling the various error estimates necessary for scheme convergence.
While the Armstrong-Vicol transport system \cite{AV23} -- on which our work is based -- also shares common ground with convex integration schemes, it innovates by employing sophisticated techniques of fractal homogenization.

Convex integration schemes can be interpreted as an  inviscid limit.
Rather than looking at the inviscid limit $Re \to \infty$ in Navier-Stokes where
the Laplacian acts as a regularization mechanism, one is
considering a family $({\bf u}_k, p_k,\mathring R_k)$ of smooth solutions of the so-called Euler-Reynolds equations:
$\partial_t {\bf u}_k + {\bf u}_k \cdot \nabla {\bf u}_k + \nabla p_k = {\rm div}~
\mathring R_k$, where $\mathring R_k$ is a symmetric trace-free tensor. The regularization mechanism is now "homogenization" conveyed by the
Reynolds-stress term (in fact, an effective averaging of the small scales is induced by the integral operator ${\rm div}^{-1}$).
By properly renormalizing the solutions, it is possible to construct a 
convergent sequence in some H\"older space $C^\alpha$, $\alpha < \frac13$:
\\
\resizebox{\columnwidth}{!}{
\begin{tikzpicture}
  \node (A0) at (0,0) {$({\bf v}_0,p_0,\mathring R_0)$};
  \node (A1) at (3,0) {$({\bf v}_1,p_1,\mathring R_1)$};
  \node (Ak) at (4.5,0) {$\cdots$};
 \node (Ak1) at (6,0) {$({\bf v}_k,p_k,\mathring R_k)$};
  \node (Ak2) at (9.5,0) {$({\bf v}_{k+1},p_{k+1},\mathring R_{k+1})$};
  \node (A) at (12.5,0) {~~$ \cdots ~~({\bf v},p,\mathring R)$};
  \draw[->, bend left] (A0) to node[above] {${\cal R}_0$} (A1);
  \draw[->, bend left] (Ak1) to node[above] {${\cal R}_{k}$} (Ak2);
\end{tikzpicture},}
%\\
where ${\rm div}~\mathring R$ is zero weakly.
The renormalization operators ${\cal R}_k$ take a complicate form, where both time and space are rescaled. They also involve Nash decompositions and the use of inverse flow maps, see e.g. \cite{ReviewConvex}. 
%Their expressions are typically of the form (\ref{RenormR}) in Section \ref{sechom}.
Obtaining convergence of the Euler-Reynolds sequence $({\bf v}_k, p_k, \mathring{R}_k)$ in a H\"older space $C^\alpha$ is a difficult task.
The closer $\alpha$ is to 1/3 the harder it is to control the various error terms.

The sequence of Euler-Reynolds equations above 
can be interpreted as some controlled regularizations of the inviscid Euler equations. To this respect, both RG approach and convex integration schemes share a common conceptual basis. The approach in \cite{AM_24end}, focusing on the Sabra shell model instead of a PDE, yields simpler renormalization operators in the autonomous form ${\cal R}_k = 
\underbrace{{\cal R} \circ \cdots \circ {\cal R}}_{\scriptscriptstyle k~\mbox{times}}$. There are few drawbacks in convex integration schemes however.
First, they do not handle the Cauchy initial value problem (see however \cite{Mengual}), second the scale separation involved
is always hypergeometric and seems mandatory in order to control the many estimates.

\subsubsection{Nonuniqueness and the inviscid limit of Navier-Stokes}
As explained before, the necessary ingredient for having Eulerian spontaneous stochasticity is to have nonunique solutions of the initial value Cauchy problem in the inviscid system. It is thus important to mention first very interesting tools and concepts which have been developed by the mathematicians to handle the lack of a known global uniqueness result for the 3-D Navier-Stokes and/or to formalize the notion of ensemble average in turbulence.
One is the notion of {\it statistical solution} introduced by Foias and Vishik \cite{Foias1972,Foias1976,Vishik1977}, see \cite{Foias2013}.
Other notions of multivalued semigroups have been developed by Sell \cite{Sell1973} and
in particular the notion of {\it generalized semiflows} by Ball \cite{Ball1997}, see also
\cite{Simsen2008,James2003}. All these concepts have a strong Eulerian flavor since they consider a phase space, dynamical system point of view.

More Lagrangian-oriented mathematical results have also been obtained, starting 
from the important DiPerna-Lions theory \cite{diperna1989ordinary},
and the extension by L. Ambrosio \cite{ambrosio2004transport}, as well as the introduction of {\it superposition solutions} by \cite{Flandoli2009}. They give
deeper insights on the well-posedness of (nonautonomous) ODEs and their
natural link with linear transport equations and conservation laws.

With respect to the inviscid limit in Navier-Stokes, the work of DiPerna and Majda \cite{DiPerna1987} explicitly addresses the inviscid limit w.r.t. to viscosity for the deterministic 
3-D Navier-Stokes. They introduce
the notion of {\it measure-valued solutions} based on Young measures.
Their main theorem is a convergence result in a rather weak sense that any weak solution
converges in subsequence to a measure-valued solution. 
Unfortunately, this result does not rule out the possibility that this measure-valued solution collapses into a Dirac mass. Only strong numerical evidence suggests that the emergence of oscillations and concentrations of the solution in the limit would disrupt any Dirac-like measure, leading to a nontrivial one. The approach of 
\cite{Brenier1989} is innovative in that it gives a probabilistic view 
of the selected
solutions in the inviscid 3-D Navier-Stokes limit. The key idea is to exploit a variational formulation of the Euler equations, as a least action principle over volume-preserving flow maps in physical space. Unlike  previous studies, the framework is inherently Lagrangian and yields a natural notion of probability measure in the space of Lagrangian flow maps -- what is referred to as 
{\it generalized flows}. These ideas have been investigated thoroughly in \cite{Thalabard2020}.

Last, it is important to mention the work of S. Kuksin on the 2-D Navier-Stokes equations \cite{kuksin} forced by additive noise.
The noise must scale like the square root of viscosity in order to obtain physically meaningful solutions in the inviscid limit.
Such inviscid limit is called the {\it Eulerian limit}.
The striking result is the emergence of a selection mechanism toward a genuinely random process $U^\omega(\cdot)$ solving the 2-D Euler equations pathwise, with energy and enstrophy
becoming random variables. If $u_\nu^\omega$ is the random Navier-Stokes field, then a subsequential limit is obtained of the form
$\lim_{\nu_j \to 0} \lim_{T \to \infty} u_\nu^\omega(T+\cdot) =
U^\omega(\cdot)$ with the limits taken in that order.
While this resembles spontaneous stochasticity, it is not quite the same. Fixing the initial condition leads to deterministic dynamics. 
Only in the long-time limit does the system lose its memory, and what is left is a stochastic outcome shaped by the Casimir invariants revealing a deep link to the statistical mechanics of the 2-D Euler flow \cite{RSM}.

The structure of the paper is as follows.  
It consists of two main parts: one focusing on spontaneous stochasticity ($\SP$ later on) for the Armstrong-Vicol passive scalar,  
and another addressing spontaneous stochasticity in a general finite-dimensional framework.

\begin{itemize}
\item[\ref{mainintro}] {\bf Introduction} \dotfill \pageref{mainintro}\\
    \begin{itemize} 
    \item[\ref{Defs}] {\bf Spontaneous stochasticity: notations and definitions}  \dotfill \pageref{Defs}
    \item[] 
    \item[\ref{introAV24}] {\bf An introduction to the passive scalar \cite{AV24}}\dotfill \pageref{introAV24}
    \item[] 
    \item[\ref{mainresults}] {\bf Main results} \dotfill \pageref{mainresults} \\
    \begin{itemize}
        \item[\ref{subsubSPAV}] {\bf Lagrangian Spontaneous Stochasticity in the AV system}  \dotfill \pageref{subsubSPAV} \\
        \begin{itemize}
            \item[\ref{par1_4_1_1}] Lagrangian-$\SP$ in the AV system\dotfill\pageref{par1_4_1_1}
            \item[\ref{par1_4_1_2}] Numerical implementation of the AV system\dotfill\pageref{par1_4_1_2}
            \item[\ref{par1_4_1_3}] Lagrangian-$\SP$: Numerical results\dotfill\pageref{par1_4_1_3}
        \end{itemize}
        \item[] 
        \item[\ref{par1_4_1_4}]  {\bf Eulerian Spontaneous Stochasticity in the AV system} \dotfill\pageref{par1_4_1_4}
        \item[] 
        \item[\ref{allprob}] {\bf All probability measures are attainable by $\SP$}
        \dotfill \pageref{allprob}
        \item[] 
        \item[\ref{RGSP}]
        {\bf $\SP$ through the lens of renormalization group (RG) formalism} \dotfill \pageref{RGSP}\\
        \begin{itemize}
            \item[\ref{par1_4_3_1}] Definitions\dotfill\pageref{par1_4_3_1}
            \item[\ref{par1_4_3_2}] Main results: RG regularizations\dotfill\pageref{par1_4_3_2}
            \item[\ref{par1_4_3_3}] Quotient topology for the Bebutov flow: a reinterpretation of Theorem \ref{M0M}\dotfill\pageref{par1_4_3_3}
        \end{itemize}
        \item[] 
        \item[\ref{CNforSP}] {\bf A necessary condition for having $\SP$}
        \dotfill \pageref{CNforSP}
    \end{itemize}
    \item[] 
    \item[\ref{perspec}]  {\bf Perspectives}\dotfill \pageref{perspec}
    \end{itemize}
    \item[] 
    \item[\ref{num_LSS}] {\bf Anomalous diffusion in \cite{AV24}: from theory to numerics} \dotfill \pageref{num_LSS}
    \item[] 
    \item[\ref{TheoESSAV_proof}]{\bf Proof of Theorem \ref{THEOESSAV}} \dotfill \pageref{TheoESSAV_proof}
    \item[]
    \item[\ref{Proof_M0M}] {\bf Proof of Theorem \ref{M0M}} \dotfill \pageref{Proof_M0M}
     \item[] 
     \item[\ref{RG_up}] {\bf $\SP$ and the RG formalism} \dotfill \pageref{RG_up}
     \item[]
     \item[\ref{Examples}] {\bf Examples} \dotfill \pageref{Examples}
     \item[]
     \item[]  Acknowledgement  \dotfill \pageref{Ack} 
     \item[]
     \item[] {\bf Appendix}
     \\
     \begin{itemize}
        \item[\ref{PFTP}] {\bf Basic Notions from Measure Theory and Dynamical Systems} \dotfill \pageref{PFTP}
        \item[] 
         \item[\ref{SP_LSP}]{\bf Equivalence of Definitions: Proposition \ref{SPDEFS}}\dotfill\pageref{SP_LSP}
        \item[] 
        \item[\ref{APP:TimeCutoff}]{\bf Choice of temporal cutoff functions}\dotfill\pageref{APP:TimeCutoff}
        \item[] 
        \item[\ref{alginv_proof}]{\bf Algebraic reparameterizations: Proposition \ref{ALGINV}}\dotfill\pageref{alginv_proof}
        \item[] 
        \item[\ref{DiniApp}]{\bf Proof of Theorem \ref{CN}} \dotfill\pageref{DiniApp}
        \item[]
        \item[\ref{Chocbar2proof}]{\bf Regularization encoding all statistical behaviors: Proposition \ref{CHOCBAR2}}\dotfill\pageref{Chocbar2proof}
        \item[] 
        \item[\ref{Kgen}]{\bf Discussion on the admissible set ${\cal K}$}\dotfill\pageref{Kgen}

         \end{itemize}
\end{itemize}
\newpage
\subsection{Spontaneous stochasticity: notations and definitions}
\label{Defs}
The goal here is to introduce a mathematical definition of the concept of {\it spontaneous stochasticity}  (we use the acronym $\SP$ later on), while remaining consistent with the interpretation commonly adopted by physicists. This task is not particularly simple, as the few rigorous definitions of $\SP$ in physics concern very specific classes of models, such as those in \cite{Drivas24}. Spontaneous stochasticity is often associated with finite-time separation of trajectories in the inviscid limit. This “Richardson-like” separation is simply a consequence of the ill-posedness of the inviscid system. A more general  approach has emerged in recent years, involving the notion of renormalization group (RG), described in Section \ref{RGSP}. It involves a group acting on the transition probabilities (Markov kernels) associated with regularizations of the inviscid problem—that is, an infinite-dimensional dynamical system defined in the space of Markov kernels.

The notion of $\SP$ introduced here is a genuine measure-theoretic formulation of a lack of selection principle (denoted $LSP$ below). It reduces to $LSP$ up to a subtlety involving asymptotically vanishing null sets relative to the ambient measure, e.g., Lebesgue; see Proposition~\ref{SPDEFS}. It can also be expressed naturally in the RG framework, showing that $\SP$ admits several mathematically equivalent formulations.

For essentially technical reasons, we restrict here to finite-dimensional dynamical systems. Nevertheless, the same definition may also be applied in the context of infinite-dimensional systems, and in particular to PDEs. In that case, one naturally speaks of Eulerian-$\SP$, as opposed to Lagrangian-$\SP$. This distinction will be discussed in Section \ref{ESS_AV}.

To fix ideas, we consider a finite-dimensional, deterministic, and autonomous setting. Let $H$ be a finite-dimensional Hilbert space, equipped with inner product $\langle x, y \rangle$ and corresponding norm $\|x\|^2 = \langle x, x \rangle$. Let $f_0 \in C_b(H; H)$. The inviscid (unregularized) system is defined by
\begin{equation} \label{P_0}\tag{${\cal P}_0$}
({\cal P}_0): \left\{
\begin{array}{ll}
\dot{x} &= f_0(x), \\
x(0) &= x_0 \in H,
\end{array} \right. \quad t \in [0, T].
\end{equation}
We are interested in the inviscid limit $\epsilon \to 0$ of a family of regularized problems:
\begin{equation} \label{P_eps} \tag{${\cal P}_\epsilon$}
({\cal P}_\epsilon): \left\{
\begin{array}{ll}
\dot{x} &= f(x, \epsilon), \\
x(0) &= x_0 \in H,
\end{array} \right. \quad t \in [0, T],
\end{equation}
where $f$ belongs to the class
\begin{equation}\label{H0}
\V = \left\{ 
f \in C_b(H\times \mathbb{R}^+; H) ~:~
\lim_{\epsilon \to 0} \|f(\cdot, \epsilon) - f_0(\cdot)\|_\infty = 0,~ 
\text{and}~ f(.,\epsilon)~\in \operatorname{Lip(H;H)}~ \forall \epsilon > 0 
\right\}.
\end{equation}
We remark that the global Lipschitz condition may be replaced by the weaker assumption of well-posedness for every $\epsilon>0$. Nevertheless, the proof of Theorem \ref{M0M} requires Lipschitz continuity to ensure that the sum of regularization in $\V$ remains
in $\V$, which is not guaranteed by well-posedness alone.
Here, well-posedness is understood in Hadamard’s sense: existence, uniqueness, and continuous dependence on initial data.
The assumption $f \in \V$ guarantees existence for all $\epsilon \geq 0$ via Peano’s theorem. We denote by
$\Phi_t[f(\cdot, \epsilon)] x_0$
the unique solution to \eqref{P_eps}  at time $t$ with initial condition $x_0$. For brevity, we sometimes write $x^\epsilon$ for the solution of \eqref{P_eps} over the interval $[0, T]$. 
Let ${\cal S}_0 \subset H$ denote the set of solutions to 
\eqref{P_0} at fixed time $t$ and initial condition $x_0$:
\be \label{S0}
{\cal S}_0 := \bigcup_{\Phi} \left\{ \Phi_t[f_0(\cdot)]x_0 \right\}.
\de
It is a classical result that ${\cal S}_0$ is compact in $H$. Moreover, by Kneser’s theorem, ${\cal S}_0$ is also connected. When $H$ is infinite-dimensional like for PDEs, compactness of ${\cal S}_0$ is no more guaranteed.
\bigskip
In the following, several definitions are introduced.
\begin{definition}[$LSP$]\label{LSPdef}
A lack of selection principle applies when there exists at least
two vanishing sequences $\epsilon_n, \epsilon_n' \to 0$ such that the corresponding solutions $x^{\epsilon_n}$ and $x^{\epsilon_n'}$ converge uniformly in $C([0,T]; H)$ to distinct limits $x_1 \neq x_2$.
\end{definition}
 We now provide another definition.
Let $t>0$ fixed and
\be \label{gamma}
\gamma: \mathbb{R}^+ \to H: \epsilon \mapsto \Phi_t[f(\cdot,\epsilon)]x_0.
\de 

In the following, we refer to such curves $\gamma$ as
\emph{regularization curves}, as they encapsulate the effect of a given regularization on the system at time $t$ and initial condition $x_0$.
\begin{definition}[$\LSPO$] \label{LSPOdef}
The family of problems $({\cal P}_\epsilon)$ is said to exhibit \emph{lack of selection principle} ($\LSPO$) if there exist $t \in (0, T)$, $x_0 \in H$, and an observable ${\cal O} \in C_b(H; \mathbb{R})$ such that
\be 
\liminf_{\epsilon \to 0} {\cal O}(\gamma(\epsilon)) < \limsup_{\epsilon \to 0} {\cal O}(\gamma(\epsilon)).
\de
\end{definition}

We now adopt a measure-theoretic perspective. We refer to Appendix \ref{PFTP} for a basic introduction to the main notions. We define first ${\cal M}_0$ as  the space of Borel probability measures supported on ${\cal S}_0$ defined in (\ref{S0}), endowed with the weak topology:
\be \label{M0}
{\cal M}_0 := {\cal P}({\cal S}_0),~{\cal S}_0 :=  \bigcup_{\Phi} \left\{ \Phi_t[f_0(\cdot)]x_0 \right\}.
\de 
Convergence $\mu_\epsilon \rightharpoonup \mu$ means
$
\langle \mu_\epsilon, F \rangle \to \langle \mu, F \rangle \quad \text{for all } F \in C_b(H; \mathbb{R}),
$
where $\langle \mu, F \rangle := \int_H F(x)\, \mu(dx)$. In particular, for every Borel set $B \subset H$, we have $\mu(B) = \int_B \mu(dx)$ and $\mu({\cal S}_0) = 1$.
We also define the normalized Lebesgue measure on the interval $[0, \epsilon]$ by
\be \label{Leb}
\mathrm{Leb}_\epsilon(B) := \frac{1}{\epsilon} \mathrm{Leb}(B \cap [0, \epsilon]), \quad \text{for every Borel set } B \subset \mathbb{R}.
\de 

We consider the pushforward of $\mathrm{Leb}_\epsilon$ by the function $\gamma$ in~\eqref{gamma}, namely $\gamma_\#\mathrm{Leb}_\epsilon$. For all test functions $F \in C_b(H;\mathbb{R})$, this is given by the Ces\`aro mean (or Birkhoff average)
$$
\langle \gamma_\# \mathrm{Leb}_\epsilon, F \rangle = \frac{1}{\epsilon} \int_0^\epsilon F(\gamma(s))\, ds.
$$
In physical terms, this corresponds to taking the Lebesgue measure on the regularization parameter $\epsilon$ as the ambient measure and examining the system’s statistics at a fixed time via the pushforward $\gamma_\#\mathrm{Leb}_\epsilon$. As $\epsilon \to 0$, this provides a precise and natural meaning for the probability of finding $\gamma(0)$ at a given location in $H$.

The family of measures $\gamma_\# \mathrm{Leb}_\epsilon$ is tight, ensured by the boundedness of $\gamma$. Consequently, as $\epsilon \to 0$, Prokhorov's Theorem guarantees convergence at least along subsequences. However, uniqueness is not necessarily assured; in other words, the weak limit may not exist in general, except along subsequences. Based on this simple remark, one can now define the concept of spontaneous stochasticity ($\SP$) which has two different colors:
\begin{definition}[$\SP$]\label{SPdef}
The family (\ref{P_eps})  is said to exhibit \emph{spontaneous stochasticity} ($\SP$) if
$$
\SP := 
\left\{
\begin{aligned}
&\text{Strong} && \text{if } \gamma_\# \mathrm{Leb}_\epsilon \rightharpoonup \mu_0 \text{ as } \epsilon \to 0, \text{ and } \mu_0 \text{ non-Dirac}, \\[6pt]
&\text{Weak} && \text{if } \gamma_\# \mathrm{Leb}_\epsilon \text{ admits only weakly convergent subsequences}.
\end{aligned}
\right.
$$
\end{definition}
\begin{remark}
The notion of Weak~$\SP$ has recently emerged in the context of shell models using a renormalization-group formalism in \cite{AM_24}, \cite{AM_24end}. An example is given in Section \ref{Ex1}. 
Perhaps unexpectedly, $\SP$ is not equivalent to $LSP$. The underlying reason lies in the mismatch between a measure-theoretic concept ($\SP$) and a classical one ($LSP$): indeed, $LSP$ may occur on a set whose relative Lebesgue  measure vanishes asymptotically. 
These are characterized by the property that there exists $x^\star \in {\cal S}_0$ such that $\gamma_\# \operatorname{Leb}_\epsilon \rightharpoonup \delta_{x^\star}$, while still allowing different subsequences to converge to distinct solutions!
An explicit example is provided in Appendix~\ref{DiracLSP}. 
\end{remark}
\bigskip

Let us denote $LSP \setminus \delta,\LSPO \setminus \delta$ the lack of a selection principle excluding convergence of the associated measures to a Dirac mass. We place no restriction on $H$ beyond being a Polish space on which the solutions of \eqref{P_eps} take values. Definitions \ref{SPdef}, \ref{LSPdef}, and \ref{LSPOdef} remain unchanged. The regularization curve $\epsilon \mapsto \gamma(\epsilon)$ is assumed bounded and continuous on $(0,\infty)$, i.e., $\gamma \in C_b((0,\infty);H)$, reflecting the (local) well-posedness of \eqref{P_eps}.

\begin{proposition}\label{SPDEFS}
Let $H$ be a Polish space. Assume $\gamma \in C_b((0,\infty);H)$ and that the family $\big\{\gamma_\# \operatorname{Leb}_\epsilon\big\}_{\epsilon>0}\subset \mathcal P(H)$ is tight. Then the following conditions are equivalent:
\begin{equation}\label{spst_lsp_equal}
LSP \setminus \delta \Longleftrightarrow \LSPO \setminus \delta \Longleftrightarrow \SP.
\end{equation}
When $H$ is finite-dimensional, tightness is automatic, and the three notions in \eqref{spst_lsp_equal} coincide unconditionally. In infinite dimension, in the absence of tightness one still has
$$
\SP \Longrightarrow LSP \Longleftrightarrow \LSPO,
$$
where the reverse implication may fail.
\end{proposition}\noindent
{\bf Proof}: see Appendix \ref{SP_LSP} and Fig. \ref{tricho}. 

\begin{figure}[htbp]
\centerline{\includegraphics[width=0.60\columnwidth]{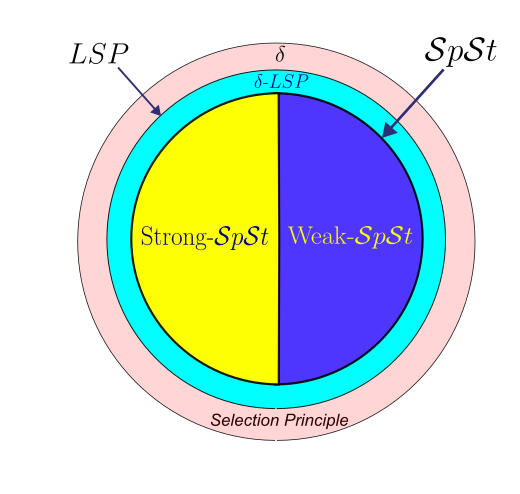}}
\caption{Mutually exclusive scenarios for the inviscid limit. 
$\delta\text{-}LSP$ refers to cases having a weak limit being Dirac but at the same time a lack of selection principle. It arises when $LSP$ occurs on a set whose relative
measure w.r.t. to the ambient one goes to zero in the inviscid limit; see Appendix \ref{DiracLSP}.
Classical
$LSP = \{ \delta\text{-}LSP\}~ \bigcup ~\{ \text{Strong-}\SP\} ~\bigcup ~\{ \text{Weak-}\SP \}$ and
Proposition \ref{SPDEFS} is $\SP\Longleftrightarrow LSP \setminus \delta \Longleftrightarrow 
\LSPO \setminus \delta$.}
\label{tricho}
\end{figure}

An important question is to determine which class of observables are eligible for $\LSPO$ when $\SP$ holds.
Assume for instance that one has Strong $\SP$ with selected non-Dirac measure $\mu_0$,
and that the system possesses certain symmetries, leading to conserved quantities.
Let $x \in H$ and let $g: H \to H$ be a group action on $H$.
If $g_{\#} \mu_0 = \mu_0$ and $\mu_0$ is ergodic with respect to $g$
(see Appendix \ref{PFTP}),
then every observable satisfying $\mathcal{O}(g(x)) = \mathcal{O}(x)$
cannot satisfy $\LSPO \setminus \delta$,
since such an observable must be constant $\mu_0$-a.e., that is,
constant on the support of $\mu_0$.
However, there always exists a nontrivial observable that satisfies $\LSPO$:
\be \label{obsuniv}
\forall x^\star \in \operatorname{supp} \mu_0,~~ {\cal O}^\star:H \to \mathbb{R}, {\cal O}^\star(x) = \| x - x^\star \|~\text{is eligible for}~
\LSPO.
\de 
This result is indeed true in any Banach space where $\SP$ holds, no matter the dimension. An obvious consequence is a sufficient condition for having anomalous diffusion: assume that one of the selected solution identically vanishes at time $t$, i.e. $0 \in \bigcup_{\mu \in \mathcal{M}(\gamma)} \operatorname{supp}(\mu)$ then $\limsup_{\epsilon \to 0} \| \gamma(\epsilon) \| > 0$.
%We have therefore as a direct consequence of \eqref{obsuniv} and %Proposition \ref{SPDEFS}:
%\begin{corollary}\label{zeroinsupp}
%Assume that $\SP$ holds. Let $\mathcal{M}(\gamma)$ denote the set of all subsequential weak limits of $\gamma_{\#} \mathrm{Leb}_{\epsilon}$ (this set is a singleton in the case of Strong-$\SP$). If
%$$
%0 \in \bigcup_{\mu \in \mathcal{M}(\gamma)} \operatorname{supp}(\mu),
%$$
%then
%\begin{equation}
%\liminf_{\epsilon \to 0} \|\gamma(\epsilon)\|
%\;<\;
%\limsup_{\epsilon \to 0} \|\gamma(\epsilon)\|.
%\end{equation}
%\end{corollary}
%This corresponds to the situation where there exists at least one selected, nonzero solution of the inviscid problem~\eqref{P_0} that vanishes at time~$t$ (it does not imply that the zero solution itself must occur). In full generality, the converse of this corollary does not hold.
%We can also indeed claim another more general corollary:
\begin{corollary}[Projection]\label{SpSt_O}
Let $G$ be some Polish space and ${\cal O}: H \mapsto G$ in $C_b({\cal S}_0;G)$ and assume that
Strong-$\SP$ holds, then provided ${\cal O}$ is not $\mu_0$-a.e. constant, one has
\be 
({\cal O} \circ \gamma)_\# \operatorname{Leb}_\epsilon \rightharpoonup
{\cal O}_\# \mu_0~\text{non-Dirac}.
\de 
One will say that system \eqref{P_eps} is ``Strong-$\SP, {\cal O}$-stable".
\end{corollary}
\begin{proof}
It is a direct consequence of the continuous mapping theorem; see Appendix \ref{PFTP}. Since ${\cal O}$ is continuous, it gives $({\cal O} \circ \gamma)_\# \operatorname{Leb}_\epsilon \rightharpoonup
{\cal O}_\# \mu_0$. From the hypothesis that ${\cal O}$ is not $\mu_0$-a.e. constant, the limit is non-Dirac.
\end{proof}
\bigskip

Definition \ref{SPdef} depends on the choice of an ambient measure, here the Lebesgue measure used to define $\operatorname{Leb}_\epsilon$.  One might instead seek a more flexible formulation.  The following example shows that Lebesgue measure need not be the most natural choice.  Let $
\gamma(s)=e^{i\log s}$, 
and consider analytic test functions 
$
F(z)=\sum_{k\ge0}F_k\,z^k,\quad 
F_k=\frac{1}{2\pi}\int_{0}^{2\pi}F(e^{i\theta})\,e^{-ik\theta}\,d\theta.
$
A direct calculation yields
$
\bigl\langle\gamma_{\#}\mathrm{Leb}_\epsilon,\;F\bigr\rangle
=\frac{1}{\epsilon}\int_{0}^{\epsilon}F\bigl(e^{i\log s}\bigr)\,ds
=\sum_{k\ge0}\frac{F_k}{1+ik}\,e^{ik\log\epsilon},
$
so that the limit as $\epsilon\to0$ exists only along subsequences.  By contrast, if one takes 
$
\mathbb{P}_\epsilon
=(e^{-1/x})_{\#}\mathrm{Leb}_\epsilon,
$
then 
$
\gamma_{\#}\mathbb{P}_\epsilon\rightharpoonup\delta_0,
$
the Haar measure on $\mathbb{S}^1$ (whose real part obeys the arcsin law).  Thus strong $\SP$ can be obtained simply by modifying the ambient measure.  Since 
$
(\gamma\circ h)_{\#}\mathrm{Leb}_\epsilon
=\gamma_{\#}\bigl(h_{\#}\mathrm{Leb}_\epsilon\bigr),
$
this adjustment is equivalent to reparameterizing the regularization curve.
\\\\
Motivated by this remark, one enlarges the class of admissible ambient measures. A relevant set is
\be 
\mathcal A
=\bigl\{ \{ \mathbb P_\epsilon \}_{\epsilon > 0}\in\mathcal P(\mathbb R^+)\colon
\mathbb P_\epsilon\ll\mathrm{Lebesgue},\ 
\mathbb P_\epsilon\rightharpoonup\delta_0\bigr\},
\de 
We consider the alternative definition \ref{SPdef}
\begin{definition}[General definition $\SP$]\label{SPDEF4}
The family \eqref{P_eps} exhibits \emph{spontaneous stochasticity} ($\SP$) if 
$$
\SP=
\begin{cases}
\text{Strong}&\displaystyle
\exists\{\mathbb P_\epsilon\}_{\epsilon>0}\in \mathcal A\colon
\gamma_{\#}\mathbb P_\epsilon\rightharpoonup\mu_0,\ 
\mu_0\text{ non‐Dirac},\\[6pt]
\text{Weak}&\displaystyle
\forall\{\mathbb P_\epsilon\}_{\epsilon>0}\in \mathcal A,\ 
\gamma_{\#}\mathbb P_\epsilon\text{ admits only subsequential weak limits}.
\end{cases}
$$
\end{definition}
\begin{remark}
 There is no general consensus on the choice of $\mathcal A$ beyond the requirement $\mathbb{P}_0=\delta_0$, which is trivially needed to measure statistics for the inviscid problem. However, if one drops absolute continuity with respect to Lebesgue, pathological situations can arise. For instance, if $\mathbb{P}_\epsilon=\delta_\epsilon$, then Strong~$\SP$ is empty; only Weak~$\SP$ is possible. Theorem~\ref{THEOESSAV} is a nontrivial instance in which Strong~$\SP$ holds for some measure family $(\mathbb{P}_\epsilon)_{\epsilon>0}\in\mathcal A$, in the sense of Definition~\ref{SPDEF4}. Assuming tightness holds, Definition \ref{SPDEF4} implies the trichotomy:
 \begin{itemize}
     \item[] (i) Strong $\SP$: there exists a family in ${\cal A}$ such that the pushforward limit is non-Dirac
     \item[] 
     \item[] (ii) Weak $\SP$: for all families in ${\cal A}$, only subsequential pushforward limits exists
     \item[] 
     \item[] (iii) There is at least one $\{\mathbb{P}_\epsilon\}_{\epsilon > 0} \in {\cal A}$ such that the pushforward family converges to a Dirac mass, and all convergent limits are necessarily Dirac (hence, excluding Strong $\SP$). It does not rule out $LSP$ with Dirac limits.
 \end{itemize}

\end{remark}

\begin{remark}
There is yet another alternative formulation using the framework of large-deviation theory, where ${\cal S}_0$ is identified as a set of "ground states". We do not discuss this interesting approach and refer to \cite{Eyink_Bandak20} for more details.
\end{remark}
\subsection{An introduction to the passive scalar \cite{AV24}}
\label{introAV24}
We consider the following Cauchy problem for the advection-diffusion 
equation:
\be \label{ad_dif0}
({\cal P}_\kappa): \left\{ 
\begin{array}{l}
\partial_t \theta^\kappa + {\bf b} \cdot \nabla \theta^\kappa = \kappa \Delta
\theta^\kappa ~~{\rm in}~ (0,\infty) \times \mathbb{T}^2 \\\\
\theta^\kappa(0,\cdot)  = \theta_0~~{\rm on}~~\mathbb{T}^2
\end{array}\right..
\de  
The energy equality for a divergence-free vector field ${\bf b}$ writes as
\be \label{energyeq}
||\theta_0||^2_{L^2} - ||\theta^\kappa(1,\cdot)||^2_{L^2} = 2 \kappa ||\nabla \theta^\kappa||_{L^2((0,1)\times \mathbb{T}^2)}^2.
\de 
In particular, if the velocity field ${\bf b}$ has enough spatial regularity, say 
in $L_t^1 C_x^{0,1}$, then $\lim_{\kappa \to 0} \kappa ||\nabla
\theta^\kappa||_{L^2}^2 = 0$, implying that the energy is conserved. The main question is to understand the limit $\kappa \to 0$ of the passive
scalar $\theta^\kappa$ when the velocity field lacks regularity, typically when it is only H\"older continuous and DiPerna-Lions theory does not apply.
\\
For the sake of consistency, we explicitly recall the main Theorem 1.1 of \cite{AV24}:
\begin{theoremav}[Anomalous dissipation of scalar variance] Let
$d \geq 2$ and $\alpha \in (0,1/3)$. There exists a vector field
\be \label{bdef}
{\bf b} \in C_t^0 C_x^{0,\alpha}([0,1] \times
\mathbb{T}^d) \cap C_t^{0,\alpha} C_x^0([0,1]\times \mathbb{T}^d)
\de  
which satisfies $\nabla \cdot {\bf b}(t,\cdot) = 0$, $\forall t \in 
[0,1]$ such that,
for every mean-zero initial datum $\theta_0 \in H^1(\mathbb{T}^d)$, the family of unique solutions $\{\theta^\kappa\}_{\kappa > 0} \in C([0,1];L^2(\mathbb{T}^d))$ of (\ref{ad_dif0}) satisfy
\be \label{anomal0}
\limsup_{\kappa \to 0} 
\kappa ||\nabla  \theta^\kappa||^2_{L^2((0,1)\times \mathbb{T}^d)}
\geq \varrho^2 ||\theta_0||^2_{L^2(\mathbb{T}^d)},
\de 
for some constant $\varrho = \varrho(d,\theta_0) \in (0,1]$ which depends only on 
$d$ and the ratio $||\theta_0||_{L^2(\mathbb{T}^d)}/||\nabla \theta_0||^2_{L^2(\mathbb{T}^d)}$.
\end{theoremav}
The loss of regularity in the velocity field ${\bf b}$ leads to a breakdown of energy conservation in the limit -- manifested as an effective anomalous dissipation term (\ref{anomal0}) in the transport equation.
Although the proof is quite involved, it yields an explicit and 'controlled' construction of the transport PDE, typical of convex integration approaches although it differs in many aspects (see Subsection \ref{subsec_convex}), making it especially valuable. 
As explained next, the strength of the derived estimates enables to state much more than what the Theorem 1.1 is claiming.
\\\\
The main idea, as developed in~\cite{AV24}, is to demonstrate the existence of a well-posed advection--diffusion PDE
that is close to~(\ref{ad_dif0}) in a well-defined sense.
This {\it effective} system is defined in the following way:
\be \label{tm}
\left\{\begin{array}{l}
\partial_t \theta_m + {\bf b}_m \cdot \nabla \theta_m = \kappa_m(\kappa) \Delta \theta_m~~{\rm in}~~
(0,\infty) \times \mathbb{T}^2, \\\\
\theta_m(0,\cdot) = \theta_0
\end{array}\right.,
\de
where $m \in \{m^\star,\cdots,M\}$ and $\kappa_m$
are \emph{renormalized diffusivities} satisfying
\begin{equation}\label{avseq}
\kappa_{m-1}  =  \kappa_m + c_0 \frac{\epsilon_m^{2\beta}}{\kappa_m},~~
\kappa_M  =  \kappa \in {\cal K}~\text{with}~{\cal K}:=\bigcup_{m \geq 1} I_m,\quad I_m =  \left[\frac12 \sqrt{c_0} \epsilon_m^{\frac{2 \beta}{q+1}},2 \sqrt{c_0} \epsilon_m^{\frac{2 \beta}{q+1}}\right],
~c_0 = \frac{9}{80},
\end{equation}
and
\be \label{2_8av}
\epsilon_m^{-1} = \big \lceil \Lambda^{\frac{q^m}{q-1}} \big \rceil ,~q = \frac{\beta}{4(\beta-1)},~\beta = \alpha + 1.
\de 
System (\ref{tm}) can  be seen as a {\it regularized}  version of (\ref{ad_dif0}) and
describes the effective behavior of the passive scalar at scales larger than $\epsilon_m$.
The key mechanism is that the scale $\epsilon_m$
can homogenize at scale $\epsilon_{m-1}$ {The key mechanism is that the problem at scale $\epsilon_m$ with diffusivity $\kappa_m$
can homogenize to the one at (larger) scale $\epsilon_{m-1}$ with (larger) diffusivity $\kappa_{m-1}$ } provided well-defined constraints are met. More explanations are given in 
Section \ref{sechom}; see also \cite{AV23,AV24,AAV25}.
In order to achieve this, a multiscale fractal vector field ${\bf b}$ must 
be defined in an ad-hoc recursive way, similar to the convex integration schemes (see Section \ref{subsec_convex}), namely
$
{\bf b}_m = \sum_{k = 0}^m {\bf v}_k,
$
with ${\bf b}_0 = 0$ and where ${\bf v}_k \in C^\infty$ consists of many oscillatory components at scale $\epsilon_k$. It is shown in \cite{AV24}
that ${\bf b}:=\lim_{m \to \infty} {\bf b}_m$ is a well-defined space-time periodic divergence free velocity field and  satisfies (\ref{bdef}). In addition, the velocity field is shown (Proposition 2.8 in \cite{AV24}) to satisfy the incompressible Euler-Reynolds equation with right-hand side, the traceless tensor field $\mathring{\bf R}$:
$$
\partial_t {\bf b} + \operatorname{div}({\bf b} \otimes {\bf b}) + \nabla p = \operatorname{div} \mathring{\bf R},~\operatorname{div}{\bf b} = 0,
$$
with
\be 
||p||_{L^\infty(\mathbb{R};C^{0,\beta'}(\mathbb{R}^2))} + ||\mathring{\bf R}||_{L^\infty(\mathbb{R};C^{0,\beta'}(\mathbb{R}^2))} \leq C
\Lambda^{-\frac{q}{q-1}(2(\beta-1)-\beta')} + C \Lambda^{-\frac{2q}{q-1}(2(\beta-1)+ \frac{2\delta}{q}-\beta')},~\beta' < 2(\beta-1) + \frac{2 \delta}{q}.
\de 
The effective system (\ref{tm}) and renormalized diffusivity sequence (\ref{avseq}) yield a proof that there is a lack of selection principle of the transport equation $({\cal P}_\kappa)$ as we will discuss later.
We introduce another set of parameters in \cite{AV24}:
\begin{table}[ht]
  \centering
  \caption{Summary of parameters used in \cite{AV24} and their definitions.}
  \label{tab:parameters}
  \begin{tabular}{|l||l|l|l|l|}
    \hline
     & definition & $\inf \alpha$ & $\sup \alpha$ & meaning \\ 
    \hline 
    $\alpha$   & $\alpha$                              & $0$           & $\tfrac13$     & velocity Hölder exponent          \\ 
    $\beta$    & $\alpha + 1$                          & $1$           & $\tfrac43$     & streamfunction regularity         \\ 
    $q$        & $\dfrac{\beta}{4(\beta - 1)}$         & $+\infty$     & $1$            & rate of scale separation          \\ 
    $\delta$   & $\dfrac{(q - 1)^2}{4(q + 1)(4q - 1)}$ & $\tfrac{1}{16}$ & $0$           &                                   \\ 
    $\gamma$   & $\dfrac{q - 1}{q + 1}\,\beta$         & $1$           & $0$            &                                   \\ 
    \hline
  \end{tabular}
\end{table}

 \begin{propositionav}[Lack of selection principle]\label{prop55}
 Fix $\beta \in [68/67,4/3)$. There exists a constant $C_\star = C_\star(\beta) \geq 1$ such that the following holds. For every parameters $A \in (0,1]$ and $B>1$, assume that $\Lambda$ is taken sufficiently large with respect to $\beta,A,B$ to ensure that
$$
C_\star \Lambda^{-\delta} \leq \min \left\{ 
A,B^{-\frac{2}{2+\gamma}(2-\beta+\frac{2\beta}{q+1})}
\right\}.
$$
 Fix the sequence $\{\epsilon_m \}_{m \geq 0}$ according to (\ref{2_8av}). Choose an initial datum $\theta_0 \in \dot{H}^2(\mathbb{T}^2)$. Define the length scale 
$L_{\theta_0}:= \frac{||\theta_0||_{\LLT}}{||\nabla \theta_0||_{\LLT}}$ 
and let $m_\star \geq 1$ be the unique integer such that 
$\epsilon_{m^\star}
\leq C_\star L_{\theta_0}^{\frac{2}{2+\gamma-2q\delta}} < \epsilon_{m^\star-1}$.
Assuming that $\theta_0$ satisfies 
\be \label{theta0set}
\frac{||\nabla \theta_0||^4_{\LLT}}{||\theta_0||^2_{\LLT} 
||\Delta \theta_0||^2_{\LLT}} \geq A,~~{\rm and}~ C_\star L_{\theta_0}^{\frac{2}{2+\gamma-2q \delta}} \leq B \epsilon_{m^\star},
\de 
there exist two sequences of diffusivities $\{\kappa_m^{(1)} \}_{m \in \mathbb{N}}$ and
$\{\kappa_m^{(2)} \}_{m \in \mathbb{N}}$, both converging to 0 as $m \to \infty$, such that the corresponding solutions $\{\theta^{\kappa_m^{(1)}} \}_m$ and $\{\theta^{\kappa_m^{(2)}} \}_m$ of the advection-diffusion equation with initial data $\theta_0$ and drift ${\bf b}$, converge as $m \to \infty$ in $C^{0,\mu}((0,1);\LLT)$ (for some $\mu > 0$)
to two distinct weak solutions of the associated transport equation.
\end{propositionav}

We stress that both Theorem 1.1 and Proposition 5.5 rely on the core Proposition 5.2 in \cite{AV24}. This proposition gives a proof that system (\ref{tm})  is such that $\theta_m$ homogenizes to $\theta_{m-1}$ with diffusivity controlled by the sequence (\ref{avseq}).

\begin{propositionavCORE}[homogenization]
  There exist constants $C(\beta) < \infty$, such that, if the minimal scale separation parameter $\Lambda$ satisfies $\Lambda \geq C$, then the following statement is valid. For every $R_{\theta_0}>0$ and $\theta_0 \in C^\infty(\mathbb{T}^2)$ having zero mean
   which satisfies the quantitative analyticity condition
  \be 
  \max_{|\alpha|=n} || \partial^\alpha \theta_0||_{\LLT} \leq ||\theta_0||_{\LLT} \frac{n!}{R_{\theta_0}^n},~\forall n \in \mathbb{N},
  \de 
  if we define
  \be 
  m_{\theta_0} := \min \left\{  m \in \mathbb{N}~:~m \geq 2, \epsilon_{m-1}^{1+\gamma/2} \leq R_{\theta_0}
  \right\},
  \de 
  then, for every $m \in \{ m_{\theta_0},\cdots,M \}$, we have the estimates
  \be 
  ||\theta_m - \theta_{m-1}||_{L^\infty((0,1);\LLT)} + \kappa_m^{1/2} || \nabla \theta_m - \nabla \tilde{\theta}_m||_{L^2((0,1) \times \mathbb{T}^2)} \leq C \epsilon_{m-1}^\delta ||\theta_0||_{\LLT}
  \de 
  \be 
  \left| \frac{\kappa_m ||\nabla \theta_m||^2_{L^2((0,1) \times \mathbb{T}^2)}}{\kappa_{m-1} ||\nabla \theta_{m-1}||^2_{L^2((0,1) \times \mathbb{T}^2)}}
  -1\right|  \leq C \epsilon_{m-1}^\delta.
  \de 
\end{propositionavCORE}
\noindent
The scalar $\tilde{\theta}_m$ is an explicit ansatz of the solution solving (\ref{tm}) than we do not describe; see Section 4.2 in \cite{AV24}. Note also that $m^\star$ in Proposition 5.5 is related to the level $m_{\theta_0}$ by the identity $m^\star=m_{\theta_0}-1$.

\subsection{Main results}\label{mainresults}

We present our principal findings. Before hand, it is essential to clarify our motivations. A sequence of recent mathematical works \cite{DeLellis2021,Drivas_Elgindi,colombo,AV23,Titi2023,burczak2023anomalous} have highlighted the absence of a selection mechanism 
($LSP$) for transport equations in the vanishing-diffusivity limit.
Notably, these phenomena are not confined to a singular
velocity field but emerge across various classes, with proof strategies that diverge markedly among the works. 
\\

The cases in \cite{AV23,AV24} and later in \cite{burczak2023anomalous} are especially notable, as they use velocity fields tied to weak solutions of the incompressible Euler equations. As a result, the absence of a selection rule feels more physical than in some other examples (see, e.g., \cite{colombo,Titi2023}). 
\begin{center}
\emph{We emphasize that these $LSP$ are direct manifestations of Eulerian spontaneous stochasticity.}
\end{center}

\begin{itemize}
\item Our first objective is to investigate the AV system numerically, both through direct numerical simulation (DNS) and by analyzing its sequence of renormalized diffusivities. The first result is a numerical illustration of Theorem~1.1, providing evidence for \emph{Lagrangian spontaneous stochasticity}; see Sections~\ref{subsubSPAV}, \ref{num_LSS}.
\item[]
\item 
We are also interested in determining whether \eqref{ad_dif0} exhibits Strong Eulerian-$\mathrm{SP}$. In general, the absence of a selection principle does not rule out either weak-$\SP$ or convergence to a Dirac mass. 
We therefore establish in Theorem \ref{THEOESSAV} that the system AV
exhibit more than a lack of selection principle, it has the Strong Eulerian-$\SP$ property; see Definition \ref{SPDEF4}. One can indeed exhibit a map $\Phi$ and an ambient probability measure
$\mathbb{P}_\kappa$ such that, as $\kappa \to 0, \kappa \in {\cal K}$
$$
\Phi_\# \mathbb{P}_\kappa \rightharpoonup \Theta_\infty~\text{non-Dirac}~
\in {\cal P}(L^2(\mathbb{T}^2)).
%LSP_{\cite{AV24}} \;\Longrightarrow\; \text{Strong Eulerian-}\SP
$$
The proof relies extensively on the proof of Proposition 5.5 \eqref{prop55} in \cite{AV24}; see Sections \ref{par1_4_1_4},\ref{TheoESSAV_proof}.
\item[] 
\item 
Independently, we study the statistical behavior in the inviscid limit of \emph{all} regularizations of an ill-posed finite-dimensional system
. It is shown that the set of probability measures on the state space of the inviscid system can be put in correspondence, in a precise sense, with the regularizations of the system (Theorem~\ref{M0M}). This provides a rigorous framework for defining universality classes of regularizations that yield the same statistical behavior. Moreover, the connection with the singularities of the system is clearly established (Theorem~\ref{CN}); see Sections \ref{allprob},\ref{CNforSP},\ref{Proof_M0M}.
\item[]
\item 
These results are further reinterpreted from the viewpoint of dynamical systems theory, establishing a precise link with renormalization-group approaches \`a la Feigenbaum, developed in recent years by A.~Mailybaev and collaborators. In particular, these results extend those of \cite{Drivas21} and \cite{Drivas24} and highlight the importance of exit times from a neighborhood of singularities. The close relationship between these exit times and rescaling is made explicit and open new perspectives; see Sections \ref{RGSP}, \ref{RG_up}, \ref{RGreg}.
\end{itemize}

\subsubsection{Lagrangian Spontaneous Stochasticity in the AV system}\label{subsubSPAV}
This numerical section is devoted exclusively to anomalous diffusion and the numerical implementation of the AV system.
\paragraph{Lagrangian-$\SP$ and anomalous diffusion are synonymous}\label{par1_4_1_1}
This property is well known for passive scalars and has been rigorously established in \cite{Drivas17}.  
For completeness, we briefly recall the reasoning here.

One considers the backward It\^o SDE associated
with the passive scalar transport equation:
\be \label{eq:BackLag}
d{\bf X}_s^\kappa = {\bf b}({\bf X}_s^\kappa,s) ds + \sqrt{2 \kappa} d{\bf W}_s, {\bf X}_t^\kappa= {\bf x}, s \leq t.
\de
Here ${\bf X}_s^\kappa := {\bf X}^\kappa(s;{\bf x},t), s \leq t$ corresponds to the particle position ${\bf X}_s^\kappa$ at a prior time $s \leq t$ such that it goes through position ${\bf x}$ at time $t$. Let us denote $\Theta_s^\kappa := \theta({\bf X}_s^\kappa,s)$ where $\theta$ satisfies the
transport equation (\ref{ad_dif0}). It satisfies a martingale property: 
$\theta^\kappa({\bf x},t) = \Theta_t^\kappa = \mathbb{E}[\Theta_s^\kappa] = \mathbb{E}[\Theta_0^\kappa] = \mathbb{E}[\theta_0({\bf X}_0^\kappa)]$.
It turns out that the fluctuations obey
$$
\Theta_t^\kappa- \Theta_0^\kappa  = \sqrt{2\kappa} \int_0^t
d {\bf W}_s \cdot \nabla \Theta_s^\kappa.
$$
It gives after squaring:
$$
{\rm Var} \left[\theta_0({\bf X}_0^\kappa)\right] = 2\kappa \int_0^t \mathbb{E}\left[~|\nabla \Theta_s^\kappa|^2 \right] ds.
$$
The fluctuation theorem of \cite{Drivas17} is the integrated version 
$\langle \cdot \rangle = \int_{\mathbb{T}^2} d{\bf x}$ over the physical domain (where one uses the divergence-free property of the velocity field):
\be
\frac12 \langle {\rm Var} \left[\theta_0({\bf X}_0^\kappa) \right]  \rangle = \kappa \int_0^t 
\langle |\nabla \theta^\kappa|^2 \rangle~ds = \kappa ||\nabla \theta^\kappa||^2_{L^2((0,t) \times \mathbb{T}^2)}.
\de

Breaking of the determinism of the Lagrangian flow map is therefore equivalent to anomalous diffusion. We therefore adopt the following definition of Lagrangian-$\SP$:
\begin{definition}[Lagrangian-$\SP$]
An advection–diffusion equation \eqref{ad_dif0} exhibits Lagrangian-$\SP$ if there exists $t > 0$ such that
$$
\limsup_{\kappa \to 0} 2\kappa \| \nabla \theta^\kappa \|_{L^2((0,t)\times \mathbb{T}^2)}^2 > 0.
$$
Equivalently, with $\gamma(\kappa) := \theta^\kappa(t,\cdot)$,
$$
\limsup_{\kappa \to 0} {\cal O}(\gamma(\kappa)) > 0, 
\qquad {\cal O}(\gamma) := \|\theta_0\|_{L^2}^2 - \|\gamma\|_{L^2}^2.
$$
\end{definition}\noindent
The distinction from  Definition \ref{LSPOdef} is that the behavior of the liminf is not specified; for instance, it may coincide with the limsup.
Theorem 1.1 then provides a proof that the AV system possesses Lagrangian-$\SP$. The relation between Eulerian and Lagrangian $\SP$ is discussed further in Section \ref{ESS_AV}.

\paragraph{Numerical implementation of the AV system}\label{par1_4_1_2}
Numerically simulating the passive scalar presents significant challenges, primarily due to extreme scale separation in the velocity field $\vect_m= \nabla \times \phi_m$. A key bottleneck arises from the hypergeometric nature of the problem. For instance, with the parameters used in the proof of anomalous diffusion in~\cite{AV24}, where $\Lambda \geq 2^7$, numerical simulations become infeasible. Relaxing the condition on $\Lambda$ enables computation but still restricts the number of accurately represented scales to at most 3--4 (see Table \ref{tab:ParamsB} of Section \ref{ALGO}).

Another major difficulty arises from the nonlocal temporal dependencies in both the past and future, preventing a straightforward forward-in-time propagation as in traditional Cauchy initial value problems. Instead, the causality structure necessitates the use of a memory-based strategy to account for these dependencies.

\begin{figure}[htpb]
%\centering
\centerline{\includegraphics[width=0.99\linewidth]{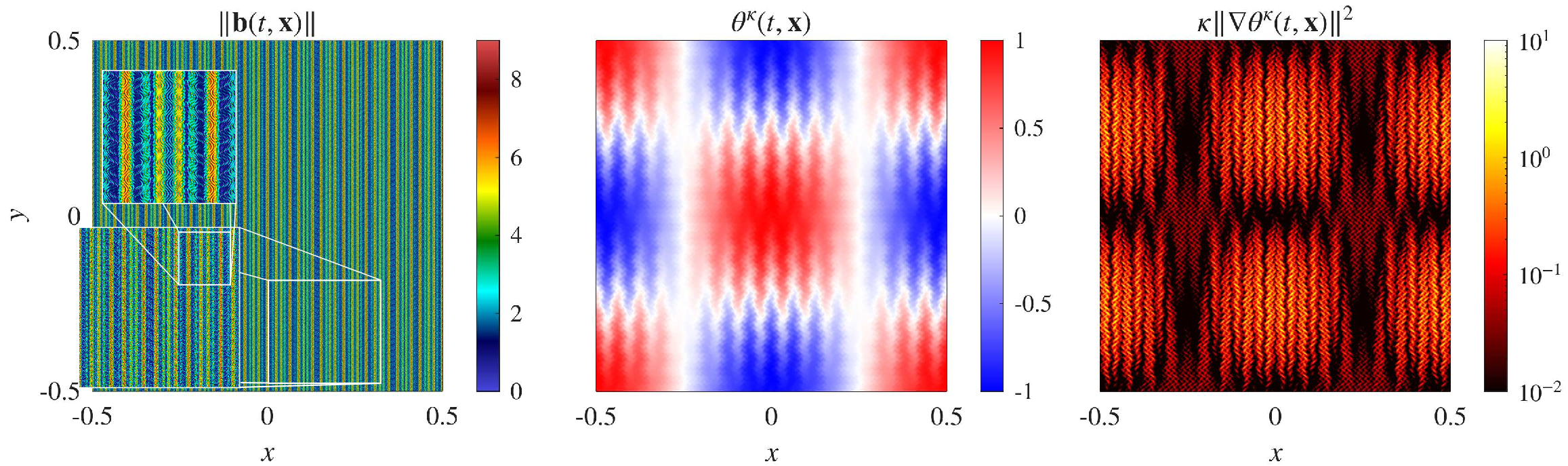}}
\centerline{\includegraphics[width=0.99\linewidth]{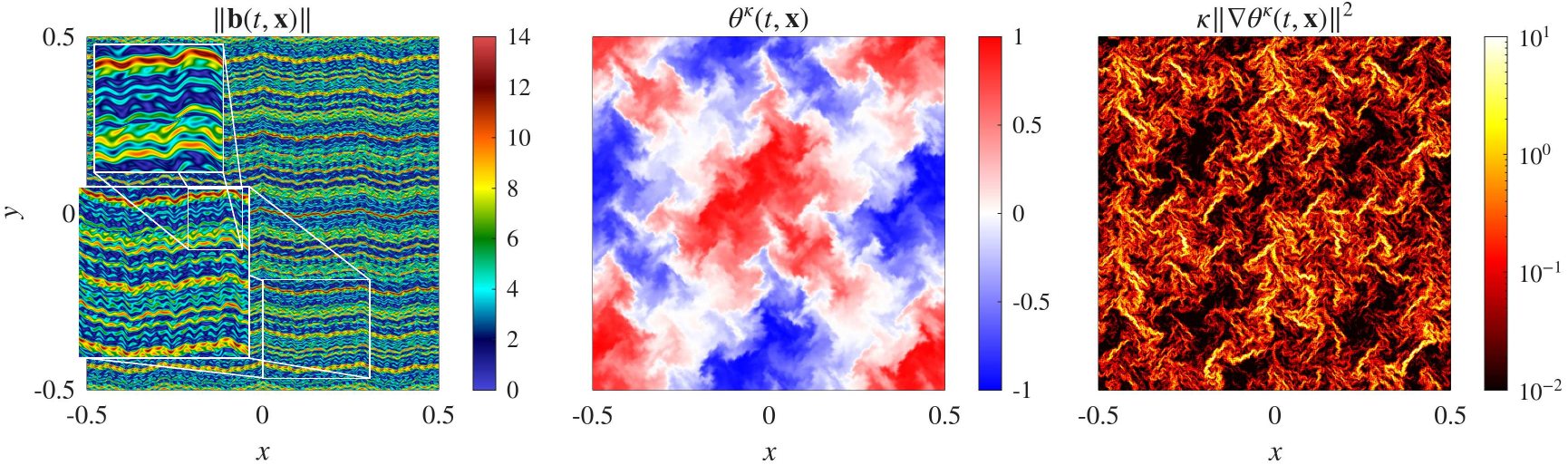}}
\caption{Illustration of the velocity and passive scalar fields, upper row is constructed with a hyper-geometric scale separation while the bottom row is the geometric case. From left to right, snapshot of: the norm of the vector field $\mathbf{b}_M$, the passive scalar field and the local dissipation rate of the passive scalar. }
\label{fig:avfield}
\end{figure}

 In Fig.~\ref{fig:avfield}, we present a typical snapshot of the multiscale vector field and passive scalar field obtained numerically.
Interestingly, all convex integration schemes encounter similar numerical implementation challenges but are subject to additional constraints (e.g. Nash decompositions). We provide an algorithmic outline of our approach, with full self-contained details Section \ref{ALGO}.

\begin{figure}[htpb]
\centering
\includegraphics[width=0.38\linewidth]{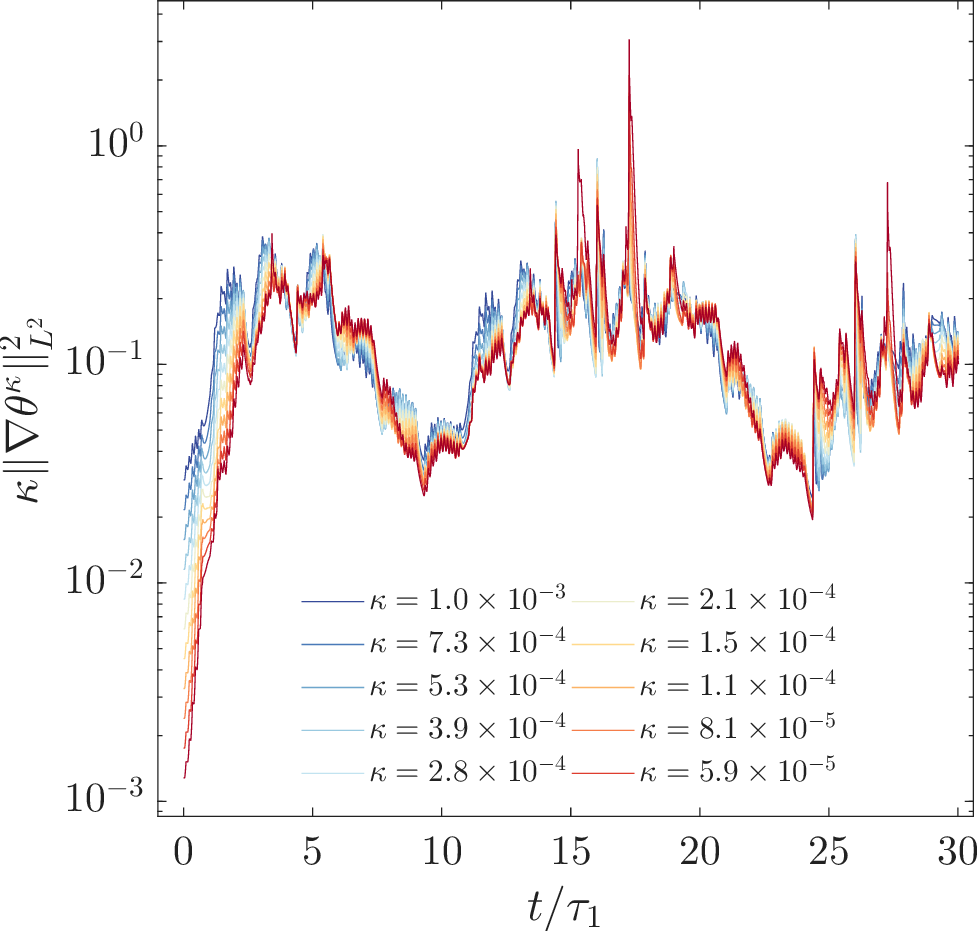}
\includegraphics[width=0.38\linewidth]{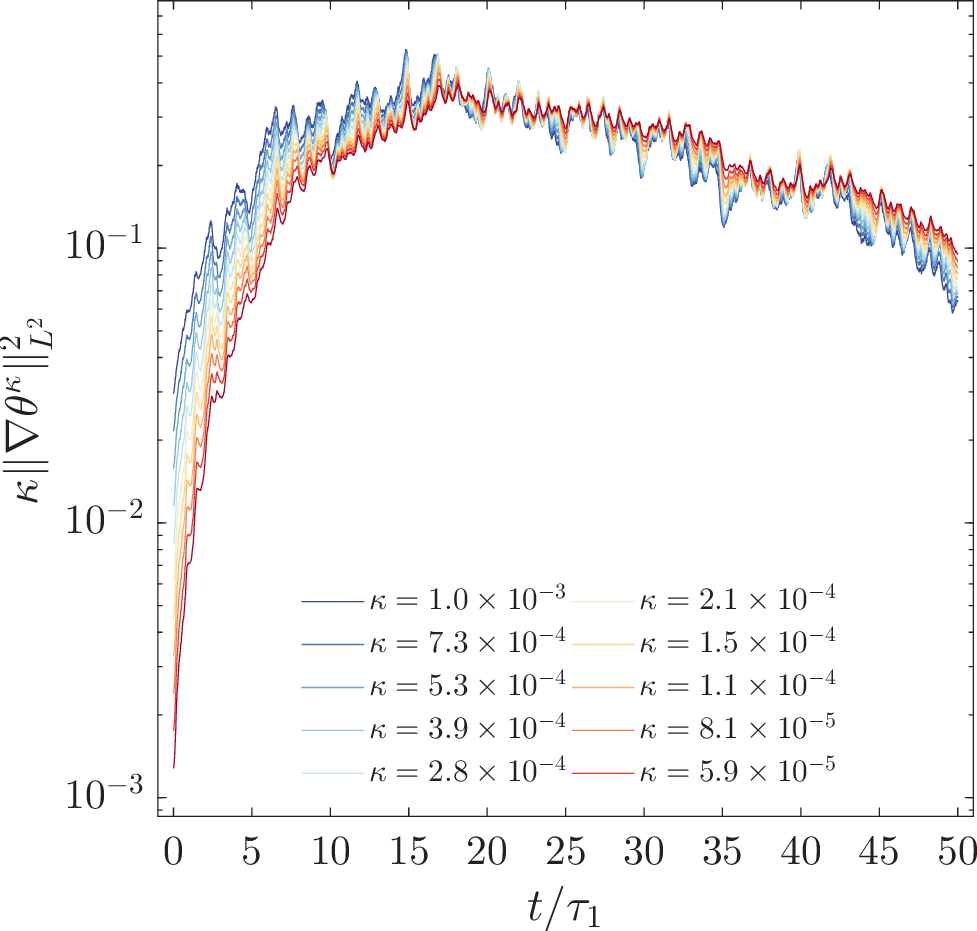}
\includegraphics[width=0.38\linewidth]{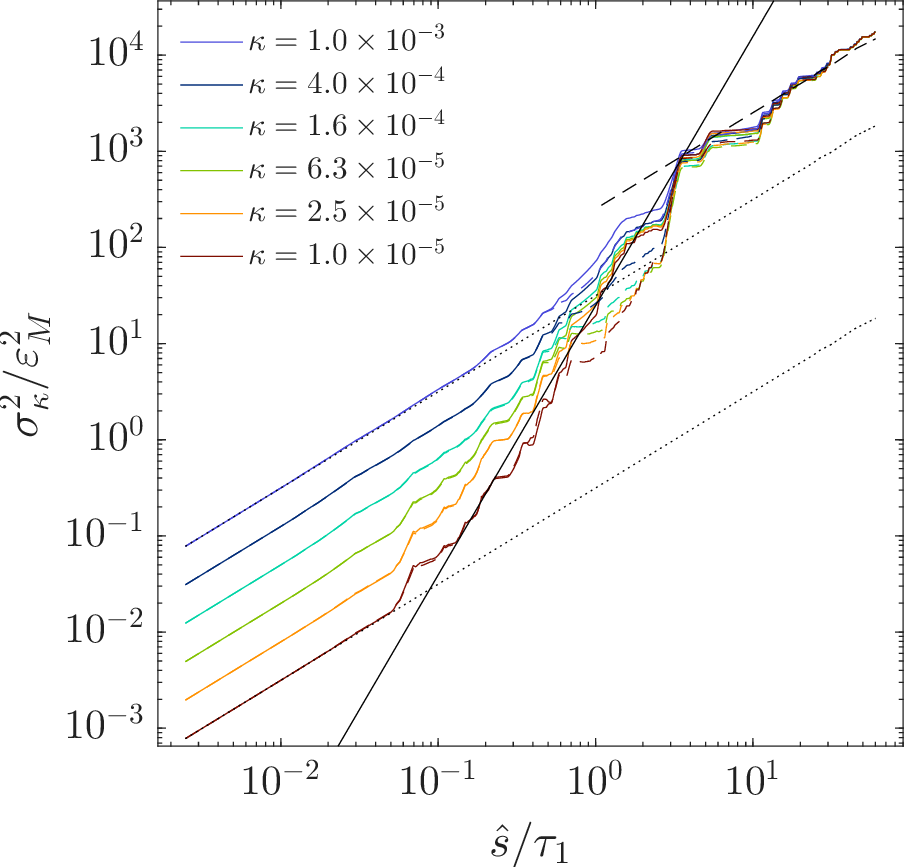}
\includegraphics[width=0.38\linewidth]{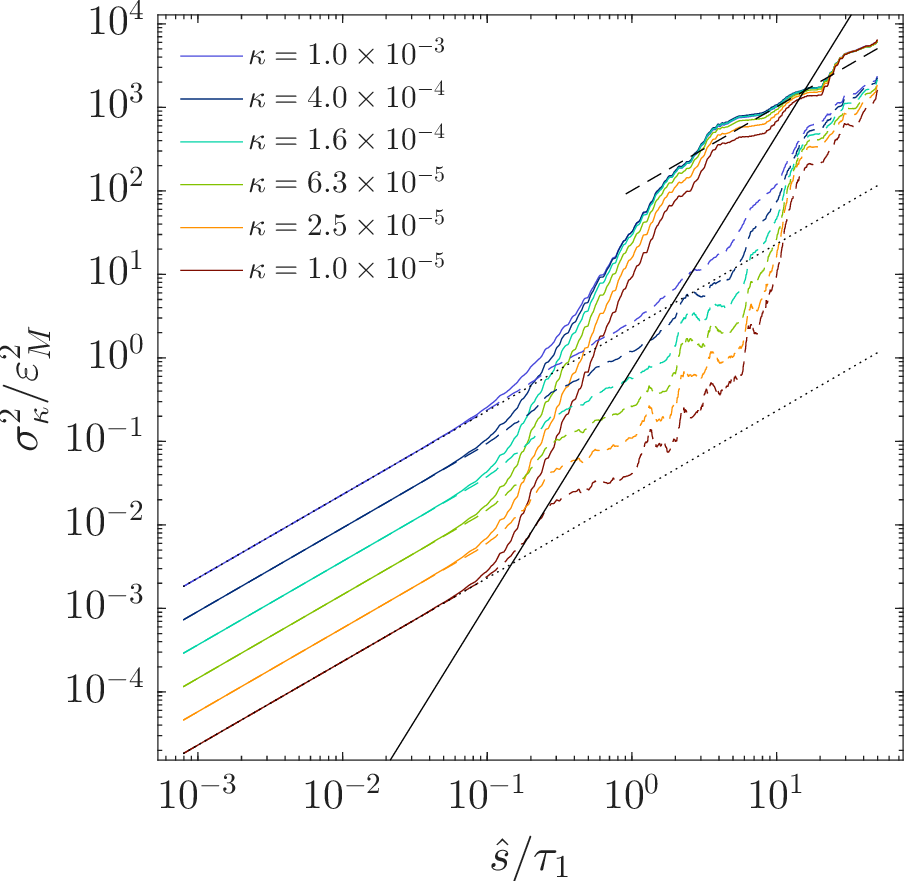}
\caption{
Top row---dissipation rate in the hypergeometric (left) and geometric (right) cases for several diffusivities $\kappa$. After a few large-eddy turnover times
$\tau_{1}$, the rate fluctuates around a finite value independent of $\kappa$; this corresponds to anomalous diffusion.
Bottom row---variance of the backward stochastic Lagrangian trajectories.
Colored curves mark different $\kappa$; solid curves are tracers that finish at
$\mathbf{x}=(0,0)$, dotted curves those ending in
$(\tfrac14,\tfrac14)$. For short times the variance of tracers position follow the behavior $8\kappa \hat{s}$ expected from diffusion alone (dotted
curves). In the inertial range they become super-diffusive; the black solid
line shows the Richardson scaling, $\hat{s}\propto t^{\nu}$ with $\nu=1 + \frac{1 +\alpha +\gamma}{1-\alpha}$ where $\alpha, \, \gamma$ are defined in Table \ref{tab:parameters}. At long times
the variance behave diffusively again;  with an affective diffusivity independent of $\kappa$.}
\label{Anom1}
\end{figure}

\paragraph{Lagrangian-$\SP$: Numerical results}\label{par1_4_1_3}
We illustrate the connection between the anomalous diffusion of the passive scalar and Lagrangian spontaneous stochasticity, as discussed in the previous section and studied in detail in~\cite{Drivas17}. 
\\\\
The upper row, left panel of Fig.~\ref{Anom1} shows the diffusion rate $2\kappa \| \nabla \theta^\kappa \|^2_{\LLT}$ as a function of time for various values of $\kappa$, in the cases of hyper-geometric scale separation (left) and geometric scale separation (right). Because the initial condition of the passive scalar equation is independent of diffusivity, the initial diffusion rate is proportional to $\kappa$. After a transient phase of order $\tau''_M$, however, the dissipation rate approaches a finite value independent of $\kappa$ in both the hyper-geometric and geometric cases, demonstrating anomalous diffusion.

The bottom row of Fig.~\ref{Anom1} (right panel) displays the variance of backward Lagrangian trajectories passing through $\xi$ at time $t_f$:
$$
\sigma^2_\kappa(s):= \mathbb{E}^{(1,2)} \left[ \left. \left| {\bf X}^{\kappa,(1)}_s - {\bf X}^{\kappa,(2)}_s \right|^2 \,\right|\, {\bf X}^{\kappa,(1,2)}_{t_f}=\xi \right], \qquad 0 \leq s \leq t_f,
$$
plotted as a function of $\hat{s}/\tau_M := (t_f-s)/\tau_M$ for different values of $\kappa$, where $t_f=0.5$ is the final time of the simulation. Numerical results are shown for hyper-geometric scale separation (left panel) and geometric scale separation (right panel). In both cases, we present results for $\xi=(0,0)$ (solid colored lines) and $\xi=(\frac14,\frac14)$ (dotted colored lines). For small values of $\hat{s}/\tau_M$, the variance scales diffusively, 
\(\sigma^2_\kappa \underset{\hat{s}\to 0}{\sim} 8 \kappa \hat{s}\),
and thus depends strongly on $\kappa$. After a few small-scale turnover times $\tau_M$, however, $\sigma^2_\kappa$ transitions into a superdiffusive regime characteristic of the Richardson regime.

As noted in \cite{AV24} (though not proven there), the expected Richardson regime depends on the regularity of the advecting vector field. Specifically, it is anticipated that 
$$
\sigma^2_\kappa \propto \hat{s}^{1+ \frac{1+\alpha+\gamma}{1-\alpha}}, 
\qquad \text{where} \quad 
\gamma = \frac{(1+\alpha)(1 - 3\alpha)}{5\alpha + 1}.
$$
The exponent of this power law equals $3$ when $\alpha=1/3$, corresponding to the classical Richardson regime of three-dimensional turbulence. The solid black line in the bottom row of Fig.~\ref{Anom1} depicts this superdiffusive scaling (up to an arbitrary multiplicative constant). 

For hyper-geometric scale separation, the numerical results show good agreement with Richardson scaling for both values of $\xi$. In the geometric case, however, the results are more contrasted: Richardson scaling does not appear to hold for certain values of $\xi$. Further experiments are needed to determine whether averaging over $\xi$ restores Richardson scaling. 

Most importantly, throughout this superdiffusive regime, the variance of Lagrangian trajectories $\sigma^2_\kappa$ becomes independent of $\kappa$, indicating that the Lagrangian trajectories are indeed spontaneously stochastic. This superdiffusive regime persists until $\hat{s}/\tau_M \simeq 10^2$, which coincides with the duration of the transient regime for the dissipation rate shown in the top row of Fig.~\ref{Anom1}.
\\

For large values of $\hat{s}/\tau_M \geq 10^2 \simeq 2 \tau_1/\tau_M$, $\sigma^2_\kappa$ once again exhibits diffusive behavior. In Fig.~\ref{Anom1}, for large $\hat{s}$, $\sigma^2_\kappa$ grows linearly with $\hat{s}$, except in the geometric case for $\xi=(\frac14,\frac14)$ (bottom row, right panel, dotted colored lines), where a clear diffusive regime is not reached within the simulation. The black dotted line in the bottom row, left panel of Fig.~\ref{Anom1} corresponds to an effective diffusivity of order $8 \times 10^{-3}$ in the hyper-geometric case, independent of $\xi$. 

In the geometric scale separation case, however, a large-time diffusive regime is reached only for $\xi=(0,0)$, with an effective diffusivity of order $4.4 \times 10^{-2}$, while for $\xi=(\frac14,\frac14)$ no clear diffusive behavior is observed at large times. We expect that with longer simulations, a diffusive regime independent of $\xi$ would be recovered in the geometric case. This question cannot be fully addressed here, since the field $\mathbf{b}_M$ is computed only up to $t_f=0.5$.

As shown in the figure, the effective diffusivity remains finite in the vanishing-diffusivity limit, consistent with a finite diffusion rate as $\kappa \to 0$. Exploring this effectively diffusive regime is essential for comparison with the renormalized diffusivity sequence of \cite{AV24}. However, due to limited spatial resolution, accurately estimating an effective diffusivity remains highly challenging. At this stage, we are unable to provide a reliable numerical measurement of these diffusivities.

In summary, our numerical results of the Armstrong–Vicol model are consistent with the prediction of~\cite{AV24} concerning the anomalous diffusion of the passive scalar. Furthermore, by examining the Lagrangian counterpart of the passive scalar, we have gained additional insight into the spontaneous stochasticity of the Lagrangian tracers.
We also provide an external link to various movies, including the advection of Wandrille's cat in action :  \url{https://www.youtube.com/playlist?list=PLH-V8wbBnQjbTkHdJVxenflTYnN5gDC7B&jct=pcAkkDgSTfYZur1uhx9sCw}.

\subsubsection{Eulerian Spontaneous Stochasticity in the AV system}\label{par1_4_1_4}
\label{ESS_AV}
From the lack of selection principle established in \cite{AV24} alone, one cannot infer much about the existence of $\SP$.  
Proposition~\ref{SPDEFS} requires the crucial hypothesis that a tight
family of the form $\{ \Phi_\# \mathbb{P}_\kappa \}_{\kappa > 0}$ in ${\cal P}(\LLT)$ exists, $\Phi(\kappa)$ being the solution of 
\eqref{ad_dif0} at some time $t$. Since ${\cal S}_0$ lacks compactness, additional information on the inviscid limit is needed.
However, the sharp estimates obtained in \cite{AV24} make it possible to establish a property much stronger than $LSP$ alone:
namely, Strong Eulerian-$\SP$ in the sense of Definition~\ref{SPDEF4}.
The proof becomes relatively simple 
once the proof strategy of Proposition~5.5 is adapted.
Moreover, in addition to being Strong Eulerian-$\SP$, the system is stable with respect to the $L^2$ norm, thereby implying $\LSPO$ for the $L^2$ norm and hence Lagrangian-$\SP$.
Theorem~\ref{THEOESSAV} thus shows that
$$
\text{Strong Eulerian-}\SP, \; L^2\text{-stable} \Longrightarrow \LSPO_{L^2} 
\Longrightarrow \text{Lagrangian-}\SP,
$$
where $\LSPO_{L^2}$ means
\begin{equation} \label{eq:LSP_L2_condition}
\liminf_{\kappa\to 0} \| \theta^\kappa(t,\cdot) \|_{\LLT} 
< \limsup_{\kappa\to 0} \| \theta^\kappa(t,\cdot) \|_{\LLT}.
\end{equation}

Finally, recall that for transport equations in general, the converse implication 
``Lagrangian-$\SP \Longrightarrow$ Eulerian-$\SP$'' does not hold. A classical counterexample is the Kraichnan model, where for a fixed realization {of the advection field}
$$
\liminf_{\kappa\to 0} \| \theta^\kappa(t,\cdot) \|_{\LLT}
= \limsup_{\kappa\to 0} \| \theta^\kappa(t,\cdot) \|_{\LLT} 
= \lim_{\kappa \to 0} \| \theta^\kappa(t,\cdot) \|_{\LLT} = \lim_{\kappa \to 0} \left( \|\theta_0\|^2 - 2 \kappa \| \nabla \theta^\kappa \|_{L^2((0,t) \times \mathbb{T}^2)}^2\right)^\frac12 < \|\theta_0 \|_{\LLT}
$$
This also shows why it is essential, when referring to Lagrangian Spontaneous Stochasticity or anomalous diffusion, to specify whether the statement concerns a full limit or only a $\limsup$ limit.

\begin{theorem}[$\text{Strong Eulerian-}\SP$]\label{THEOESSAV}
Let $\theta_0,\Lambda,t$ satisfying the conditions of Proposition 5.5 \cite{AV24}.
Let $\Phi:{\cal K} \to \LLT$ such that $\Phi(\kappa)$ is
the solution of \eqref{ad_dif0} at time $t$ and initial condition $\theta_0$, with diffusivity $\kappa$. Then 
$\Phi$ is continuous on ${\cal K}$ and there exists
a measure $\mathbb{P}_\kappa \in {\cal P}({\cal K})$, with
$\mathbb{P}_0 = \delta_0$ and which is a.c. w.r.t. Lebesgue for all $\kappa \in {\cal K} > 0$, such that
\be 
\lim_{\kappa \to 0,\kappa \in {\cal K}} \Phi_\# \mathbb{P}_\kappa = \Theta_\infty \in {\cal P}(\LLT),~\text{and}~\Theta_\infty~\text{is non-Dirac}.
\de 
Therefore \eqref{ad_dif0} is Strongly Eulerian-$\SP$ in the sense of Definition \ref{SPDEF4}. Moreover, $\LSPO_{L^2}\setminus \delta$ holds.
\end{theorem}
\begin{remark}
   The proof exhibits not just one ambient measure, but many (see Lemma \ref{mylemma56}) all depending crucially on the diffusivity sequence \eqref{avseq} (in fact \eqref{trueavseq}), a diagonal extraction, and the choice of a probability measure $\pi \in {\cal P}(I_\tau)$ such that it is $\sim $ Lebesgue (this condition can be weakened). Here, $I_\tau$ is some fixed interval $I_\tau:= [\frac12,2]$
   The explicit complicated expression of the ambient measure is  given in \eqref{AVambient}.
   One can also remove the condition $\kappa \in {\cal K}$ in the formulation, by extending the ambient measure to have zero probability whenever $\kappa \notin {\cal K}$. 
\end{remark}
\begin{proof}[Idea of the proof]
It crucially relies on the proof of Proposition 5.5 \ref{prop55}. We describe the main steps.
In all the following, when we write $\kappa \to 0$, it always mean for $\kappa \in {\cal K}$.
\begin{itemize}
\item The first step is technical but necessary to proceed. It amounts
to slightly adapt Lemma 5.6 of \cite{AV24} (see \ref{lemma56}). 
We can then exhibit a family of sequences $\kappa_m$ close to \eqref{avseq}. These sequences are parametrized by $\tau \in I_\tau$ and are denoted $\kappa_m(\tau)$. They have a nearly explicit expression $\kappa_m(\tau) \approx \tau^{(-1)^m} \sqrt{c_0} \epsilon_m^{\frac{2\beta}{q+1}}$ up to negligible terms going to zero when $m \to \infty$. From Lemma 5.6, there exists a
set ${\cal J} \subset I_\tau \times I_\tau$ such that the ratio $\frac{\kappa_m(\tau)}{\kappa_m(\tau')}$ is bounded away from one, and for all
$(\tau,\tau') \in {\cal J}$. By introducing a measure $\pi \in {\cal P}(I_\tau)$ which is $\sim$ Lebesgue, e.g. a uniform measure on $I_\tau$, one obtains $(\pi \otimes \pi)({\cal J}) > 0$.
This will be used to show that $\Theta_\infty$ is non-Dirac.
\item[] 
\item We then define the family of probability measures in $L^2$ as
$$
\Theta_m:= \Phi_\# (\kappa_m)_\# \pi \in {\cal P}(L^2).
$$
In order to justify this, one needs to show that $\Phi$ is continuous on ${\cal K}$. This follows from standard Schauder estimates
using the fact that the solution of \ref{ad_dif0} belongs to 
$C_t^0 C_x^{2,\alpha} \cap C_t^{1,\alpha/2} C_x^0$. 
Of course, the main difficulty is that the map $\Phi$ is not continuous at $0 \notin {\cal K}$.
\item[] 
\item We now proceed to the limit $m \to \infty$. One can shortcut tightness proof by directly showing that $\{\Theta_m\}_{m \geq m^\star}$ is a Cauchy sequence. At this stage, the results of \cite{AV24} are crucial since
$\{\theta^{\kappa_m}\}_m$ is shown to be Cauchy sequence in $C^{0,\mu/2}([0,1];\LLT)$ for some $\mu > 0$. As a direct consequence, one can show that the
limit $m \to \infty$ yields a well-defined probability measure on $L^2$ denoted $\Theta_\infty$. 
\item[] 
\item One then show that $\Theta_\infty$ is necessarily non-Dirac. Again, \cite{AV24} are able to establish a uniform lower bound such that
in the limit $m \to \infty$, $\theta^{{\kappa_m}(\tau)}$ and 
$\theta^{{\kappa_m}(\tau')}$ are distinct provided $(\tau,\tau') \in {\cal J}$, and as a direct consequence there is a lack of selection principle. The conclusion is immediate by proceeding  by contradiction
using Lemma \ref{mylemma56}.
The final step is to build an ad-hoc ambient measure $\mathbb{P}_\kappa$. 
This is achieved by remarking that $\Theta_\infty$ is a double limit 
$\lim_{m \to \infty} \lim_{M \to \infty} \Phi_\# (\kappa_m^{(M)})_\# \pi$, 
where $\kappa_m^{(M)}$ arises from the proof of Lemma 5.6 (see \eqref{smartseq}), 
and using a classical diagonal argument.
 \end{itemize}
The proof can be found in Section \ref{TheoESSAV_proof}.
\end{proof}

\subsubsection{All probability measures are attainable by SpSt}
\label{allprob}
This part is independent of the previous discussion on the passive scalar. We are interested in the statistical behavior of regularizations of \eqref{P_0}, $\dot x = f_0(x)$, of the form \eqref{P_eps}, $\dot x = f(x,\epsilon)$ with $f \in \V$, in the inviscid limit $\epsilon \to 0$, for a fixed time $t$ and initial condition $x_0 \in H$. We assume that $H$ is finite dimensional and take the ambient measure to be $\operatorname{Leb}_\epsilon$ \eqref{Leb}. Definition \ref{SPdef} is adopted as the definition of spontaneous stochasticity $\SP$. The main result asserts that, if the inviscid system \eqref{P_0} is ill-posed, then any non-Dirac probability measure can be obtained through a suitable regularization, yielding strong $\SP$. 
	This flexibility stems from the large class of admissible regularizations $f \in \V$.

Before proceeding, we reinterpret the singular regime $\epsilon \to 0$ for \eqref{P_eps} as an asymptotic limit $\tau \to \infty$, where $\tau$ may be viewed as a Reynolds-type parameter rather than a viscosity-like parameter. This is primarily a matter of convention: the regularized system \eqref{P_eps} is now denoted $({\mathcal{P}}_\tau)$, and the normalized Lebesgue measure \eqref{Leb} is replaced by $\mathrm{Leb}_\tau$.
However, this reformulation also reveals that the problem in its native formulation $\epsilon \to 0$ can indeed be reduced to the study of genuine Birkhoff averages when $\tau \to \infty$ under algebraic rescaling.

\begin{proposition} \label{ALGINV}
Let $\gamma(s) := \Phi_t[f(\cdot, s)] x_0$ and let $g \in \alg$, where $\alg$ is the class of admissible algebraic reparameterizations defined by
\begin{equation} \label{AlgDef}
\alg := \left\{
\begin{aligned}
g \in C^2  :   g > 0, ~ g' < 0, ~ \lim_{r \to \infty} g(r) = 0, \exists\, a, b > 0 \text{ such that } g(r) = r^{-a} e^{h(r)}, 
 \sup_{r \geq r_0} |h(r)| < \infty, ~ |h'(r)| \leq \frac{C}{r^{1 + b}}
\end{aligned}
\right\}.
\end{equation}
Then the following invariance property holds:
\begin{equation}
\lim_{\epsilon \to 0} \gamma_\# \mathrm{Leb}_\epsilon 
= \lim_{\tau \to \infty} (\gamma \circ g)_\# \mathrm{Leb}_\tau
= \mu.
\end{equation}
Moreover, this convergence equality holds for all subsequential limits as well. In addition, 
the class 
\begin{equation} \label{BlgDef}
\blg := \left\{
\begin{aligned}
f \in C^2  :   f > 0, ~ f' > 0, ~\exists\, a, b > 0 \text{ such that } f(r) = r^{a} e^{h(r)}, 
 \sup_{r \geq r_0} |h(r)| < \infty, ~ |h'(r)| \leq \frac{C}{r^{1 + b}}
\end{aligned}
\right\}.
\end{equation}
is such that 
\be 
\alg  \circ \blg = \alg.
\de 
\end{proposition}
\begin{proof}
    See Appendix~\ref{alginv_proof}.
\end{proof}
As an illustration, one has for instance
$\lim_{\tau \to \infty} \frac{1}{\tau} \int_0^\tau F(e^{i \theta^\alpha}) d\theta = \frac{1}{2\pi} \int_0^{2\pi} F(e^{i\theta}) d\theta = F(0)$ independently of $\alpha$.
This is just a manifestation of the Riemann-Lebesgue and/or stationary phase lemma.
\\

The goal is to investigate the properties of the class $\V$ with respect to $\SP$ (Definition \ref{SPdef}). 
To this end, we adopt a measure-theoretic perspective, focusing on the probability measures generated by the full family of regularization curves. We recall that we have mapped the problem to the study of an asymptotic limit $\tau \to \infty$. We define the set of curves in $H$ induced by vector fields in $\V$ after being reparametrized:
\begin{equation}\label{Gamma}
\Gamma := \left\{ \gamma: \mathbb{R}^+ \to H \,:\, \exists f \in \V, g \in \alg \text{ such that } \gamma(\tau) = \Phi_t[f(\cdot,g(\tau))] x_0 \right\} \subset C_b(\mathbb{R}^+;H).
\end{equation}
We equip $\Gamma$ with the compact-open topology of uniform convergence on compact sets.
Given some $\gamma \in \Gamma$, we define the set of probability measures obtained as subsequential limits (in the weak topology):
\begin{equation}\label{Mgam}
{\cal M}(\gamma) := \left\{ \mu \,:\, \exists \tau_n \to \infty\text{ such that } \gamma_\# \mathrm{Leb}_{\tau_n} \rightharpoonup \mu \right\}.
\end{equation}
Note that this set is invariant by reparameterization by algebraic transformations in $\blg$, see Proposition \ref{ALGINV}, namely ${\cal M}(\gamma \circ g) = 
{\cal M}(\gamma)$  for all $g \in\blg$.
Moreover, this induces the straightforward equivalence relation on $\Gamma$: $\gamma_1 \sim \gamma_2 \Longleftrightarrow \exists g \in \blg $ such that $\gamma_1 = \gamma_2 \circ g$. 
\\\\
Definition \ref{SPdef} can then be reformulated as follows:
\be \label{SPdefbis}
\SP := 
\left\{
\begin{aligned}
&\text{Strong} && \quad {\cal M}(\gamma) \text{ is a singleton containing a non-Dirac probability measure in } {\cal M}_0= {\cal P}({\cal S}_0);\\
% tout a fait
&\text{Weak} && \quad {\cal M}(\gamma) \text{ is not a singleton, in which case } \gamma \text{ is said to be non-renormalizable (NR)}.
\end{aligned}
\right.
\de 
\begin{remark}
    Since one adopts a  measure-theoretic approach,  $\delta\text{-}LSP$ cannot be discriminated from a genuine selection principle. 
    As we will see next, they live at the "border" of $\cup_\gamma {\cal M}(\gamma)$ as extreme points. We also emphasize that the notion of Weak $\SP$ does not need to specify whether the accumulation measures are Dirac masses or not.
\end{remark}

\bigskip

We recall the standard notation for the convex hull:
$
\operatorname{co}(A) := \left\{ \sum_{i=1}^k \theta_i a_i \,:\, k \in \mathbb{N},~ \theta_i \geq 0,~ \sum_{i=1}^k \theta_i = 1,~ a_i \in A \right\}.
$
We obtain the following result for the set of all subsequential limits arising from regularization curves in $\Gamma$:
\begin{theorem}\label{M0M}
Let ${\displaystyle {\cal M} := \bigcup_{\gamma \in \Gamma} {\cal M}(\gamma)}$, and ${\cal E} := \{\delta_x\}_{x \in {\cal S}_0}$. Then ${\cal M}$ is compact, convex, and
\begin{equation}
{\cal M} = {\cal M}_0 = \overline{\operatorname{co}}({\cal E}).
\end{equation}
Moreover, $\forall \mu \in {\cal M}_0$, $\exists \gamma_\mu \in \Gamma$ such that
\be \label{singleton}
{\cal M}(\gamma_\mu) = \{ \mu \}.
\de 
\end{theorem}
\noindent
\begin{proof}[Idea of the proof]
The inclusion ${\cal M} \subset {\cal M}_0$ is immediate; the main difficulty lies in proving ${\cal M}_0 \subset {\cal M}$. The argument proceeds in several steps.
\begin{itemize}
\item First,  we construct, for each $x \in {\cal S}_0$, an explicit vector field in $\V$ such that the associated regularization curve $\gamma_x \in \Gamma$ satisfies ${\cal M}(\gamma_x) = \{\delta_x\}$, i.e., ${\cal E} \subset {\cal M}$. In other words, for such regularizations, one has indeed a classical selection principle. 
\item[] 
\item We then show that regularizations can be constructed to attain arbitrary finite convex combinations of these Dirac measures, so that ${\operatorname{co}}({\cal E}) \subset {\cal M}$. We thus have
$$
\operatorname{co}(\mathcal{E}) \subseteq \mathcal{M} \subseteq \mathcal{M}_0 = \overline{\operatorname{co}}(\mathcal{E}) \Longrightarrow \overline{\cal M} = {\cal M}_0 =\overline{\operatorname{co}}(\mathcal{E}). 
$$
The measures in ${\cal E}$ are often referred to as the \emph{extreme points}, or \emph{pure states}, of ${\cal M}_0$. 
\item[]
\item We then show that for all $\mu \in \overline{\cal M}$, one can construct some vector field and corresponding curve $\gamma_\mu \in \Gamma$ such that ${\cal M}(\gamma_\mu) = \{\mu\}$. The proof is more difficult and amounts to renormalize any sequence in ${\rm co}({\cal E})$ in a controlled way. The immediate consequence is that ${\cal M}$ is closed.
\item[] 
\item Since ${\cal M}_0$ is compact, we conclude that ${\cal M}$ is compact.
The detailed proof is given in Section \ref{Proof_M0M}.
\end{itemize}
\end{proof}
Two straightforward consequences of Theorem \ref{M0M} can be inferred.
\begin{corollary}\label{MND}
Assume (\ref{P_0}) admits non unique solutions. Then for every non-Dirac probability measures $\mu \in {\cal M}_0$, there exists a regularization \eqref{P_eps} (or (${\cal P}_\tau)$) such that it is strongly $\SP$ with selected measure $\mu$.  
\end{corollary}
\begin{proof}
Note that (\ref{P_0}) has a unique solution if and only if ${\cal M}_0 \setminus {\cal E} = \emptyset$, we therefore assume that non unique solutions exist.
We then take some arbitrary non-Dirac measure $\mu$ in ${\cal M}_0 \setminus {\cal E}$ so that a corresponding $\gamma_\mu$ exists such that ${\cal M}(\gamma_\mu) = \{ \mu \}$ by Theorem \ref{M0M}. Note that
$\mu$ by definition cannot belong to ${\cal E}$.
By the definition of strong $\SP$ (see Definition \ref{SPdefbis}) and \eqref{P_eps}, the Corollary follows. 
\end{proof}
Of interest is that the set ${\cal M}_0 \setminus
{\cal E}$ is indeed open, convex and 
$$
\overline{{\cal M}_0 \setminus  {\cal E}} = {\cal M}.
$$
Another consequence is the possibility to write a partition of $\Gamma$ (and $\V$) into regularizations which are strongly $\SP$, weakly $\SP$ and having a general selection principle with measure convergence to some Dirac mass. 
\begin{definition}\label{defgenesets}
Let $\mu \in \mathcal{M}_0$. The associated \emph{Birkhoff classes} are defined as follows:
\begin{equation}\label{genesets}
\begin{aligned}
\mbox{Strong class:} \quad 
\mathcal{B}_\mu &:= \left\{ \gamma \in \Gamma \;:\; \mathcal{M}(\gamma) = \{ \mu \} \right\}, \\\\
\mbox{{Weak class:}} \quad
\mathcal{B}_{\mathrm{NR}} &:= \left\{ \gamma \in \Gamma \;:\; \left| \mathcal{M}(\gamma) \right| > 1 \right\}.
\end{aligned}
\end{equation}
Any regularization in ${\cal B}_{\rm NR}$ is said {\it non-renormalizable}.
\end{definition}
From Theorem \ref{M0M}, these classes are never empty. Then, one is allowed to write the partitions:
\be \label{s_w_s}
\Gamma = \Gamma_{{\rm strong}} \bigsqcup \Gamma_{{\rm weak}} \bigsqcup \Gamma_{{\rm select}}:~
\left\{
\begin{array}{llcc}
\Gamma_{\rm strong} & := & {\displaystyle \bigsqcup_{\mu \in {\cal M}_0 \setminus {\cal E}} {\cal B}_\mu}\\\\
\Gamma_{\rm weak} & := & {\displaystyle {\cal B}_{\rm NR}} \\\\
\Gamma_{\rm select} & := & {\displaystyle \bigsqcup_{\mu \in {\cal E}} {\cal B}_\mu}
\end{array}\right..
\de 
Note that one has also invariance by algebraic reparameterizations, e.g. ${\cal B}_\mu \circ \blg = {\cal B}_\mu$ (see Proposition \ref{ALGINV}).
As expected, $\Gamma_{\rm select}$ also contains $LSP$ with Dirac limits
and can therefore be split into two subclasses, we do not write it.
This partition applies as well for the set $\V$ from the definition of $\Gamma$ (see Eq. (\ref{Gamma})), namely
\be
\V = V_{0,{\rm strong}} \bigsqcup V_{0,{\rm weak}} \bigsqcup V_{0,{\rm select}}.
\de 

Of independent interest, the following result highlights the inherent flexibility of $\SP$, showing that a single regularization can indeed realize the full spectrum of statistical behaviors.
\begin{proposition}\label{CHOCBAR2}

There exists a non-renormalizable regularization $\gamma_{\rm un}$ such that 
\be 
{\cal M}(\gamma_{\rm un}) = {\cal M}_0.
\de 
\end{proposition}\noindent
\textbf{Proof}. See Appendix~\ref{Chocbar2proof}.

\subsubsection{$\SP$ through the lens of renormalization group (RG) formalism}\label{RGSP}
This section enlarges our perspective by providing a dynamical system interpretation of $\SP$.

\paragraph{Definitions}\label{par1_4_3_1}

Renormalization-group (RG) dynamics studies the evolution of the system as the regularization parameter $\tau$ varies, rather than the physical time $t$. We represent it by a continuous semigroup $(\mathcal{R}_\tau)_{\tau\ge0}$ acting on the functional space $\mathcal{H}$ of flow maps for the regularized problem $(\mathcal{P}_\tau)$. However, in our approach, the choice of the initial condition plays a crucial role (see Section \ref{CNforSP}) which would not appear when considering flow maps.
We therefore fix $t$ and $x_0$ and restrict to the state space $H$, a choice that is both simpler and better suited to our purposes; we shall write $H$ from now on. Each operator $\mathcal{R}_\tau$ sends a state $y \in H$ to the state $\mathcal{R}_\tau y$ produced by evolving $y$ at regularization level~$\tau$.

This framework arises naturally in the study of systems as $\epsilon\to0$ (equivalently $\tau\to\infty$), where one investigates the emergent behavior in the inviscid limit.  In practice, different regularization mechanisms yield distinct semigroups $\mathcal{R}_\tau$.  By “mechanisms” we refer to dynamical rules that generate entire families of regularizations.  
For example, in shell models of turbulence, an explicit renormalization procedure with
(discrete) semigroup property is explicitly infered only if the regularization shares common symmetries with 
the inviscid system \eqref{P_0} \cite{AM_24}. Within that class, each choice of initial data for the semigroup $\mathcal{R}_\tau$ thus determines a specific regularization of \eqref{P_0}.
\\
To make things more explicit, we will consider two vector fields
$(g,f) \in H \times H$ such that, for
$t > 0,x_0 \in H$ given, 
the following holds
\begin{itemize}
\item the RG generator $g:H \to H$ is Lipschitz.
\item[] \item $f:(x,\tau,y) \in H \times \mathbb{R}^+ \times H \mapsto 
f(x,\tau,y) \in H$ is Lipschitz in the first variable $x$ for all $(\tau,y) \in \mathbb{R}^+ \times H$.
\item[] \item Consider the system: 
\begin{subequations}\label{RG_system}
  \begin{empheq}[left=\empheqlbrace]{alignat=2}
    \frac{\partial X}{\partial s} 
      &= f(X,\tau,y), 
      &\quad X(0,x,\tau,y) &= x.
      \label{RG_Phi1} \tag{\theparentequation a}\\
          \frac{\partial Y}{\partial \tau} 
      &= g(Y), 
      &\quad Y(0,y) &= y, 
      \label{RG_Phi2} \tag{\theparentequation b} 
  \end{empheq}
\end{subequations}
where $X=X(s,x,\tau,y)$ and $Y=Y(\tau,y)$, then one imposes
\be \label{RGconstraint}
 X(t, x_0,\tau,y) = Y(\tau,y).
\de 
\item[] \item The regularized vector field $f$ must be compatible with the inviscid system \eqref{P_0}:
\be 
\lim_{\tau \to \infty} \| f(\cdot,\tau,y) - f_0(\cdot) \|_\infty = 0,~\forall y \in H.
\de 

\end{itemize}
The choice of an autonomous vector field $g$ automatically makes 
the solution of \eqref{RG_Phi2} having the semigroup property w.r.t. the regularization parameter $\tau$, and we use instead the notation:
\be 
{\cal R}_\tau y:= Y(\tau,y)
\de 
The set $\V$ in \eqref{H0} becomes, still calling it $\V$:
\begin{equation}\label{H0_RG}
\V := \left\{ 
f \in C_b(H\times \mathbb{R}^+ \times H; H),
\lim_{\tau \to \infty}  \|f(\cdot,\tau,y)-f_0(\cdot)\| = 0
,\ 
\text{and } \eqref{RG_Phi1} ~\text{ is well-posed for all } \tau \geq 0, y \in H
\right\}.
\end{equation}
Similarly, the set $\Gamma$ in \eqref{Gamma} is replaced by
\be \label{Gamma_RG}
\Gamma := \{ \gamma:\mathbb{R}^+ \to H, \exists f \in \V,~y \in H~:~\gamma: \tau \mapsto {\cal R}_\tau y = X(t,x_0,\tau,y) \}.
\de 
Note that all quantities also depend on $t$ and $x_0$, but this dependence is omitted for clarity of notation.
The vector field $f$ must explicitly depend on $y \in H$, if not it gives some quantity $X$ becoming independent of $y$ and in particular, at time $t$, the fixed slice $Y$ does not depend on $y$ either. In fact, one can identify $y$ as parameterizing the whole set of regularized vector fields $f$.
Finally, system \eqref{RG_system} is well-posed and non-empty: the weak constraint \eqref{RGconstraint} is only at a given fixed slice $(t,x_0)$. 
Explicit examples are discussed in Section \ref{RGreg}.

\begin{definition}[RG flow]\label{RGflow}
A family of strongly continuous maps $({\cal R}_\tau)_{\tau \ge 0}$, ${\cal R}_\tau:{H}\to{H}$, is called an \emph{RG flow} associated with \eqref{P_0} if it satisfies
\begin{enumerate}
  \item[(i)] {\bf Semigroup}: ${\cal R}_0={\rm Id}$ and $\mathcal{R}_{\tau+\tau'}=\mathcal{R}_\tau\circ\mathcal{R}_{\tau'}$.
  \item[]
  \item[(ii)] {\bf Regularization}: The solution of \eqref{RG_Phi1} at time $t$ and initial condition $x_0$ is ${\cal R}_\tau y$. It therefore becomes increasingly singular (less regularized) as $\tau\to\infty$. 
  We will say that the regularization is {\it autonomous} and call \emph{orbit}, the curve $\gamma:\tau \mapsto {\cal R}_\tau y \in \Gamma \subset C_b(\mathbb{R}^+;{H})$.
\end{enumerate}
\end{definition}
\begin{remark}\label{RGRemarks}
\leavevmode
\begin{enumerate}

\item There is a continuum of possible semigroups 
${\cal R}_\tau$ depending on the choice of a generator $g$ in \eqref{RG_Phi2}. 

\item 
It is perfectly possible to work with \emph{discrete} dynamical systems by iterating a map: for a given ${\cal R}:H \to H$, define ${\cal R}^n := \underbrace{{\cal R} \circ \cdots \circ {\cal R}}_{n~\text{times}}$. This approach is frequently employed, for example in the study of shell models, and it allows for a broader range of dynamical behaviors. One should note, however, that in this setting, by definition,  the inviscid limit is obtained only along a given subsequence, rather than as a continuous limit.

\item

Once the generator $g$ in \eqref{RG_Phi2} of the semigroup is specified, one obtains a family of regularizations indexed by the initial condition $y \in H$, and ${\cal M}(\gamma)$ in \eqref{Mgam} can be written as ${\cal M}(y)$. We define the set of all RG regularizations for a given dynamical system ${\cal R}$ by
\be \label{MR}
\mathcal{M}_{\mathcal{R}}:=\bigcup_{y \in{H}}\mathcal{M}(y), \qquad
\mathcal{M}(y):=\Bigl\{\mu\in\mathcal{M}_0\ :\ \exists\,\tau_n\to\infty,\;\frac{1}{\tau_n}\int_0^{\tau_n}\delta_{\mathcal{R}_\tau y}\,d\tau\rightharpoonup\mu\Bigr\}.
\de

\item 
We emphasize that “RG” here refers to a direct generalization of the Feigenbaum renormalization procedure.  The historical example is the period‐doubling cascade in one‐dimensional maps, which admits a nontrivial RG fixed point with a dominant eigenvalue $
\delta \approx 4.6692\ldots$
known as the Feigenbaum constant \cite{Feigenbaum1976}.  This should not be confused with the Wilson RG scheme: the Feigenbaum transformation rescales toward finer scales (higher frequencies), whereas the Wilson method integrates out fine‐scale fluctuations to produce an effective description at coarser scales.  In other words, the two approaches run in opposite directions in scale space.
\end{enumerate}
\end{remark}

%\paragraph{Statistical Attractors and Pushforward Dynamics}

Although $(\mathcal{R}_\tau)$ acts deterministically on individual states $y \in {H}$, it induces a corresponding evolution on probability measures via its pushforward; see Definitions in Appendix \ref{PFTP}.
It allows one to study the evolution not just of individual regularizations, but of entire ensembles, thus providing a natural statistical framework.
Within this setting, we define:
\begin{definition}[Statistical Attractor]\label{DefstatAtt}
Let \(X\) be a Polish space.  Denote by ${\cal P}(X)$
the space of all Borel probability measures on \(X\), endowed with the topology of weak convergence.  Suppose
\((S_\tau)_{\tau\ge0}\) is a continuous semigroup on \(\mathcal P(X)\).
A set \(\mathcal A\subset\mathcal P(X)\) is called a \emph{statistical attractor} for \((S_\tau)\) if:
\begin{enumerate}
  \item[(i)] {\bf Compacity}: \(\mathcal A\) is nonempty and compact in the weak topology.
  \item[]
  \item[(ii)] {\bf Invariance}: It is \emph{forward-invariant}: 
    $$
      S_\tau(\mathcal A)\;\subset\;\mathcal A
      \quad\forall\,\tau\ge0.
    $$
  \item[(iii)] {\bf attraction}: Its \emph{Birkhoff basin of attraction} satisfies ${\cal B}({\cal A}) \neq \emptyset$, with
    $$
      \mathcal B(\mathcal A)
      \;:=\;
      \bigl\{\nu\in\mathcal P(X)\colon\omega(\nu)\subset\mathcal A\bigr\},~\text{with}~
      \omega(\nu)
      \;:=\;
      \bigcap_{R>0}
      \overline{\Bigl\{
        \tfrac1{R'}\!\int_0^{R'} S_\tau\nu \,d\tau 
        \;\colon\; R'\ge R
      \Bigr\}}
      \;\subset\;\mathcal P(X),
    $$
\end{enumerate}
\end{definition}
In this work, we will be concerned in particular with the choice $S_\tau := ({\cal R}_\tau)_\#$. Definition \ref{DefstatAtt} differs from the classical definition in~\cite{KARABACAK_ASHWIN_2011} by omitting the minimality assumption, making it strictly weaker than the notion of \emph{minimal attractors} introduced by Ilyashenko~\cite{arnold1999bifurcation}. This broader formulation allows us to interpret emergent statistical laws as attractors of the dynamics, encompassing not only the asymptotic behavior of deterministic regularizations as defined in~\eqref{RG_system} and the set~$\Gamma$ in~\eqref{Gamma_RG}, but also that of random regularizations.
\\

As noted in Remarks~\ref{RGRemarks}, the RG framework is a priori more restrictive and does not, in general, capture the full flexibility of Theorem~\ref{M0M}. There is, however, a method -- reminiscent of skew-product constructions -- that makes it possible to always work with a semigroup property, at the cost of lifting the problem from $H$ to the larger space $\Gamma$. The natural dynamical system acting on the full space of curves is the \emph{Bebutov flow}, a semigroup $(\mathscr{S}_\tau)$ acting by $\tau$ translation.
\begin{definition}[Bebutov]\label{Bebutov}
The \emph{Bebutov flow} $(\mathscr{S}_\tau)_{\tau \in \mathbb{R}^+}$ is the semigroup acting on the topological space $\Gamma$, endowed with the compact-open topology. It is defined by left translations
\be 
\mathscr{S}_\tau: \Gamma \to \Gamma, \quad (\mathscr{S}_\tau \gamma)(\tau') := \gamma(\tau + \tau').
\de 
In addition, define the evaluation map
\be \label{evalmap}
\Psi_\tau: \Gamma \to H, \quad \Psi_\tau(\gamma) := \gamma(\tau).
\de 
One has $\gamma(\tau+\tau_0) = \Psi_{\tau_0}(\mathscr{S}_\tau \gamma)$, and for brevity we write $\Psi := \Psi_{\tau_0}$ for some arbitrary $\tau_0 \geq 0$. The curve $\gamma$ is thus regarded as the \emph{trace} of the dynamical system $(\mathscr{S}_\tau)$. Moreover, $(\mathscr{S}_\tau)$ is a continuous semigroup on $\Gamma$ for the compact-open topology.
\end{definition}

This construction corresponds to Theorem 7.2.1 in \cite{Lasota1994} and is a classical result in ergodic theory.
This shift acts on the space $\Gamma$ of regularization curves and induces a corresponding pushforward semigroup $(\mathscr{S}_\tau)_\#$ on the space of probability measures $\mathcal{P}(\Gamma)$.

Finally, since one does not expect $\Gamma$ to be a Polish space care must be taken when considering ${\cal P}(\Gamma)$. However, since only the inviscid limit $\tau \to \infty$ ($\epsilon \to 0$ for \eqref{P_eps}) matters, one introduces the \emph{$\Psi$-topology}:
\begin{definition}[$\Psi$-topology]
Let $X$ and $Y$ be topological spaces with $Y$ Polish, let $\Psi: X \to Y$. We define the $\Psi$-topology, the coarsest topology that makes the pushforward map $\Psi_\#:{\cal P}(X) \to 
{\cal P}(Y)$ continuous:
 $$
 \nu_n \overset{\scriptscriptstyle \Psi}{\rightharpoonup} \nu \Longleftrightarrow \Psi_\# \nu_n \rightharpoonup \Psi_\# \nu.  
 $$ 
 It induces a quotient topology on ${\cal P}(X)\big/\!\sim$ via the equivalence relation:
 $$
 \nu_1 \sim \nu_2 \Longleftrightarrow \Psi_\# \nu_1 = \Psi_\# \nu_2.
 $$
\end{definition}

In our case $X=\Gamma$, $Y=H$, and $\Psi$ is the evaluation map
in (\ref{evalmap}).
In other words, this topology corresponds to convergence in expectation against the class of observables
$$
\mathcal{O}_\Psi := \left\{ F \circ \Psi \;:\; F \in C_b({H};\mathbb{R}) \right\},
$$
i.e., observables that depend only on the evaluation $\gamma(0)$ (or any $\gamma(\tau_0)$ for fixed $\tau_0$).

\paragraph{Main results: RG regularizations}
\label{par1_4_3_2}
We now make precise the idea that the sets ${\cal M}(\gamma)$ in Theorem~\ref{M0M} are genuine \emph{statistical attractors}. The intuition is straightforward: when regularizations are generated by an RG flow, the sets ${\cal M}(\gamma)$ of Theorem \ref{M0M}—now written as ${\cal M}(y)$—attract trajectories for the pushforward dynamics. For example, if some ${\cal M}(y)$ consists of a single non-Dirac measure $\mu$, then Strong-$\SP$ holds for the regularization indexed by $y$ as a consequence of Theorem \ref{M0M}. Moreover, it holds for all regularizations indexed by $y$ in the corresponding basin of attraction, since the Birkhoff average yields the same $\mu$.

In the definition of ${\cal M}(y)$ in~\eqref{MR} one observes that
\be\label{Diracid}
\delta_{{\cal R}_\tau y}\;=\;({\cal R}_\tau)_\#\delta_y,
\de 
where the pushforward of a measure is defined in Appendix \ref{PFTP}.
Indeed, for any bounded continuous $F:{H}\to\mathbb{R}$, one has 
$
\langle({\cal R}_\tau)_\#\delta_y,F\rangle
=\langle\delta_y,F\circ{\cal R}_\tau\rangle
=F({\cal R}_\tau y)
=\langle\delta_{{\cal R}_\tau y},F\rangle
$. This motivates lifting the RG flow to its pushforward
$({\cal R}_\tau)_\#$ acting on general probability measures
$\mu \in {\cal P}({H})$.
Note that family $({\cal R}_\tau)_\#$ is a Markov semigroup: it preserves mass and positivity and satisfies the Chapman--Kolmogorov relation
\be \label{Chapman}
({\cal R}_{\tau+\tau'})_\# \;=\;({\cal R}_\tau)_\#\circ({\cal R}_{\tau'})_\#.
\de 
While ${\cal R}_\tau$ acts pointwise, the pushforward semigroup tracks the statistical evolution of ensembles, so the orbit $\tau\mapsto({\cal R}_\tau)_\#\mu$ captures emergent measures in the inviscid limit. 
Putting all these remarks and definitions together, we obtain
\begin{theorem}[Statistical attractors for $({\cal R}_\tau)_\#$]\label{RGattractors}
Let $(\mathcal R_\tau)_{\tau\ge 0}$ be an RG flow acting continuously on a (possibly infinite-dimensional) space $ H$ endowed with the supremum norm $\|\cdot\|_\infty$.
Assume there exists a \emph{compact, absorbing set} $K$ such that
${\cal R}_\tau K \subset K,~\forall \tau \geq 0$.
Define the Birkhoff $\omega$-limit set for $\mu \in {\cal P}(K)$
$$
\omega(\mu):=
\bigcap_{\R>0}
\overline{\left\{\, \mu_{\R'}(\mu)
  \;:\;R'\ge R
\right\}},~\text{with}~\mu_R(\mu) :=\frac{1}{\R}\int_0^\R ({\cal R}_{\tau'})_\# \mu\, d\tau',
$$
Then:
\begin{enumerate}
\item[(i)] 
$\omega(\mu)$ is non–empty, compact,
forward–invariant under $({\cal R}_\tau)_\#$, and therefore a
statistical attractor; see Definition~\eqref{DefstatAtt}.
\item[] 
\item[(ii)] the following inclusions hold
%Define
%\[\mathcal M_{\mathcal R}\ :=\ \bigcup_{\mu\in\mathcal P(K)}\omega(\mu).\]
\be \label{inclusionsRG}
{{\cal E}}_{\cal R} \subset {\cal M}_{\cal R} \subset 
{\cal M}_{\cal R}^\# 
 = \operatorname{Inv}({\cal R}) = \overline{\operatorname{co}}\bigl({\cal E}_{\cal R}\bigr),
\de 
where 
$$
{\cal M}_{\cal R} := \bigcup_{y \in H} \omega(\delta_y) = \bigcup_{y \in H} {\cal M}(y),
\quad
{\cal M}_{\cal R}^\# := \bigcup_{\mu \in {\cal P}(H)} \omega(\mu),\quad 
\operatorname{Inv}({\cal R}) = \bigl\{\eta \in \mathcal{P}(K) : (\mathcal{R}_\tau)_{\#} \eta = \eta,\ \forall\, \tau \ge 0\bigr\}.
$$
The set ${\cal E}_{\cal R}$ of ergodic measures of ${\cal R}$ are those containing all singleton sets ${\cal M}(y)$ for $\mu$-a.e.~$y$:
$$
{\cal E}_{\cal R} := \bigl\{ \mu : \omega(\delta_y) = \{\mu\} \ \ \mu\text{-a.e. } y \bigr\}.
$$
Moreover ${\cal M}_{\cal R}^\#$ is a compact and convex subset of ${\cal P}(H)$ and ${\cal M}_{\cal R}$ is neither closed nor convex in general. 
\end{enumerate}
\end{theorem}
\begin{proof}
see Section \ref{RG_up}.
\end{proof}
A simple illustration of these results is given in Section \ref{RGreg}.
Inclusions \eqref{inclusionsRG} take different flavors depending on which dynamical system is involved. As a corollary, one has (see dynamical systems Definitions in Appendix \ref{PFTP}) some non-exhaustive list of scenarios
\begin{corollary}\label{coroDS}
Let $({\cal R}_\tau)_{\tau \geq 0}$ be an RG flow as in Theorem \ref{RGattractors}, then provided there is an ergodic measure $\mu$ such that it is non-Dirac then the system is Strongly-$\SP$ for $\mu$-a.e. $y$. 
The following scenarios occur:
\begin{enumerate}
    \item {\bf Uniquely ergodic}: ${\cal E}_{\cal R} = {\cal M}_{\cal R} = {\cal M}_{\cal R}^\# = \operatorname{Inv}({\cal R}) = \{ \mu \}$.     \item[] 
    \item {\bf Axiom A}: Let the nonwandering set $\Omega$ decomposes into finitely many basics sets $\Lambda_i$ (compact, invariant and topologically transitive), then
    ${\displaystyle {\cal E}_{\cal R} \subset
    {\cal M}_{\cal R} = \bigcup_i \operatorname{Inv}(\left.{ \cal R}\right|_{\Lambda_i}) \subset \operatorname{Inv}({\cal R}) = 
    \overline{\operatorname{co}} \left(\bigcup_i  \operatorname{Inv}(\left.{ \cal R}\right|_{\Lambda_i})
    \right)}$. 
    \item[] 
    \item {\bf Morse-Smale systems}: 
    ${\cal E}_{\cal R} = {\cal M}_{\cal R} \subset {\cal M}_{\cal R}^\# = \operatorname{Inv}({\cal R})$. If the system is gradient (no limit cycle), one cannot obtain $\SP$. 
\end{enumerate}

\end{corollary}
It is natural to ask whether the same degree of flexibility as in Theorem~\ref{M0M} can be achieved.  
Dissipative dynamical systems cannot produce measures that are absolutely continuous with respect to the Lebesgue measure.  
Conversely, since ${\cal S}_0$ is a compact subset of $H$ with a nonempty boundary, it is unclear whether the presence of the boundary prevents conservative systems from producing ergodic measures on ${\cal S}_0$ with dense orbits.  
The standard example of a linear flow on the torus ($\dot{x} = 1, \ \dot{y} = b$, $b \in \mathbb{R} \setminus \mathbb{Q}$) cannot be used here, as the torus has no boundary.  
In dimensions $d \geq 3$, it is unknown whether one can construct a flow with dense orbits in ${\cal S}_0$. 
At the very least, such a vector field must satisfy $\operatorname{div} g = 0$ in ${\cal S}_0$ and be tangent to the boundary, i.e., $g \cdot \mathbf{n} = 0$ on $\partial {\cal S}_0$. 
Producing dense orbits within ${\cal S}_0$ is a considerably more challenging problem. 
More generally, we leave open the question
\be 
\bigcup_{\cal R} {\cal M}_{\cal R} = {\cal M}_0,
\de
where ${\cal R}$ denotes an RG flow in the sense of Definition~\ref{RGflow}.

\paragraph{Quotient Topology for the Bebutov Flow: A reinterpretation of Theorem \ref{M0M}}\label{par1_4_3_3}
The Bebutov flow defined in \eqref{Bebutov} acts as a shortcut against RG flows, by bypassing/relaxing totally the semigroup property on ${H}$. It does so by considering another semigroup property in a larger space of regularization orbits (the set $\Gamma$) no matter it is inherently autonomous or not.
As explained below, it is {\it universal} in the sense that it yield a complete upgrading of Theorem \ref{M0M}.

Although the Bebutov flow $(\mathscr{S}_\tau)$ also defines a semigroup, it is different from $\mathcal{R}_\tau$, since it acts on a space of curves $\Gamma \subset C_b(\mathbb{R}^+; {H})$. 
Let $i: {H} \to \Gamma$ denotes a (non-unique) embedding from the state space to the space of regularization curves, namely
for $\phi \in H$, $i(\phi) = \gamma_\phi \in \Gamma$.
Moreover, in order for ${\cal R}_\tau$ to be a semigroup on $H$, one imposes the lift to form autonomous curves:
$$
\Psi \circ i = {\rm Id}~\text{and}~i(\phi)(\tau_1+\tau_2)=i(i(\phi)(\tau_2))(\tau_1).
$$
This is indeed the more abstract rephrasing of \eqref{RG_system}.
The relationship between these two semigroups is expressed by the following diagram:
$$
\begin{tikzcd}[column sep=large, row sep=large]
{H} \arrow[r, hook, "i"] \arrow[d, "\mathcal{R}_\tau"']
&
\Gamma \arrow[d, "\mathscr{S}_\tau"] \arrow[r, "\Psi"']
&
{H} \\
{H} \arrow[r, hook, "i"]
&
\Gamma \arrow[r, "\Psi"']
&
{H}
\end{tikzcd}
$$
They both decrease the level of regularization of (\ref{P_0}) with
\be \label{RPSi}
{\cal R}_\tau = \Psi \circ \mathscr{S}_\tau \circ i.
\de 
 Given a solution $\phi$ of $(\mathcal{P}_\tau)$ for some fixed $\tau$, multiple regularizations can arise, corresponding to different vector fields $f \in \V$ generating flows that agree at that point.
One can interpret this relation as parameterizing the set of RG semigroups through the embeddings $i$, namely given
${\cal R}_\tau$, it is defined as $i(\phi)(\tau) := {\cal R}_\tau \phi$ and conversely given some embedding $i$, ${\cal R}_\tau$
is just defined through the relation (\ref{RPSi}).

In a similar way than previously, rather than $\mathscr{S}_\tau$, the natural object to consider is its pushforward $(\mathscr{S}_\tau)_\#$. The identity (\ref{Diracid}) is replaced by:
\be \label{Bebuid}
\delta_{\Psi \circ \mathscr{S}_\tau \gamma} = \Psi_\# (\mathscr{S}_\tau)_\# \delta_\gamma = \delta_{\gamma(\tau)},~\gamma \in \Gamma.
\de 
\begin{theorem}[Statistical attractors for $(\mathscr S_\tau)_\#$]\label{thm:M0Mupup}
Let $\Gamma\subset C_b(\mathbb R^{+};H)$ be equipped with the compact–open
topology and $(\mathscr S_\tau)_{\tau\ge0}$ the Bebutov left–shift on $\Gamma$.
Write $\Psi(\gamma):=\gamma(0)$ and endow $\mathcal P(\Gamma)$ with the
$\Psi$–topology.  We moreover assume that $\Psi(\Gamma)$ is closed.
For $\nu\in\mathcal P(\Gamma)$ set 
$$
\mathscr M(\nu):=\omega(\nu):=
\bigcap_{\R>0}
\overline{\left\{\, \nu_{\R'}(\nu)
  \;:\;R'\ge R
\right\}}^{\scriptscriptstyle \Psi},~\text{with}~\nu_{\R}(\nu) :=\frac{1}{\R}\int_0^\R (\mathscr{S}_{\tau'})_\# \nu\, d\tau'.
$$
Then:
\begin{enumerate}
\item[(i)] $\mathscr M(\nu)$ is non–empty, compact for the $\Psi$–topology,
forward–invariant under $(\mathscr S_\tau)_\#$, and therefore a
statistical attractor with $\nu\in\mathcal B(\mathscr M(\nu))$.
\item[] 
\item[(ii)] $\Psi_\#\mathscr M(\nu)\subset\mathcal M_0:=\mathcal P(\mathcal S_0)$.
If $\nu=\delta_\gamma$ with $\gamma\in\Gamma$, then
$\Psi_\#\mathscr M(\delta_\gamma)=\mathcal M(\gamma)$.
\item[] 
\item[(iii)] Define
$$
\mathscr M:=\bigcup_{\nu\in\mathcal P(\Gamma)}\mathscr M(\nu).
$$
The set $\mathscr M$ is a global (non-minimal) statistical attractor for the $\Psi$–topology, 
and satisfies
$$
\Psi_\#\mathscr M=\mathcal M_0
=\overline{\operatorname{co}}({\cal E}),
\qquad
   \mathscr{M}
   \;=\;
   \operatorname{Inv}(\mathscr{S})
   \;:=\;
   \bigl \{\eta\in\mathcal P(\Gamma)\;:\;(\mathscr{S}_\tau)_\# \eta=\eta,\ \forall\tau\ge0\bigr\}.
%{\cal E}:=\{\delta_x:x\in{\cal S}_0\}.
$$
\end{enumerate}
\end{theorem}
\begin{proof}
see Section \ref{BEBU_auto}.
\end{proof}

\paragraph{Discussion}
In our deterministic setting \eqref{P_eps}, Theorem 
 \ref{thm:M0Mupup} is restricted to $\nu = \delta_\gamma, \gamma \in \Gamma$. Therefore, it provides a straightforward new interpretation of Theorem \ref{M0M} and \eqref{SPdefbis}.
\begin{corollary}[Theorem \ref{M0M} and Theorem \ref{thm:M0Mupup}]
Let ${\cal A}_\gamma:= \mathscr{M}(\delta_\gamma)\big/\!\sim$.
Then ${\cal A}_\gamma$ is a statistical attractor for the Bebutov flow under the quotient topology and
$$
  {\cal A}_\gamma ~\subset \operatorname{Inv}(\mathscr{S})\big/\!\sim ~~\cong \mathcal{M}_0 = \overline{\operatorname{co}}(\mathcal{E}) .
$$
Moreover, the following characterizations hold:
$$
\gamma \in \begin{cases}
\text{\emph{Strong $\SP$}} & \text{if } {\cal A}_\gamma \cong \{\mu \}  \text{ and } \mu \text{ is non-Dirac}, \\
\text{\emph{Weak $\SP$}} & \text{if } {\cal A}_\gamma \text{ is not a singleton}.
\end{cases}
$$
\end{corollary}

Working at the level of $(\mathscr{S}_\tau)$ offers also the possibility to upgrade deterministic regularizations \eqref{P_eps} to random ones. Let us define the random differential equation (RDE):
\be 
\dot x = f_\Theta(x,\tau), x(0)=x_0 \in H,~\Theta:(\Omega,{\cal F},\mathbb{P}) \to \mathscr{E}.
\de 
The solution at time $t$ and initial condition $x_0$ is now a random variable $X_\Theta(t;x_0;\tau)$ and we define a Markov kernel as
$$
K(\tau,A) := \Pr(X_\Theta(t;x_0;\tau) \in A)~A~\mbox{Borel of $H$ i.e.}~K(\tau,dx) \in {\cal P}(H).
$$
The natural question is therefore to ask when $\{K(\tau,\cdot)\}_{\tau \geq 0}$ define a measure $\nu \in {\cal P}(\Gamma)$?
The answer is simply: 
\begin{itemize}
    \item for $\mathbb{P}$-almost $\omega \in \Omega$, 
$f_{\Theta(\omega)} \in \V$; see \eqref{H0}.
\item[] 
\item $\tau \mapsto X_{\Theta}(t;x_0;\tau)$ is continuous and bounded and $\omega \mapsto X_{\Theta(\omega)}$ is measurable.
\end{itemize}

Under these conditions, the assumptions of Theorem \ref{thm:M0Mupup} hold true without modification, so that the conclusions of this theorem remain unchanged. In particular, the introduction of randomness into the regularizations has no effect on the asymptotic results, since only the behavior in the limit $\tau\to\infty$ matters. This fact highlights the relevance of the $\Psi$-topology, which captures precisely the asymptotic distribution. As Theorem \ref{thm:M0Mupup} is essentially a quotient-topological reformulation of Theorem \ref{M0M}, the conclusions of Theorem \ref{M0M} also remain valid, up to interpreting the statistical attractors via the quotient topology induced by~$\Psi$. In this sense, we have provided a true upgrading to randomized regularizations.

In principle, upgrading these results to SDE is largely feasible but more technical. One must work pathwise and relax the boundedness hypothesis. We do not intend to provide this extension and let it for future works.

\subsubsection{A necessary condition for having $\SP$}
\label{CNforSP}
We have established in Corollary \ref{MND} that provided 
(\ref{P_0})  has nonunique solutions then there exists a regularization \eqref{P_eps} such that a strong form of $\SP$ holds.
It is thus natural to ask when (\ref{P_0})  has nonunique solutions. The goal here is to establish a sufficiently sharp necessary condition.

There exists a class of systems with an isolated H\"older singularity that can be renormalized in a suitable way, allowing for sharp conditions under which $\SP$ may occur (see \cite{Drivas21}, \cite{Drivas24}). As we will see, these systems satisfy the criteria for $\SP$ to arise. An example is given in Subsection \ref{Ex3}.

We therefore aim at identifying a necessary condition for nonuniqueness in the inviscid system (\ref{P_0}). The anticipated scenario is that the inviscid flow hits a singular set in finite time. However, the precise definition of such sets must be clarified. A typical example is a non-Lipschitz isolated singularity $x^\star \in \mathbb{R}^n$, where $f(x^\star) = 0$ lacks Lipschitz continuity, though this is not a sufficient condition. There are also cases where $f(x^\star) \neq 0$, yet $x^\star$ still leads to a loss of uniqueness for flows going through it (see Subsection \ref{Ex3}). 

In all the following, we assume $f$ at least continuous to ensure Peano’s classical solution existence.
Numerous theorems addressing uniqueness exist for general nonautonomous systems of the form $\dot x = f(x,t),~x(t_0) = x_0$ (see, in chronological order \cite{Walter1970,Agarwal_Lak,Hartman2002,Bahouri2011}). These results 
typically take the form:
{\it if condition $(A)$ holds then the solution to the Cauchy problem is unique}.
Consequently by contraposition, there are many possible necessary conditions of the form
$\neg{(A)}$ for nonuniqueness. In order to obtain some sharper condition, one must look at 
a "disjunction" of different uniqueness theorems. The 1-D autonomous case is well understood.
First the singular set, denoted $\Gamma^\star$, is necessarily reduced
to an isolated critical point $x^\star$ i.e. such that $f(x^\star) = 0$ (by a straightforward use of the implicit function theorem, see e.g. Theorem 1.2.7 in 
\cite{Agarwal_Lak}). Second, $f$ must increases sufficiently rapidly in the neighborhood of
$x^\star$ either to the left or right, in particular it cannot be decreasing or being Lipschitz at $x^\star$. The optimal growth rate condition is convergence of at least one of the integral $\int_{{x^\star}^\pm} \frac{dx}{f(x)}$. With respect to the initial condition, it is highly constrained in 1-D, either $x(0)=x^\star$ or if $x_0 < x^\star$, then $f$ must be positive so that $x^\star$ is reached in finite time (and conversely $x_0 > x^\star$, $f$ must be negative). Typical examples giving nonunique solutions are $\dot x = {\rm sign}(x) |x|^\alpha$, $\dot x = \pm |x|^\alpha$, $\alpha \in (0,1)$. But systems like $\dot x = |x|^\alpha + cst$
or $\dot x = -{\rm sign}(x) |x|^\alpha$ have unique solutions no matter the initial condition. 

The situation in higher dimension is much less favorable, in particular 
$f(x^\star) = 0$ is no more necessary which is the main reason this problem becomes nontrivial. 
The aim here is thus to modestly characterize regions where 1) $f$ has some expanding behavior
2) $f$ is not Lipschitz. 

We consider here a similar idea to that of Dini derivatives (see Appendix \ref{DiniApp}). 
They correspond to a generalized notion of derivatives in contexts where $f$ is not even continuous. Dini derivatives share many desirable properties with classical derivatives.
For instance, being Lipschitz continuous is equivalent to having
finite Dini derivatives. They 
therefore provide a natural and practical way to express non-Lipschitz behavior.
We slightly extend the notion to the Osgood property.
To simplify the presentation, we will consider
$$
H = \mathbb{R}^n.
$$
Let us introduce the following definitions:
\begin{definition}\label{DiniME}
Let $f: \mathbb{R}^n \to \mathbb{R}^n$, 
Let $v \in \mathbb{S}^{n-1}$, then
\be 
\Lambda_\Omega^+(x,v) := \limsup_{t \to 0^+} 
\left \langle \frac{f(x+t v)-f(x)}{\Omega(t)},v
\right\rangle .
\de 
where $\Omega$ is a modulus of continuity: $\Omega: \mathbb{R}^+ \to \mathbb{R}^+, \Omega(0) = 0$ and $\Int_{0^+} \frac{dz}{\Omega(z)} = \infty$.

We also introduce a notion of stable finite-time set:
Let $\Phi: (t,x_0) \mapsto \Phi_t x_0$ be the flows associated to 
$\dot x = f(x), x(0)=x_0$, and consider some arbitrary set $\Gamma \subset \mathbb{R}^n$
\begin{equation}\label{Wm_G}
{\cal W}^-(\Gamma) \equiv  \bigcup_{\Phi} \left\{ x \in \mathbb{R}^n
 \left| \right. \exists t^\star < \infty,~t^\star \geq  0, \lim_{t \to +t^\star} {\rm dist}(\Phi_t x,\Gamma) = 0 \right\} 
\end{equation}
\end{definition}
From this definition, one can derive a useful necessary condition for breaking uniqueness
and thereby guaranteeing (strong) $\SP$:
\begin{theorem}\label{CN}
Assume that (\ref{P_0}) has nonunique solutions, then one 
can find some $\Omega$ and a nonempty set $\Gamma^\star \subset \mathbb{R}^n$ such that
$x_0 \in {\cal W}^-(\Gamma^\star)$ 
where
$$
{\displaystyle \Gamma^\star = \big\{ x \in \mathbb{R}^n ~|~ \exists v, ||v||=1, 
 ~\Lambda^+_\Omega(x,v)= +\infty \big\}.}
$$
\end{theorem}
\noindent {\bf Proof}: see Appendix \ref{DiniApp}.\\
The idea is to replace the classical notion of Jacobian of $f$
by some weaker notion involving Dini (directional) derivatives 
(say with $\Omega(z)=z$). The unit vectors $v$ indeed plays the role of an eigenvector
and $\Gamma^\star$ simply detects regions in phase space where the eigenvalues blow up to $+\infty$. 
 
In addition, one can also characterise for free the set of initial conditions giving
spontaneous stochasticity by a straightforward generalization of a stable set.
Note that such result is not sharp and cannot be sufficient. However, there is still room to improve this condition. We let it for future work.

The 1-D case is rather particular and is of less interest, in particular the critical 
condition $f(x^\star) = 0$ is not distinguished. This is expected
since Theorem \ref{CN} is aimed at situations where nonuniqueness arises from 
more general situations than those due solely to the presence of non-Lipschitz 
critical points.

\subsection{Perspectives}\label{perspec}
From the results obtained, we first emphasize the following important points:
%\emph{
\begin{enumerate}
    \item {\bf Randomization:} Random regularizations are not required to observe $\SP$. Its emergence is a consequence of the ill-posedness of the inviscid problem itself, regardless of whether the regularizations are random. What matters is the ergodic behavior of a given regularization in the inviscid limit. For instance, mixing properties are sufficient but not necessary.
    \item[] 
    \item {\bf Flexibility:} 
    In finite dimensions, any probability measure in the space ${\cal P}({\cal S}_0)$ can be obtained as the inviscid weak limit of a suitable family of regularizations. For such regularizations, the system is therefore strongly $\SP$. In contrast, regularizations generated by a single RG flow are more constrained, with limiting probability measures containing at least the ergodic measures of the RG flow.
    \item[] 
    \item {\bf Singularities:}
  $\SP$ is fundamentally linked to the presence of singular sets in phase space responsible for nonuniqueness in the inviscid problem. Local well-posedness persists only until the trajectory encounters such a singular set, at which point ill-posedness (post-blowup or “life after death” \cite{Drivas21}) emerges. In our framework, $\SP$ manifests exactly when the set of probability measures ${\cal P}({\cal S}_0)$ undergoes a \emph{bifurcation set} where time $t$ plays the role of a bifurcation parameter. At the critical blowup time $t=t^\star$, the Dirac singleton regime of well-posedness is destroyed enabling spontaneous stochasticity.
  Since all quantities are indexed by time $t$, the measures $\mu = \mu_t$ involved are indeed \emph{Young measures}.
\end{enumerate}
%}
Several open questions remain, though the following list is not exhaustive:
\begin{enumerate}
\item {\bf Weak \boldmath{$\SP$}:} In light of Definition~\ref{SPDEF4}, it is natural to ask whether this class is empty or not. Can one always construct an ambient measure for which Strong~$\SP$ holds?
\item[] 
\item {\bf Universality:} This concept is tighted to the size of the basins of attraction; see Definitions \ref{defgenesets}, \ref{DefstatAtt}. The presence of chaos in the inviscid system might enlarge them significantly as suggested \cite{AV24}. Consider the following formal scenario: suppose one has some RG flow being chaotic with a compact attractor ${\cal A}$ and an SRB measure $\mu_{\rm SRB}$ supported on it. Then the basin of attraction ${\cal B}_{\mu_{\rm SRB}}$ is expected to contain a large class -- measure-theoretically -- of regularizations $\gamma$ that are attracted to ${\cal A}$ (in the compact-open topology). Non-generic orbits are expected to form much smaller sets or may fail to be renormalizable.
\item[] 
\item {\bf Numerics: } 
Despite considerable effort, we have not yet succeeded in designing a numerical method that significantly increases the number of iterations \(M\) that can be reached. We believe that our algorithm can be slightly improved by enhancing the efficiency of the backward flow map computation. However, this component is not the numerical bottleneck, and improving it alone will not suffice to access larger values of \(M\). Achieving higher resolution would require the development of a fundamentally different numerical method, adapted to the intricate temporal structure of the problem -- a direction we leave for future work. 

An alternative strategy to increase the number of active scales would be to consider a numerical toy model for the vector field, for instance using logarithmic lattices as in~\cite{Campolina_2021,Ortiz2025Spontaneous}. This, however, would raise significant questions regarding the interpretation of the results. Lastly, the question of the genericity of the vector field \(\vect\) is also relevant and could be readily explored numerically by modifying the base flow \(\psi_{m,k}\); for example, one might consider shear flows with random orientations.
\item[]
\item {\bf PDEs:} It is possible that Eulerian-$\SP$ could be established for other PDEs, including the three-dimensional Euler and Navier–Stokes equations. In light of Property \ref{SPDEFS}, proving $\SP$ alone -- without attempting to classify it as Strong or Weak -- may be within reach. Indeed, the existence of uniform estimates like Obukhov-Corrsin (see Section 
\ref{OCpar}) in sufficiently regular spaces already ensures tightness via compact embeddings. We also believe that the special limiting case where $LSP$ leads to a Dirac measure can often be ruled out without difficulty. 
The main difficulty in establishing Strong-$\SP$ lies in deriving the required deterministic PDE estimates and in obtaining quantitative bounds for fundamental mechanisms such as homogenization and intermittency.

\item[] 
\item{\bf Sweeping: }
The role of \emph{sweeping} -- that is, working in a Lagrangian reference frame at a given scale to allow homogenization at the next scale -- plays a central role in~\cite{AV24}. Whether this mechanism is necessary to achieve Eulerian-$\SP$ in transport equations in general, remains unclear.\\ 

\end{enumerate}

\section{Anomalous diffusion in \cite{AV24}: from theory to numerics}\label{num_LSS}
\subsection{Lagrangian description: constraints on the scales}
\label{sechom}
The vector field ${\bf b}$ of \cite{AV24} is constructed in a very involved way and consists of space-time fractal structures with periodic shear flows alternating in time.
It is a multiscale flow with infinitely many scales separated hyper-geometrically.
Our aim is to draw a simple picture of the homogenization mechanism of the vector field
constructed in \cite{AV24}. For this, it is useful to adopt a Lagrangian viewpoint
and to express the dynamics of particles transported by a given shear flow. In addition, 
it gives for free the various scaling constraints for the velocity field in \cite{AV24}.
Let us consider a toy version of the vector field \cite{AV24}. Take $a_m>0$ and $\epsilon_m^{-1}$ a positive integer and define the incompressible vector field $\mathbf{v}_m(x,y)= (2\pi a_m \epsilon_m \cos(2\pi/\epsilon_m x),0)$.
The diffusion process associated with $\mathbf{v}_m$ with diffusivity $\kappa$ yields the following SDE:
\begin{align}
&dX_t= 2\pi a_m \epsilon_m \cos\left( \dfrac{2\pi}{\epsilon_m} Y_t\right) +\sqrt{2\kappa} \, d W^1_t,~~
dY_t= \sqrt{2\kappa} \, d W^2_t,~~  (X_0,Y_0) = (0,0).\nonumber 
  \nonumber
\end{align}
Where $W_t^{1,2}$ are independent Brownian motions, $a_m$ is the Lipschitz norm of $\mathbf{v}_m$ and $\epsilon_m$ its space period. In fact, in order for the vector field $\sum_m \mathbf{v}_m$ to be H\"older continuous with exponent strictly less than $\alpha$, one must take $a_m = \epsilon_m^{\alpha-1}$. We chose the initial condition to be $(0,0)$ for the sake of simplicity but this heuristic arguments holds independently of the initial position. Some standard
calculations, further simplified by our choice of initial condition, give
\begin{align} \label{eq:effdiff_shear}
\begin{array}{llll}
\mathbb{E}X_t^2  & =  2\left(\kappa+\dfrac{1}{2}\dfrac{a_m^2 \epsilon_m^4}{\kappa} \right) t + \left(   \dfrac{a_m \epsilon_m^3}{2\pi\kappa} \right)^2 \left( e^{- \omega_m t} -1 \right) \left. +\dfrac{1}{12}\left(e^{-4 \omega_m t}-1\right) \right),  ~~~ \omega_m := \frac{4\pi^2 \kappa}{\epsilon_m^2},  \\
\mathbb{E}Y_t^2 &=  2\kappa t.
\end{array}
\end{align} 

\begin{figure}[htbp]
\centerline{\includegraphics[width=0.8\columnwidth]{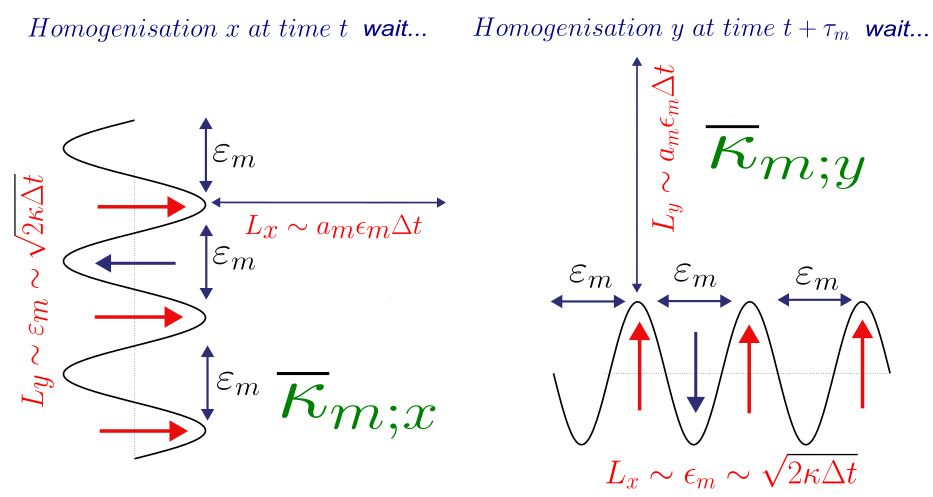}}
\caption{A sketch of homogenization at scale $\epsilon_m$.
The time $\Delta t$ is the time needed to see the shear at scale $\epsilon_m$, namely 
to have a displacement along $y$ of order $\epsilon_m$
giving $\epsilon_m \sim \sqrt{2 \kappa \Delta t}$. 
It gives the effective diffusivity along $x$: $\overline{\kappa}_{m,x} = \frac{\langle L_x^2 \rangle}{\Delta t}
= \kappa + a_m^2 \epsilon_m^2 \Delta t = \kappa + \frac{a_m^2 \epsilon_m^4}{\kappa}$.
Then one rotates the field after waiting some time $\tau_m \gg \Delta t$,
that is waiting long enough for homogenization to occur. Then one proceeds again in the
$y$ direction.
}\label{LagHom}
\end{figure}

By direct inspection, the particle displacement in the $x$ direction yields an effective diffusivity
$\bar\kappa_x= \kappa +  \dfrac{a_m^2 \epsilon_m^4}{2\kappa}$ in the $x$ direction (to be compared with (\ref{avseq})).
Let us call $\tau_m$ the typical time for homogenization to occur. It must be much larger than the relaxation time $\omega_m^{-1}$. The first constraint
is therefore $\tau_m \gg \frac{\epsilon_m^2}{\kappa}$. Let us call $\epsilon_{m-1}=\sqrt{\mathbb{E}X_t^2}$
the typical particle displacement in the $x$ direction during this homogenization process: it is
$\epsilon_{m-1}^2 \propto \left( \kappa +  \dfrac{a_m^2 \epsilon_m^4}{2\kappa} \right) \tau_m \gg \left( \kappa +  \dfrac{a_m^2 \epsilon_m^4}{2\kappa} \right) \omega_m^{-1}
= \frac{\epsilon^2_m}{4\pi^2} + \frac12 \left( \frac{a_m \epsilon_m^3}{2\pi \kappa}
\right)^2$ giving the second constraint: $\epsilon_{m-1} \gg \frac{a_m \epsilon_m^3}{ \kappa}$. Those constraints correspond exactly to the constraints used in \cite{AV24}.
This homogenization process is illustrated in a simplified form in Fig.~\ref{LagHom}.

In summary, the process $(X_t,Y_t)$ can be interpreted, at large times, as an effective diffusive process with diffusivity $(\bar\kappa_x,\kappa)$. 
From the expression of $\bar \kappa_x$, it is clear that even in the limit of vanishing diffusivity, the effective diffusivity $\bar\kappa_x$ 
remains strictly positive and can even diverge. However, this does not imply that the passive scalar undergoes anomalous diffusion in the sense of \cite[Theorem 1.1]{AV24}. 
Indeed, the effective diffusive description is only valid on time scales much larger
than $\tau_m$ and length scales much larger 
than $\epsilon_{m-1}$, preventing the observation of finite effective diffusivity on time scales of order unity in the vanishing $\kappa$ limit.

A last remark is that, in order for the homogenization process to cascade to larger scales, much care is required. Take for instance a naive zonal superposition of shear flows: 
${\bf b}_M = \sum_{m=0}^M  \mathbf{v}_m$. Using the previous calculation, one can show that the mean particle displacement is 
\begin{align*}
\mathbb{E}X_t^2  = 2 {\bar \kappa}_{M,x} t +\mathcal{O}_{t \to \infty}(1), ~~ \mathbb{E}Y_t^2  = 2  \kappa t 
\end{align*}
with ${ {\bar \kappa}_{M,x} = \kappa + \frac{1}{2\kappa} \sum_{m=0}^M a_m^2 \epsilon_m^4}$.

This computation demonstrates that superimposing shear flows at different length scales increases the effective diffusivity ${\bar \kappa}_{M,x}$ beyond that of a single shear $ \mathbf{v}_m$ flow obtained earlier. However, for the same reasons as before, this superposition alone does not lead to anomalous diffusion. Furthermore, the expression for ${\bar \kappa}_{M,x}$ reveals that the problem at scale $m$ is not simply homogenized into the problem at scale $m-1$; instead, all scales are homogenized simultaneously. As such, this straightforward superposition of shear flows does not enable anomalous diffusion. To address this issue, the authors of~\cite{AV24} introduced a crucial modification: they added the shear flow at scale $m$ in the reference frame of the one at scale $m-1$. This adjustment allows for anomalous diffusion, as proven in \cite{AV24}. Interestingly, this modification also constrains the H\"older regularity of the limiting vector field to be strictly less than $1/3$. 

Additionally, in the framework discussed earlier, the effective diffusive regime is achieved only in the $x$-direction. To extend this homogenization to the $y$-direction, one must introduce an advection field that alternates in time between the $x$ and $y$ directions. The time period of alternation between $x$- and $y$-directed shear flows must be sufficiently large to ensure that the effective diffusive regime is fully realized as depicted in Fig.~.\ref{LagHom}.

\subsection{How to numerically implement the Armstrong-Vicol model}\label{ALGO} 
In the following section, we present the numerical methods used to study the Armstrong-Vicol model, with particular emphasis on the numerical computation of the vector field, which is the most challenging aspect. Before proceeding, we briefly recall the construction of the streamfunction as detailed in \cite{AV24}. The streamfunction $\phi$ of the two dimensional vector field $\vect= \nabla 
^\perp\phi$ is defined as the limit of a sequence
of streamfunctions $( \phi_m)_{m\geq 0}$ which are generated recursively according to the induction relation for $m \geq1$:
\begin{equation}
\begin{cases}\label{eq:DefAVField}
& \hspace*{-0.25cm} \phi_{m}(t, \pos)=\displaystyle\phi_{m-1}(t,\pos)+ \sum \limits_{l\in \mathbb{Z}} \sum \limits_{k\in \mathbb{Z}} \zeta_{m,k}(t) \hat{\zeta}_{m,l}(t)\psi_{m,k}(\flow^{-1}_{m-1,l}(t,\pos,l \tau_m'')), \\
& \hspace*{-0.25cm} \phi_0(t,\pos)=0,\quad (t,\pos) \in [0,1]\times \mathbb{T}^2.
\end{cases}
\end{equation}
The functions $\psi_{m,k}$ are the building blocks of the flow and are chosen in~\cite{AV24} to be,
$$\psi_{m,k}(\pos)= a_m \epsilon_m^2 \left[\mathbbm{1}_{k\in 4 \mathbb{Z}+1 } \sin\left( \dfrac{2\pi x_1}{\epsilon_m} \right) +  \mathbbm{1}_{k\in 4 \mathbb{Z}+3 } \sin\left(  \dfrac{2\pi x_2}{\epsilon_m} \right) \right]. $$
This streamfunction corresponds to a cosine shear flow with Lipschitz norm $a_m$
and period $\epsilon_m$, this cosine shear flow is the one we just used in the toy example of the previous section with slight modifications allowing the shear flow to be along $x_1$, $x_2$ or turned off. By definition, $\psi_{m,k}$ itself is time-independent.
The smooth alternation between shear flows in the $x_1$ and $x_2$-directions is controlled by the time cutoff function $\zeta_{m,k}$, which is smooth and compactly supported:
$$ \mathrm{Supp}\, \zeta_{m,k}  \subset \left[ \left(k -\frac{2}{3}\right) \tau_m, \left(k +\frac{2}{3}\right) \tau_m \right].$$
This implies that for any given value of $t$, there is at most one $k\in \mathbb{Z}$ such that $\zeta_{m,k}(t) \psi_{m,k}(\pos) \neq 0$. As mentioned in the Lagrangian description~\ref{sechom}, simply adding shear flows is not sufficient to prove anomalous diffusion. Therefore, the perturbation added to $\phi_{m-1}$ in \eqref{eq:DefAVField} 
is introduced within the reference frame of the vector field $\vect_{m-1}$. This is the role of $\flow^{-1}_{m-1,l} $ entering \eqref{eq:DefAVField}, which is the inverse flow map of the field $ \vect_{m-1}$ (on a given time interval).
The inverse flow map is obtained as solution of:  \be \label{eq:InverseFlow}
\displaystyle \partial_t \flow^{-1}_{m-1,l}+ \vect_{m-1}\cdot \nabla \flow^{-1}_{m-1,l}=0,~~
\flow^{-1}_{m-1,l_k}( l \tau_m'',\pos)=\pos,
\de 
with $(t,\pos) \in \left[\left(l -\frac{1}{2} \right) \tau_m'', \left(l +\frac{1}{2} \right) \tau_m'' \right] \times \mathbb{T}^2$.

The inverse flow map of the field $\vect_{m-1}$ is not computed on the whole $[0,1]$ time interval but rather on sub-intervals of length $\tau_m''\ll a_{m-1}^{-1}$ ensuring that the inverse flow maps are not deformed too much (see \cite{AV24} for a precise meaning). These inverse flow maps are refreshed every $\tau_m''$ unit of time, the time cutoff $\hat{\zeta}_{m,l}$ ensures that this refreshing is done smoothly in time. The new timescale $\tau_m''$ must be homogenized as well, necessitating the introduction of a third timescale $\tau_m'\ll \tau_m''$. This time cutoff $\hat{\zeta}_{m,l}$ is therefore smooth and compactly supported, such that:
\begin{equation*}
      \mathbbm{1}_{t\in [(l-\frac{1}{2})\tau_m'' +2\tau_m',(l+\frac{1}{2})\tau_m'' -2\tau_m']} \le \hat{\zeta}_{m,l}\le \mathbbm{1}_{ t \in [(l-\frac{1}{2})\tau_m'' +\tau_m',(l+\frac{1}{2})\tau_m'' -\tau_m']}. \label{eq:supportzetahat}
\end{equation*}
In practice, for a given $t\in[0,1]$ and $k\in \mathbb{Z}$, there is only one value of $l\in \mathbb{Z}$ such that $\zeta_{m,k}(t) \hat{\zeta}_{m,l}(t) \neq 0$. This implies that the sums over $k,l$ appearing in~\eqref{eq:DefAVField} can effectively be written as a sum over $k$ only. For the sake of completeness, we give our choice for the temporal cutoff functions $\zeta_{m,k}, ~ \hat{\zeta}_{m,l}$ in Appendix~\ref{APP:TimeCutoff}

With this construction, the authors of~\cite{AV24} are able to prove that, roughly speaking, the $L^2$ norm drop of the passive scalar advected by $\vect_m$ and with diffusivity $\kappa_m$ is of the same order as that of the passive scalar advected by $\vect_{m-1}$ and with diffusivity $ \kappa_{m-1}= \kappa_m + c (a_m^2 \epsilon_m^4)/\kappa_m$. We emphasized in section~\ref{sechom} that this homogenization occurs only if the scales $(\epsilon_m,a_m,\tau_m,\tau_m',\tau_m'')$ are carefully chosen. In particular, the following constraints must be satisfied:
\be
\begin{array}{llll}
\tau_m\ll \tau_m'\ll \tau_m''\ll a_{m-1}^{-1}, &
\dfrac{a_m\epsilon_m^3 }{\kappa_m} \ll \epsilon_{m-1}, & 
\dfrac{\epsilon_m^2}{\kappa_m} \ll \tau_m.
\end{array}
\de 
The goal of this procedure is to obtain a large-scale effective diffusivity $\kappa_0$ of order unity. We refer the reader to~\cite{AV24} for a more detailed discussion on these constraints. In particular, the H\"older regularity of the velocity field $\vect$ cannot be greater than $1/3$ for these constraints to be satisfied. 

In~\cite{AV24}, the sequence of length scales $\epsilon_m$ is chosen to decrease hyper-geometrically, and all other scales are built from $\epsilon_m$ such that the set of constraints is met. However, with the sequence used in~\cite{AV24}, the spatial resolution required to numerically compute $\phi_2$ would already be prohibitive, while $\phi_1$ is explicit and thus not numerically interesting. 
Therefore, we adopt a different choice of sequences, which retains the hyper-geometric decay of $(\epsilon_m)_{m\geq1}$, namely:
\be 
\begin{array}{llllllll}
\epsilon_m^{-1}= \left\lceil \Lambda^{  \frac{q^m}{q-1}} \right\rceil, &
a_m=\epsilon_m^{\beta-2}, & 
\tau_m= \tau_m', 
\tau_m'= \frac{\tau_m''}{3(4\left\lceil \epsilon_m^{-\delta} \right\rceil +1) }, & 
\tau_m''=2^{-2} \left\lfloor\frac{a_{m-}}{ \epsilon_m^{2\delta}}\right\rfloor^{-1}.
\end{array}
\de 
where $q$ and $\delta$ are defined in Table~\ref{tab:parameters} and the choice of parameters $\Lambda$, $\beta$ and $M$ are detailed in Table~\ref{tab:ParamsB}, where $M$ is the maximum number of iterations we are able to reach numerically. 

With this choice of sequences, the hypotheses of Theorem 1.1 (\cite{AV24}) are not strictly satisfied. However, our numerical simulations still show important features of the model such as anomalous diffusion, spontaneous stochasticity of Lagrangian tracers and Obukhov-Corrsin regularity of the passive scalar as we discuss in the following. We also give an algorithm below meant to make reproduction of our results easier. The purpose of the Algorithm~\ref{alg:cap} is to compute the streamfunction $\phi_m$ given $\phi_{m-1}$ according to the recursion \eqref{eq:DefAVField}. For this we need to integrate the backward flow map equation~\eqref{eq:InverseFlow}. We do this using a de-aliased pseudo spectral method. However, the field $\flow^{-1}_{m-1,l}(l\tau_m'',\pos)=\pos$ is not periodic in space. In order to tackle this issue we work with $\mathbf{Y}:=\flow^{-1}_{m-1,l}-\mathbbm{1}$ which is periodic and satisfies 
\be\label{eq:PeriodicBackFlow}
\partial_t\mathbf{Y}=-\vect_{m-1}\cdot\nabla \mathbf{Y}-\vect_{m-1},~~\mathbf{Y}(l\tau_m'',\pos)=0,
\de 
where $(t,\pos)\in [(l-\frac12 )\tau_m'',(l+\frac12 )\tau_m'']\times \mathbb{T}^2$.
The variable $\mathbf{Y}$ in Algorithm~\ref{alg:cap} is therefore the numerical approximation of the solution of~\eqref{eq:PeriodicBackFlow} and the function $\mathrm{RHS}$ is the numerical approximation of the right hand side of~\eqref{eq:PeriodicBackFlow} computed using a pseudo-spectral dealiased differentiation.
\begin{algorithm}[hptb]
\caption{Compute $\phi_m$ given $\phi_{m-1}$}\label{alg:cap}
\begin{algorithmic}
\Require $\phi_{m-1}$, $m$, $\epsilon_m$, $\tau_m$, $\tau^{'}_m$, $\tau^{''}_m$
\State find all $n_{0}^l$ such that $t^{n_{0}^l}=l \tau_m^{''}, \, l \in \left \lbrace 0, \dots , 1/\tau_m^{''} \right \rbrace.$
\State find all $n_{\pm}^l$ such that $t^{n_{\pm}^l}=(l\pm1/2) \tau_m^{''}, \, l \in \left \lbrace 0, \dots , 1/\tau_m^{''} \right \rbrace.$

\For{$l$ ranging from $0$ to $1/\tau_m^{''}$}
%\State $n_m \gets n_0^l$, $n_{\text{inf}} \gets n_{-}^l$, $n_{\text{sup}}\gets n_{+}^l$
\If{ $l\neq0$ and $l\neq1/\tau_m^{''}$ }
\State $ \mathbf{Y}_+, \mathbf{Y}_-  \gets \mathrm{zeros}(N_x,N_x) $  \Comment{Inverse flow maps going forward  $\mathbf{Y}_+(t> l_k \tau_m^{''})$ and backward $  \mathbf{Y}_- (t< l_k \tau_m^{''})$  }
%\State $\vect^{n_0^l}= \nabla^\perp \phi_{m-1}^{n_0^l}$
\State $\phi_{m-1}^{n_0^l} \gets  \phi_{m-1}^{n_0^l} + \sum \limits_{k \in \mathbb{Z}} \hat{\zeta}_{m,l}^{n_0^l} \zeta_{m,k}^{n_0^l} \psi_{m,k} \circ \mathbf{Id} $ \Comment{In practice the sum contains at most one non-zero element } 
\State $f^{\pm}_0 \gets $ RHS$(\mathbf{Y}_\pm,\vect^{n_0^l})$
\State $\mathbf{Y}_\pm \gets  \mathbf{Y}_\pm  \pm \delta t (c_{0,0} f_0^\pm) $ \Comment{The $c_{i,r}$ are the coefficient of A-B order $r$ scheme}
\State $f^{\pm}_1 \gets $ RHS$(\mathbf{Y}_\pm,\vect^{n_0^l\pm 1})$
\State $\phi_{m-1}^{n_0^l\pm 1} \gets  \phi_{m-1}^{n_0^l \pm 1} + \sum \limits_{k \in \mathbb{Z}} \hat{\zeta}_{m,l}^{n_0^l\pm 1} \zeta_{m,k}^{n_0^l \pm 1} \psi_{m,k} \left(\mathbf{Y}_\pm + \mathbf{Id} \right)$ 
\State $\mathbf{Y}_\pm \gets  \mathbf{Y}_\pm  \pm \delta t (c_{0,1} f_1^\pm +c_{1,1} f_0^\pm) $

\State $f^{\pm}_2 \gets $ RHS$(\mathbf{Y}_\pm,\vect^{n_0^l\pm 2 })$
\State $\phi_{m-1}^{n_0^l\pm 2} \gets  \phi_{m-1}^{n_0 ^l\pm 2} + \sum \limits_{k \in \mathbb{Z}} \hat{\zeta}_{m,l}^{n_0^l\pm2} \zeta_{m,k}^{n_0^l \pm 2} \psi_{m,k} \left(\mathbf{Y}_\pm + \mathbf{Id} \right)$ 
\State $\mathbf{Y}_\pm \gets  \mathbf{Y}_\pm  \pm \delta t (c_{0,2} f_2^\pm +c_{1,2} f_1^\pm +c_{2,2} f_0^\pm) $

\State $f^{\pm}_3 \gets $ RHS$(\mathbf{Y}_\pm,\vect^{n_0^l\pm 3 })$

\State $\phi_{m-1}^{n_0^l\pm 3} \gets  \phi_{m-1}^{n_0^l \pm 3} + \sum \limits_{k \in \mathbb{Z}} \hat{\zeta}_{m,l}^{n_0^l\pm3} \zeta_{m,k}^{n_0^l \pm3} \psi_{m,k} \left(\mathbf{Y}_\pm + \mathbf{Id} \right)$ 

\For{$n_t$ ranging from $4$ to $n_+^l-n_0^l$}
\State $n_\pm \gets n_0^l \pm n_t$
\State $\mathbf{Y}_\pm \gets  \mathbf{Y}_\pm  \pm \delta t (c_{0,3} f_3^\pm +c_{1,3} f_2^\pm +c_{2,3} f_1^\pm  +c_{3,3} f_0^\pm) $
\State $ f_0^\pm \gets f_1^\pm$, $ f_1^\pm \gets f_2^\pm$, $ f_2^\pm \gets f_3^\pm$
\State $f^{\pm}_3 \gets $ RHS$(\mathbf{Y}_\pm,\vect^{n_\pm })$
\State $\phi_{m-1}^{n_\pm} \gets  \phi_{m-1}^{n_\pm} + \sum \limits_{k \in \mathbb{Z}} \hat{\zeta}_{m,l}^{n_\pm} \zeta_{m,k}^{n\pm } \psi_{m,k} \left(\mathbf{Y}_\pm + \mathbf{Id} \right) $
\EndFor
\ElsIf{$l=0$}
\State Integrate only forward in time
\Else
\State Integrate only backward in time
\EndIf
\EndFor
\end{algorithmic}
\end{algorithm}

\begin{figure}[htbp]
\centerline{\includegraphics[width=0.7\columnwidth]{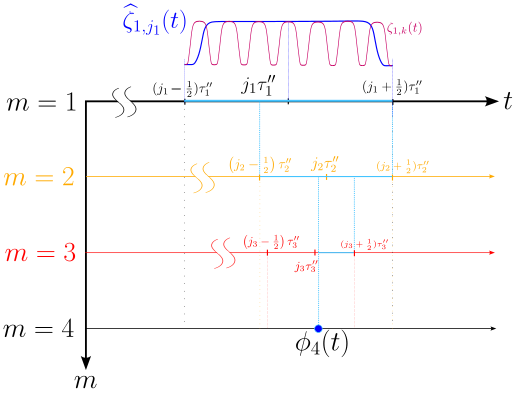}}
\caption{
Schematic representation of the non-local in time dependence of streamfunction $\phi_m$ with respect to $\phi_{p}, \, p< m$. In this example, one wish to compute the value of $\phi_4(t,\pos)$ at a given time. At each sub levels $m$, the light blue line represent the interval of time on which one need to know $\phi_{m<4}$  in order to compute $\phi_4(t)$. It is clear from this picture that the value of $\phi_4(t)$ depends on both posterior and ulterior values of $\phi_{m<4}$, making its computation difficult.}
\label{fig:RecurPhi}
\end{figure}

From a numerical perspective, the computation of the inverse flow map is the main bottleneck. This is due to its role in equation \eqref{eq:DefAVField}, which causes the streamfunction $\phi_m$ at a given time $t$ to depend not only on $(\phi_{p})_{p<m}$ at the same time but also on its values at earlier and later times. In other words, the inverse flow map makes the recursion relation \eqref{eq:DefAVField} non-local in time. Figure~\ref{fig:RecurPhi} provides a schematic illustration of this temporal non-locality.

This complex dependence of $\phi_m$ on $(\phi_{p})_{p<m}$, together with the requirement to integrate equation~\eqref{eq:InverseFlow} both forward and backward in time, represents the primary limitation on the numerical resolution we have been able to attain. Among the various methods explored, the simplest approach proved to be the most effective in achieving the highest spatial resolution.
 
Our procedure is as follows. First, we choose $M$ the number of iterations of \eqref{eq:DefAVField} we are computing and construct a space-time grid fine enough to resolve $\tau_M$ and $\epsilon_M$. We then compute $\phi_1$ explicitly on this fine grid and write it in a file. Since $\phi_1$ is now available over the entire space-time grid, we can compute $\phi_2$, store it, and continue this process iteratively until $\phi_M$ is obtained. The main constraint is the size of the file containing $\phi_M$. In our numerical experiment, $\phi_M$ is computed over the time interval $[0,1/2]$ on a $2048^2$ grid, resulting in a file several hundred gigabytes in size. This presents a significant limitation, preventing us from reaching higher number of iteration $M$. Nevertheless, this approach has the advantage of making the full field immediately accessible, which is particularly useful for studying passive scalar dynamics and Lagrangian trajectories.

The backward flow map equation~\eqref{eq:InverseFlow} is integrated in time using a third-order Adams-Bashforth scheme, with dealiased pseudo-spectral differentiation. 
For further details on the implementation of this scheme, we provide a pseudo-code in Algorithm~\ref{alg:cap}. All the numerical results presented here are obtained using a vector field $\vect_M$ computed on a Cartesian spatial grid with resolution $\delta x$ in each direction and a temporal resolution $\delta t$. The parameters used to compute 
$\vect_M$ in all our numerical simulations are listed in Table~\ref{tab:ParamsB}.
\begin{table}[h!]
\centering
\begin{tabular}{||c c c c c||} 
 \hline
 $M$ & $\Lambda$ & $\beta$ & $\delta x$ & $\delta t$ \\ [1ex] 
 \hline\hline
 $ 3$& $2.5$ & $1.2$ & $ \frac{1}{4}2^{\left\lfloor  \log_2\left(1/ \epsilon_M\right) \right\rfloor }  $ & $\tau_M/8$ \\  [1ex] 
 \hline
\end{tabular}
\caption{Parameters used in numerical simulations.}
\label{tab:ParamsB}
\end{table}
Theses parameters correspond to a space grid of $2048^2$ points and a time step $\delta t \simeq 4 \times 10^{-5} $. 

To determine whether anomalous diffusion can be observed numerically, we need to integrate the passive scalar equation. This equation is solved using a third-order exponential Adams-Bashforth scheme. This exponential integration method enables us to better handle the numerical stiffness of the Laplace operator. From the numerical integration of the passive scalar equation, we extract several observables, such as $\|\theta^\kappa\|_{L^2}$ and $ \|\nabla\theta^\kappa\|_{L^2}$ to assess whether the system exhibits anomalous diffusion. The passive scalar field is computed on the same space-time grid as $\vect_M$, and in the results we present the initial condition for the passive scalar equation is taken to be
\be
\theta_0(\pos)= \cos\left(2\pi x_1 \right) \sin \left( 2\pi x_2\right).
\de
We have explored other choices of initial condition, without significantly changing the results exposed later. 

In order to link anomalous diffusion and spontaneous stochasticity of Lagrangian tracers, we integrate the backward in time stochastic Lagrangian trajectories~\eqref{eq:BackLag}. Following \cite{Drivas17} The SDE is integrated using an Euler-Maruyama scheme with the same time step as 
that used for computing $\vect_M$. The variance is then estimated using the unbiased estimator
$$ \sigma^2_\kappa(s) :=  \dfrac{2}{N-1} \sum_{n=1}^N  | \mathbf{X}_n(s) -\overline{\mathbf{X}}_N(s) |^2 $$
where $\mathbf{X}_n$ are independent realizations of the process \eqref{eq:BackLag} and 
$$ \overline{\mathbf{X}}_N(s)= \dfrac{1}{N} \sum_{n=1}^N   \mathbf{X}_n(s),$$
is an estimation of the mean tracers position. 
\subsection{Further results}
\paragraph{Backward probability transition for passive tracers}

Let us recall from Section~\ref{subsubSPAV} that the anomalous diffusion of the passive scalar is related to the backward It\^o process \eqref{eq:BackLag} via
$$2\kappa \| \nabla \theta^\kappa\|_{L^2((0,t),\mathbb{T}^2)}= \int_{\mathbb{T}^2}\mathrm{Var}\!\left[ \theta_0(\flow^\kappa_0(\mathbf{x})) \right] d\mathbf{x}.$$
Let $p^\kappa(s,\mathbf{y}\,|\,t,\mathbf{x})\,d\mathbf{y}$, for $s<t$, denote the probability that a Lagrangian tracer was at position $\mathbf{y}$ at time $s$ given that it is at $\mathbf{x}$ at time $t>s$. Then
$$\mathbb{E}\!\left[ \theta_0(\flow^\kappa_0(\mathbf{x})) \right] = \int_{\mathbb{T}^2} \theta_0(\mathbf{y})\, p^\kappa(0,\mathbf{y}\,|\,t,\mathbf{x})\, d\mathbf{y}.$$
Using this identity, the integrated variance can be written as
$$2\kappa \| \nabla \theta^\kappa\|_{L^2((0,t),\mathbb{T}^2)}= \int_{(\mathbb{T}^2)^4} (\theta_0(\mathbf{y})-\theta_0(\mathbf{z}_1))(\theta_0(\mathbf{y})-\theta_0(\mathbf{z}_2))\, p^\kappa(0,\mathbf{y}\,|\,t,\mathbf{x})\, p^\kappa(0,\mathbf{z}_1\,|\,t,\mathbf{x})\, p^\kappa(0,\mathbf{z}_2\,|\,t,\mathbf{x})\, d\mathbf{z}_1\, d\mathbf{z}_2\, d\mathbf{y}\, d\mathbf{x}.$$

As long as $p^\kappa$ does not converge toward a Dirac mass in the vanishing–diffusivity limit, the dissipation rate remains finite. In addition to the variance of backward Lagrangian tracers presented in paragraph~\ref{par1_4_1_3}, we also computed the backward transition probability $p^\kappa(0,\mathbf{y}|t,\pos)$ for different values of $\kappa$ in the geometric and hyper-geometric cases, using histograms.

\begin{figure}[htpb]
\centerline{\includegraphics[width=0.98\linewidth]{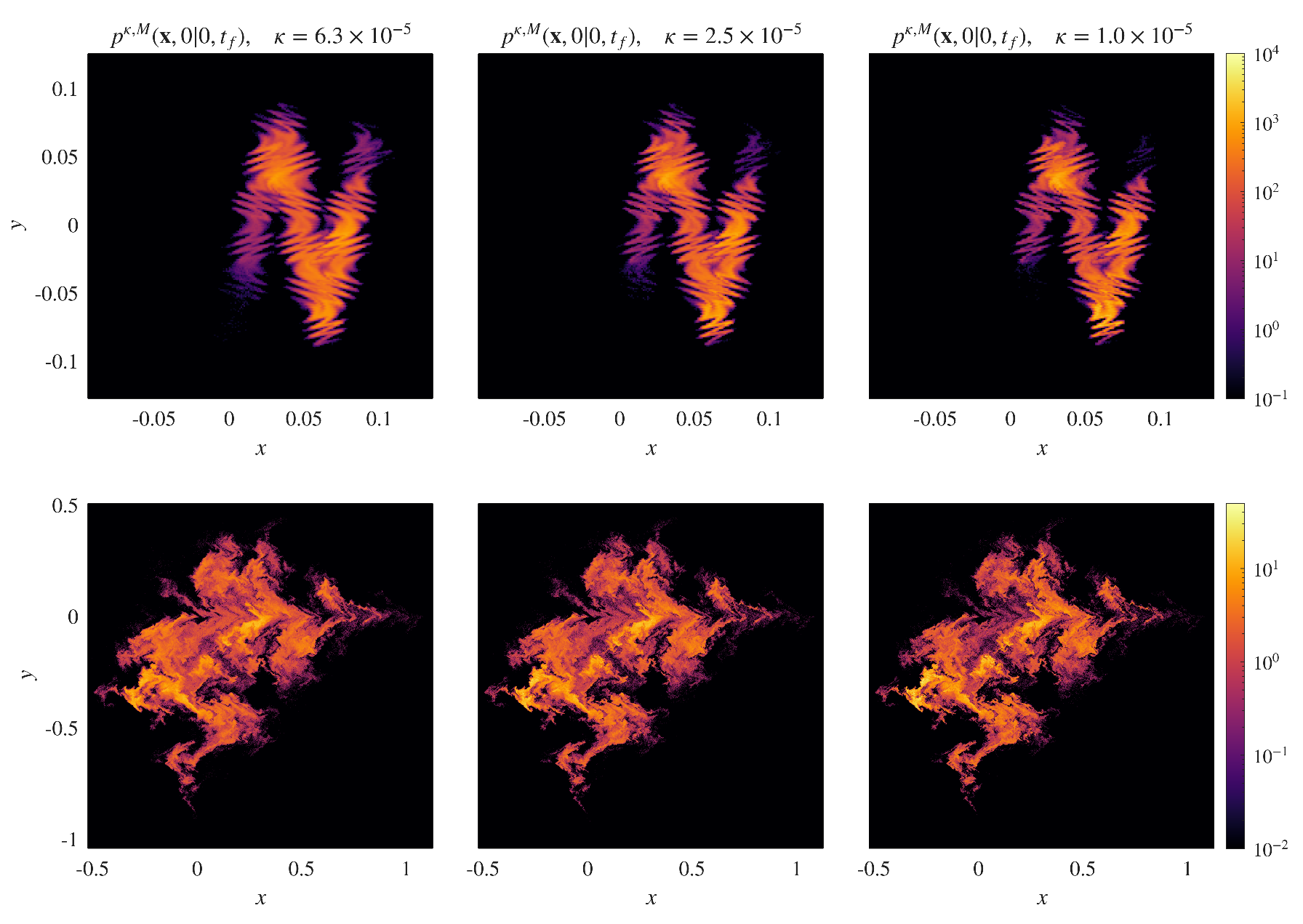}}
\caption{Backward transition probabilities in the hyper-geometric (top row), geometric (bottom row) cases. From left to right, the value of $\kappa$ becomes smaller, illustrating the convergence toward a non Dirac distribution and therefore the spontaneous stochasticity of the Lagrangian tracers. In the geometric case the strong interplay between scales leads to a more 'turbulent' picture than in the hyper-geometric scale separation.
}
\label{fig:BackwardProbs}
\end{figure}
 Figure~\ref{fig:BackwardProbs} shows the results of numerical simulations, which indeed confirm the lack of convergence of $p^\kappa$ toward a Dirac measure for both geometric and hyper-geometric scale separations.

\paragraph{Obukhov-Corrsin regularity of the passive scalar }
\label{OCpar} 
Beyond anomalous diffusion for the passive scalar, a central question concerns the regularity of the scalar field in the vanishing diffusivity limit. Based on scaling arguments from Kolmogorov’s (K41) phenomenology, Obukhov and Corrsin \cite{corrsin1951,Obukhov49} independently predicted that a passive scalar advected by a monofractal velocity field with regularity near $1/3$ should enjoy the same regularity in the vanishing diffusivity limit, provided the scalar exhibits anomalous diffusion. More generally, one may consider a passive scalar advected by a vector field of arbitrary regularity. For $1 \leq p,q \leq \infty$ with $1/p + 2/q = 1$, $\alpha \in (0,1)$, and $\alpha_{\rm oc} = (1-\alpha)/2$, the Obukhov–Corrsin conjecture can be stated, in a loose sense, through the following dichotomy:

\begin{enumerate}[label=(\roman*)]
        \item \label{conj:rigid} If $\mathbf{b} \in L^p_tC^\alpha_x$ and $\lbrace\theta^\kappa\rbrace_{\kappa >0}$ is bounded in $L^q_tC_x^\sigma $ uniformly in $\kappa$ for some $\sigma> \alpha_{\rm oc} $, then there is no anomalous diffusion. In addition, the underlying advection equation admits a unique solution, which coincides with the vanishing diffusivity limit of the advection diffusion equation.
        \item \label{conj:flexible}  There exists a divergence free $\mathbf{b} \in L^p_tC^\alpha_x$ such that $\lbrace\theta^\kappa\rbrace_{\kappa >0}$ is bounded in $L^q_tC_x^\sigma $ uniformly in $\kappa$ for some $\sigma < \alpha_{\rm oc} $ and for \emph{all} initial condition, then the passive scalar exhibits anomalous diffusion. In addition, the underlying advection admits infinitely many solutions, paving the way for Eulerian spontaneous stochasticity.
\end{enumerate}
Part~\ref{conj:rigid} of the conjecture can be derived via a Constantin-E-Titi \cite{CET94} type argument; see \cite{colombo,Drivas_Elgindi,AAV25} for a detailed discussion.
The precise formulation of Part \ref{conj:flexible}, however, is still debated.  In \cite{colombo}, the authors construct an incompressible vector field exhibiting anomalous diffusion and a uniform-in-$\kappa$ bound for the passive scalar, but only for certain initial conditions and with anomalous diffusion occurring in discrete time. Subsequently, \cite{ElgLiss24} established the existence of a divergence-free vector field for which Part~\ref{conj:flexible} holds, though again only with discrete-time anomalous diffusion. In \cite{AAV25}, Armstrong and Vicol suggest that Part~\ref{conj:flexible} should additionally require the dissipation measure to be non-atomic, a characteristic feature of turbulence. Imposing this extra condition leaves the conjecture unresolved. In \cite{AV24}, Armstrong and Vicol claim (in a yet-unpublished proof) to have established a uniform-in-$\kappa$ bound for the scalar field, which, in their construction -- where anomalous diffusion occurs continuously in 
time -- would settle Part~\ref{conj:flexible} of the Obukhov–Corrsin dichotomy. We complement their claim with numerical evidence showing that $\{\theta^\kappa\}_{\kappa>0}$ remains bounded in $L^\infty_t C^{\sigma}_x$ uniformly in $\kappa$ for all $\sigma < \alpha_{\rm oc} = (1-\alpha)/2$. To obtain reliable numerical results, it is necessary to sample the $(\kappa,\sigma)$ plane densely enough. For each sample point, one must evaluate the associated H\"older seminorm of the passive scalar at each (or most) time steps. From a computational perspective, this is very costly, as H\"older seminorms cannot be computed efficiently using the Fast Fourier Transform (FFT) algorithm. Calculating them for all time steps and for a large set of $(\kappa,\sigma)$ values would be entirely prohibitive. 

To circumvent this, we compute an equivalent Besov norm via Littlewood-Paley decomposition, which allows the use of the FFT algorithm. Precise definitions of these norms are given in \cite{Bahouri2011}, and the equivalence between the relevant Besov and H\"older norms is stated in \cite[Theorem~2.36]{Bahouri2011}. For convenience, we introduce here a loose formulation of the Besov norm to facilitate the exposition.

Let $\varphi$ be a smooth, compactly supported function on an annulus, to be used as a Fourier multiplier. We define the operator
$$
\Delta_j u(x) = \int_{\mathbb{R}^2} e^{i \xi \cdot x} \, \varphi(2^{-j} \xi) \, \widehat{u}(\xi) \, d\xi,
$$
where $\widehat{u}$ denotes the Fourier transform of $u : \mathbb{R}^2 \to \mathbb{R}^2$. This operator acts as a band-pass filter, retaining only Fourier modes with modulus of order $2^j$. 

For $\sigma \in \mathbb{R}$, we numerically compute the norm
\be \label{eq:DefBesNorm}
\|u\|_{\dot{B}_{\infty,\infty}^\sigma}
:= \sup_{j \in \mathbb{Z}} \, 2^{j\sigma} \, \sup_{x \in \mathbb{R}^2} \left| \Delta_j u(x) \right|.
\de
This norm is computationally efficient, since it can be evaluated using the FFT algorithm. As stated in \cite[Theorem~2.36]{Bahouri2011}, $\|\cdot\|_{\dot{B}_{\infty,\infty}^\sigma}$ is equivalent to the usual H\"older norm $\|\cdot\|_{C^\sigma}$. 

Using \eqref{eq:DefBesNorm}, we define
$$
\mathcal{N}(t,\kappa,\sigma) := \|\theta^\kappa(t,\cdot)\|_{\dot{B}_{\infty,\infty}^\sigma},
\quad
\overline{\mathcal{N}}(\kappa,\sigma) := \sup_{t \in [0, t_s]} \|\theta^\kappa(t,\cdot)\|_{\dot{B}_{\infty,\infty}^\sigma}.
$$

\begin{figure}[htpb]
\centering
\includegraphics[width=0.58\linewidth]{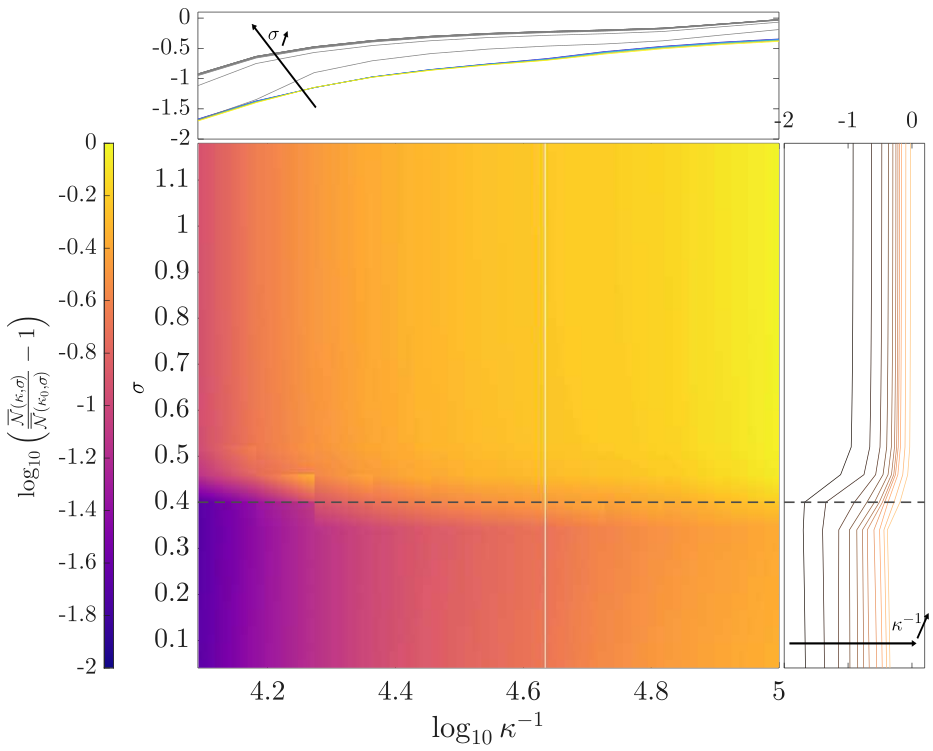}
\includegraphics[width=0.38\linewidth]{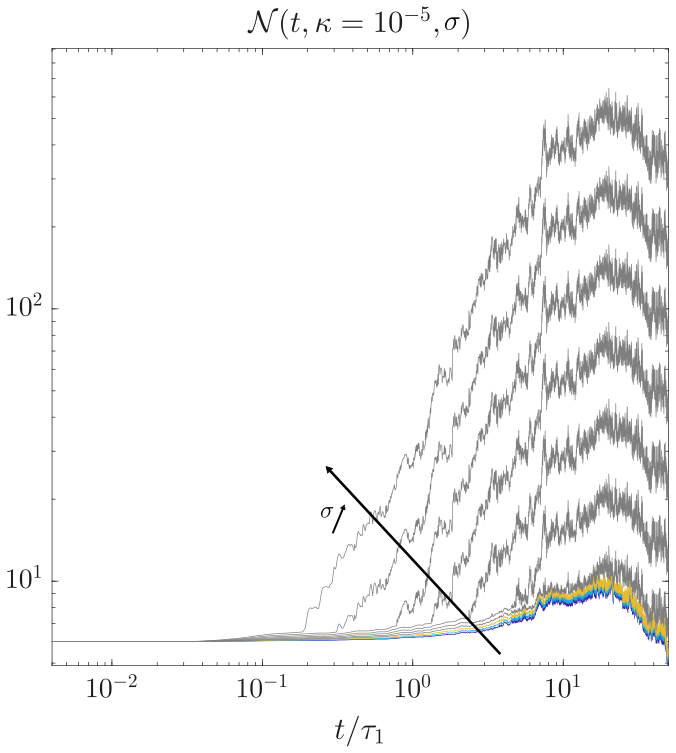}
\caption{ Left -- norm $\overline{\mathcal{N}}(\kappa,s)$ (normalized by $\overline{\mathcal{N}}(\kappa=10^{-4},s)$ and with logarithmic scale) as a function of $s$ and $\kappa$, the dotted line corresponds to $s=s_{\text{oc}}$. The top plot is a one dimensional visualization of the color plot showing the same information as function of $\kappa$ for different values of $s$. In the top plot, colored lines correspond to $s<s_{\text{oc}}$ while gray ones are for  $s\geq s_{\text{oc}}$. The right plot of the right panel correspond. Right -- Besov norms $\mathcal{N}(t,\kappa=10^{-5},s)$ as a function of time for different values of $s$. Gray solid lines are obtained for $s\geq s_{\text{oc}}=0.4$ while colored ones are for $s< s_{\text{oc}}$. }
\label{fig:BesovNorms}
\end{figure}

We present the results in Figure~\ref{fig:BesovNorms} for the case of geometric scale separation only.  

The right panel shows $\mathcal{N}(t,\kappa,\sigma)$ for $\kappa = 10^{-5}$ and various values 
$\,\alpha_{\rm oc}/10 \leq \sigma \leq 3\,\alpha_{\rm oc}$, with $\alpha_{\rm oc} = 0.4$ in our setting.  
In this panel, gray curves correspond to $\sigma \geq \alpha_{\rm oc}$, while colored curves represent the remaining $\sigma$ values.  
For $\sigma \geq \alpha_{\rm oc}$, $\mathcal{N}(t,\kappa,\sigma)$ stays roughly constant until a $\sigma$-dependent time, after which the norm grows rapidly.  
The onset of this growth is delayed as $\sigma$ approaches $\alpha_{\rm oc}$, and for $\sigma < \alpha_{\rm oc}$ (colored lines), the norm remains close to its initial value throughout the simulation.  
This transition suggests that, for fixed $\kappa$, the time evolution of the Besov norm changes qualitatively when $\sigma$ crosses the critical 
Obukhov-Corrsin regularity $\alpha_{\rm oc}$.  
While indicative, this observation alone does not fully resolve the question.

The left panel of Figure~\ref{fig:BesovNorms} offers further insight.  
Here, the main plot shows $\overline{\mathcal{N}}(\kappa,\sigma)$ as a function of both parameters,  
allowing us to examine uniform boundedness in $\kappa$ and to identify the critical regularity $\sigma_c$.  
Our simulations are naturally restricted to a finite set of $\kappa$ values, and numerically we can only treat finite $M$,  
so that the associated vector field $\vect_M$ remains smooth and, in particular, Lipschitz continuous.  
Lipschitz regularity guarantees uniqueness of solutions to the advection equation and ensures that these solutions are themselves Lipschitz  
(see \cite{alberti2019loss} for a striking counterexample when the vector field fails to be Lipschitz).  
In this setting, we thus expect $\overline{\mathcal{N}}(\kappa,\sigma)$ to remain uniformly bounded in $\kappa$ for any $\sigma \in (0,1)$ in our numerical experiments.

Our numerical simulations reveal a sharp transition in the behavior of $\overline{\mathcal{N}}(\kappa,\sigma)$ at the Obukhov–Corrsin regularity $\alpha_{\rm oc} = 0.4$.  
Beyond the H\"older regularity analysis of the passive scalar, we also examined its Sobolev regularity (data not shown).  
These simulations display a power-law decay in the squared modulus of the Fourier modes of the solution, observed at sufficiently large times and wavenumbers, and in the vanishing diffusivity limit.  
The decay exponent, equal to $\alpha - 2$, is consistent with a Sobolev regularity strictly below $\alpha_{\rm oc}$ and not exceeding it.  
We conjecture, however, that this observed power law is influenced by the limited small-scale separation accessible in our computations,  
and would not occur under the assumptions considered in \cite{AV24}.

\section{Proof of Theorem \ref{THEOESSAV}}\label{TheoESSAV_proof}
As explained in the introduction, the strategy  is to take advantage of the proof of Proposition 5.5 \cite{AV24} by showing that one has
Strong Eulerian-$\SP$ in the sense of Definition \ref{SPDEF4}.
We recall that when we say $\kappa \to 0$, it is implicitly assumed $\kappa \in {\cal K}$. We introduce the map $\Phi$:
\be \label{phidef}
\Phi: {\cal K} \to \LLT, \Phi(\kappa) = \theta^\kappa(t,\cdot)~\text{solution of \eqref{ad_dif0}}.
\de 

\subsection{The sequence}

Before hand, it is useful to recall some technical aspects which appear
in \cite{AV24}, in particular the Section 3.3. Rather than the sequence
(\ref{avseq}), the "true" sequence of renormalized diffusivities is
in fact a much more complicate expression involving space-time averages of flux correctors (see (3.15)-(3.16) in \cite{AV24}). This sequence is denoted for $m \in \{1,\cdots,M\}$:
\begin{equation}\label{trueavseq}
\kappa_{m-1} = \overline{{\bf K}}_m^{\kappa_m},~\kappa_M = \kappa.
\end{equation}
The sequence defined in (\ref{avseq}) is close to (\ref{trueavseq})
 in the sense of Lemma 3.3 \cite{AV24}: there exists a constant $C=C(\beta) \geq 1$
such that for $\kappa > 0$ and all $m \in \mathbb{N}$
\begin{equation} \label{trueavesti}
\left| \overline{\bf K}_m^\kappa -
 f_m(\kappa) {\bf I}_2 \right|
\leq \frac{\epsilon_m^{2\beta}}{\kappa} \left( C \frac{\epsilon_m^2}{\kappa \tau_m}
+ C \epsilon_{m-1}^\delta \right),~~
f_m(\kappa) := \kappa + c_0 \frac{\epsilon_m^{2\beta}}{\kappa}.
\end{equation}
In particular, to ensure that the two map iterates $\kappa \mapsto \overline{\bf K}_m^\kappa$
and $\kappa \mapsto f_m(\kappa)$ stay close to each other, one must have $\frac{\epsilon_m^2}{\kappa \tau_m} \ll 1$. This is the condition of Lemma 3.4
$(3.46_\text{\cite{AV24}})$: $c \epsilon_{m-1}^{2\delta} 
\leq \frac{\epsilon_m^2}{\kappa_m \tau_m} \leq C \epsilon_{m-1}^{2\delta}$ with universal
constants $0 < c < C < \infty$. Here $\tau_m$ is a timescale such that
$2^{-33} \epsilon_{m-1}^{2-\beta+4\delta} \leq \tau_m \leq 2^{-28} \epsilon_{m-1}^{2-\beta+4\delta}$ (see Eqs. (2.12--2.16)$_\text{\cite{AV24}}$). 

Then, Lemma 3.4 \cite{AV24} shows that, (\ref{trueavesti}) remains bounded on the \emph{admissible} set of diffusivities
% ok en fait j'avais opte pour le terme utilise dans AV mais franchement je prefere largement "admissible"
${\cal K}$ (see Eqs. (\ref{avseq})).  In more details,
\be  \label{avseqstaybounded}
\kappa \in \left[ \frac12 \sqrt{c_0} \epsilon_M^\frac{2\beta}{q+1},2 \sqrt{c_0} \epsilon_M^\frac{2\beta}{q+1}\right] \Rightarrow 
\kappa_m \in \left[ c \epsilon_m^\frac{2q \beta}{q+1}, C \epsilon_m^\frac{2 q\beta}{q+1} \right].
%\subset \left[ \frac12 \epsilon_m^\frac{2\beta}{q+1},2 \epsilon_m^\frac{2\beta}{q+1}\right],~\mbox{for $m$ large enough}.
\de 
For self-consistency, we provide a proof in Appendix \ref{Kgen}.

We then quote the Lemma 5.6 of \cite{AV24} explicitly, as it plays a key role in the proof of Proposition 5.5 and for our purpose:
\begin{lemma5_6}\label{lemma56}
There exists $C(\beta)<\infty$ and $\rho(\beta)>0$ such that, if $\Lambda \geq C$, then, for each $\tau \in \left[\frac12,2 \right]$, there exists an infinite sequence $\{\kappa_m \}_{m \in \mathbb{N}}$ satisfying
$$
\kappa_{m-1} = \overline{{\bf K}}_m^{\kappa_m}, \forall m \geq 1,
$$
such that
\be \label{kappam_tau}
\left| \frac{\kappa_m}{\sqrt{c_0} \epsilon_m^{\frac{2\beta}{q+1}}} - \tau^{(-1)^m} 
    \right| \leq C \epsilon_m^{\rho} \leq \frac{1}{100}, \forall m \in \mathbb{N}.
\de 
\end{lemma5_6}
In fact, one
builds a sequence (see (5.115)$_\text{\cite{AV24}}$):
\be \label{smartseq}
\kappa_{m-1}^{(M)} = \overline{{\bf K}}_m^{\kappa_m^{{(M)}}}~\text{and}~
\kappa_M^{(M)}:= \sqrt{c_0} \epsilon_M^{\frac{2\beta}{q+1}} \tau^{(-1)^M}.
\de 
In order for this sequence to stay bounded in the limit $M \to \infty$, one must have $\frac12 \sqrt{c_0} \epsilon_M^{\frac{2\beta}{q+1}} \leq \kappa_{M}^{(M)} \leq 
2 \sqrt{c_0} \epsilon_M^{\frac{2\beta}{q+1}}$; see \eqref{avseqstaybounded}. From the choice of $\kappa_{M}^{(M)}$, this is precisely $\frac12 \leq \tau \leq 2$.
\paragraph{New Lemma 5.6}
We provide a slightly more general version of Lemma 5.6:
\begin{lemma}\label{mylemma56}
Denote $\kappa_m(\tau)$, the sequence \eqref{smartseq} as $M \to \infty$.
Let $I_\tau = \left[ \frac12,2 \right]$, and $\pi \in {\cal P}(I_\tau)$ such that $\pi \sim \operatorname{Lebesgue}$ (strictly positive density a.e.). Then, there exists
a subdomain ${\cal J} \subset I_\tau \times I_\tau$ such that for all $m \in \mathbb{N}$
\be \label{secureratio}
\forall (\tau,\tau') \in {\cal J},~ 
\frac54 \leq \max \left\{ \frac{\kappa_m(\tau)}{\kappa_m(\tau')},
\frac{\kappa_m(\tau')}{\kappa_m(\tau)} \right\} \leq 2,~\text{and}~(\pi \otimes \pi)({\cal J}) > 0 .
\de 
\end{lemma}
\begin{proof}
As already noticed, the choice of $\kappa_M^{(M)}$ implies that 
\eqref{smartseq} stays bounded. The set ${\cal J}$ can be characterized explicitly: let $I_\tau$ be written as $I_\tau = [\tau_m,\tau_M]$. Let $\tau < \tau'$, for all $\tau \in [\tau_m,\frac45 \tau_M]$, and $\tau' \in [\frac54 \tau, \min(2 \tau,\tau_M)]$,  the condition \eqref{secureratio} holds. Of course, neither the definition of ${\cal K}$ nor the constants $5/4$ and $2$ are optimal. The goal is to have a ratio bounded away from 1.
\end{proof}
Let $m$ be fixed, and $\kappa_m$ the limit when $M \to \infty$ of \eqref{smartseq}. It defines a function $\tau \mapsto \kappa_m(\tau)$ such that \eqref{kappam_tau} is satisfied for all $\tau \in I_\tau$. In particular, one can write $\kappa_m(\tau) = \tau^{(-1)^m} \sqrt{c_0} \epsilon_m^{\frac{2\beta}{q+1}}
\left(1 + O(\epsilon_m^\rho) \right)$. Let $\pi \in {\cal P}(I_\tau)$ a probability measure on $I_\tau$, we define the probability measure for $\kappa$ to be in the interval $I_m$ (see \eqref{avseq}) under Lemma 5.6:
\be 
\mathscr{K}_m:= (\kappa_m)_\# \pi \in {\cal P}(I_m).
\de 
In particular, $\{0 \} \notin \operatorname{supp} \mathscr{K}_m$.
\subsection{Probability measure in $\LLT$ and the limit $m\to \infty$}
\paragraph{Definition of $\Theta_m \in {\cal P}(\LLT)$}
We now look at the solution of \ref{ad_dif0} from a measure-theoretic point of view. We consider the function
\be 
\tau \in I_\tau \mapsto \Phi(\kappa_m(\tau)) \in L^2.
\de 
where $\Phi$ is defined in \eqref{phidef}, namely 
$\Phi(\kappa_m(\tau))$ is the solution of
\eqref{ad_dif0} at time $t$ with diffusivity $\kappa_m(\tau)$.
We define the probability measure for a solution to be in the domain $\Phi(I_m)$:
\be 
\Theta_m := \Phi_\# \mathscr{K}_m \in {\cal P}(\Phi(I_m)) \subset
{\cal P}(\LLT).
\de 
In order to fully justify that $\Theta_m$ is indeed a probability measure on $L^2$, one must show that $\Phi$ is measurable. We have more than measurability, since $\Phi$ is continuous on ${\cal K}$.
\begin{proof}
 Since $\{0 \} \notin \operatorname{supp}(\mathscr{K}_m)$ it suffices to show that
the map $\kappa \mapsto \Phi(\kappa)$ is continuous on 
${\cal K}$. 
Let $\kappa_0 > 0$ fixed, we want to show that $\| \theta^\kappa(t,\cdot) - \theta^{\kappa_0}(t,\cdot) \|_{\LLT} \to 0$ as
$\kappa \to \kappa_0$. The difference $w = \theta^\kappa-\theta^{\kappa_0}$ satisfies
$$
\partial_t w + {\bf b} \cdot \nabla w = \kappa \Delta w + (\kappa-\kappa_0) \Delta \theta^{\kappa_0},~w(0,\cdot) = 0. 
$$
We recall that, by standard Schauder estimates, the solution
belongs to $C_t^0 C_x^{2,\alpha} \cap C_t^{1,\alpha/2} C_x^0$, see \cite{AV24}.
We then use a classical energy estimate together with Poincar\'e inequality with constant $C$ (using the fact that $\theta^\kappa,\theta^{\kappa_0}$ have zero mean):
$$
\frac{d}{dt} \| w \|_{L^2}^2 \leq -2 C^2 \kappa \|w \|^2_{{L^2}} + 2|\kappa- \kappa_0 | 
\|w \|_{L^2} \| \Delta \theta^{\kappa_0} \|_{L^2}.
$$
We obtain
$$
\|w(t) \|_{L^2} \leq |\kappa-\kappa_0| \int_0^t \|\Delta \theta^{\kappa_0}(s) \|_{L^2}~ds \leq |\kappa-\kappa_0| C_{t,\kappa_0}.
$$
One has therefore $\| w(t) \|_{L^2} \to 0$ as $\kappa \to \kappa_0$
and where $C_{t,\kappa_0}< + \infty$ showing that the map $\kappa\mapsto \Phi(\kappa)$ is continuous.   
We conclude, since $\{0 \} \notin \operatorname{supp}(\mathscr{K}_m)$ that
$\Theta_m = \Phi_\# \mathscr{K}_m \in {\cal P}(\LLT)$.
\end{proof}
As a consequence $\Phi(I_m)$ is a compact set of $L^2$. It shows
tightness of the individual measures $\Theta_m$ but not for the whole family $\{ \Theta_m \}_{m \geq m^\star}$. In fact, the next
paragraph shows that these probability measures converge of as $m \to \infty$ using Cauchy property, i.e. without the need to show tightness of the family prior, but tightness
following as a consequence.
\paragraph{$\Theta_m \rightharpoonup \Theta_\infty$}
\begin{proof}
We now use a key result from \cite{AV24}, Proposition 5.5 (5.90)+(5.141)+(5.108), which shows that for some $\mu(\beta) > 0$
$$
\theta^{\kappa_m}~\text{is Cauchy in}~C^{0,\mu/2}([0,1];\LLT).
$$
In more details, one fixes the time $t$ and uses the uniform bound
(5.141):
\be \label{phi_phim_bound}
\| \theta^{\kappa_m}(t,\cdot) - \theta_m(t,\cdot) \|_{L^2} \leq C \epsilon_m^\sigma \|\theta_0 \|_{L^2},~\sigma = 2 \beta (q - \frac{2}{q+1}) > 0,
\de 
where $\theta_m$ is solution of \eqref{tm} with ${\bf b}_m$ instead of ${\bf b}$. It is Cauchy thanks to the hyper-geometric decay of $\epsilon_m$, this is (5.90):
$$
\| \theta_{m+1}(t,\cdot) - \theta_m(t,\cdot) \|_{L^2} \leq C \epsilon_m^{\delta/2} \|\theta_0 \|_{L^2}
$$
One has therefore using our notations, uniformly in $\tau$, 
$$
\| \Phi(\kappa_{m+1}(\tau)) - \Phi(\kappa_m(\tau)) \|_{\LLT} \leq r_m \to 0,~\text{with}~r_m= C (\epsilon_m^\sigma + \epsilon_{m+1}^\sigma + \epsilon_m^{\delta/2}) \|\theta_0 \|_{L^2}.
$$
and thus
\be 
\| \Phi(\kappa_{m'}(\tau)) - \Phi(\kappa_m(\tau)) \|_{L^2}
\leq \sum_{j=m}^{m'-1} \| \Phi(\kappa_{j+1}(\tau)) - \Phi(\kappa_j(\tau)) \|_{L^2} \leq \sum_{j=m}^{m'-1} r_j \to 0,
\de 
using again the hypergeometric decay of $r_j$. As a consequence of the Cauchy property, one has the pointwise convergence w.r.t. $\tau$:
$$
\forall \tau \in I_\tau, \lim_{m \to \infty} \Phi(\kappa_m(\tau)) = \phi[\tau]~\text{strongly in}~\LLT.
$$
We recall that for all test functions in $C_b(L^2;\mathbb{R})$
$$
\langle \Theta_m,F \rangle =  \int_{I_m} F(\Phi(\kappa)) \mathscr{K}_m(d\kappa) = \int_{I_\tau} F(\Phi(\kappa_m(\tau)))~\pi(d\tau).
$$
Since $F$ is bounded and continuous, and one has pointwise convergence, we can use the dominated convergence theorem:
\be 
\lim_{m \to \infty} \langle \Theta_m,F \rangle = \int_{I_\tau} 
\lim_{m \to \infty} F(\Phi(\kappa_m(\tau)))~\pi(d\tau) = 
\int_{I_\tau} F(\phi[\tau]) ~\pi(d\tau) =: \langle \Theta_\infty,F \rangle.
\de 
\end{proof}
\subsection{$\Theta_\infty$ is non-Dirac}
\begin{proof}
A direct consequence found in the proof of Proposition 5.5 (see (5.142)$_\text{\cite{AV24}}$) is that, for all $m$ sufficiently large and for some time $t \in [0,1]$, there exists a positive constant $C(\beta,L_{\theta_0})$, where $L_{\theta_0} = \|\theta_0\|_{L^2} / \| \nabla \theta_0 \|_{L^2}$, such that for all $(\tau,\tau') \in {\cal J}$, {$\tau \neq \tau'$}
% yes {\cal J} does not contain the diagonal
\be 
\left| \| \Phi(\kappa_m(\tau)) \|_{L^2} - \| \Phi(\kappa_m(\tau'))\|_{L^2}
\right| \geq \frac12 C(\beta,L_{\theta_0}) \|\theta_0 \|_{L^2} > 0.
\de 
In particular, it holds as well in the limit $m \to \infty$:
\be \label{phitauND}
\left| \| ~\phi[\tau]~ \|_{L^2} - \| ~\phi[\tau']~\|_{L^2}
\right| \geq \frac12 C(\beta,L_{\theta_0}) \|\theta_0 \|_{L^2} > 0.
\de 
This is indeed the main result of Proposition 5.5 showing a lack of selection principle. We then proceed by contradiction, assuming $\Theta_\infty = \delta_{\phi^\star}$. We have, using Lemma \ref{mylemma56} and \eqref{phitauND}
$$
\int_{I_\tau} \int_{I_\tau} \left| \|~\phi[\tau]~\|_{L^2} - \|~\phi[\tau']~\|_{L^2}\right| (\pi \otimes \pi) (d\tau d\tau') \geq 
\frac12 C(\beta,L_{\theta_0}) \|\theta_0 \|_{L^2} (\pi \otimes \pi) ({\cal J}) > 0.
$$
However, since $\phi[\tau] = \phi^\star~\pi-a.e.~\tau $, the left-hand side is zero showing a contradiction.
\end{proof}
\subsection{Strong Eulerian-$\SP$}
\begin{proof}
From \eqref{smartseq} and the above construction, one has
\be 
\Theta_\infty = \lim_{m \to \infty} \lim_{M \to \infty} \left( \Phi \circ
\kappa_m^{(M)}\right)_\# \pi.
\de 
By a classical diagonal argument, there exists a function $g:\mathbb{N} \to \mathbb{N}$, such that 
$$
\lim_{M \to \infty} \left(\Phi \circ \kappa_{g(M)}^{(M)}\right)_\# \pi =  \Theta_\infty.
$$
Let $\mathfrak{M}:{\cal K} \to \mathbb{N}, \kappa \mapsto \mathfrak{M}(\kappa)$ such that for $\kappa \in I_M, \mathfrak{M}(\kappa) = M$.
One can therefore express the limit as
\be \label{AVambient}
\Phi_\# \mathbb{P}_\kappa \underset{\kappa \to 0}{\rightharpoonup{}} \Theta_\infty,~\text{with}~
\mathbb{P}_\kappa :=  
\left( \kappa_{g({\mathfrak{M}}(\kappa))}^{({\mathfrak{M}}(\kappa))} \right)_\# \pi.
\de 
Since $\Phi$ is continuous (except at 0) and by construction $\mathbb{P}_\kappa \in {\cal P}({\cal K})$ is a.c. w.r.t. Lebesgue, the proof is complete.
\end{proof}
\subsection{$\LSPO_{L^2}\setminus \delta$}
In order to fully conclude, we prove that one has in addition $\LSPO_{L^2}\setminus \delta$. Note that, since one has Strong-$\SP$, it already rules out $\delta$-$LSP$.
\begin{proof}
Let $\phi_1,\phi_2$ denote two weak solutions satisfying \eqref{phitauND}.
In particular, they belongs to $\operatorname{supp}(\Theta_\infty)$
so that for all $\delta>0$, one has $\Theta_\infty(B_i) > 0,~i=1,2$ for open balls $B_i:= B_\delta(\phi_i)$ centered at $\phi_i$ of radius $\delta$. From the Portmanteau Theorem; see Appendix \ref{PFTP}, 
one has $\liminf_{\kappa \to 0} \Phi_\# \mathbb{P}_\kappa(B_i)
\geq \Theta_\infty(B_i) > 0$. It implies that for all $\epsilon > 0$,
one can pick two values  $\kappa_1,\kappa_2 \in (0,\epsilon)$ such that 
$\Phi(\kappa_1) \in B_1$ and $\Phi(\kappa_2) \in B_2$. Therefore, by reverse triangle inequality
$$
\limsup_{\kappa \to 0} \| \Phi(\kappa) \|_{L^2} - \liminf_{\kappa \to 0}
\| \Phi(\kappa) \|_{L^2} \geq \left| \|\phi_1 \|_{L^2} - \| \phi_2 \|_{L^2} \right| -2\delta
\geq \frac12 C \|\theta_0 \|_{L^2} - 2 \delta > 0,
$$
for the choice   $0 < \delta < \frac14 C \| \theta_0 \|_{L^2}$.
\end{proof}

\subsection{Numerical results for the renormalized diffusivities}
This section aims to numerically illustrate the behavior of the renormalized diffusivities \eqref{avseq}. 
We recall the definition: let 
$I_m :=\left[
\frac12 \sqrt{c_0} \epsilon_m^{\frac{2 \beta}{q+1}},2 \sqrt{c_0} \epsilon_m^{\frac{2 \beta}{q+1}}\right]$, $\beta \in (1,\frac43)$, $q = \frac{\beta}{4 (\beta -1)} > 1$, and given $\kappa$, define $M=M(\kappa)$ such that $\kappa \in I_M$, the sequence of diffusivities is
$$
\kappa_{m-1} = F_m(\kappa_m),~~\kappa_M = \kappa,~~{\rm with}~~
F_m(\kappa) := \kappa + 
c_0 \frac{\epsilon_m^{2\beta}}{\kappa}, c_0 = \frac{9}{80}.
$$
The diffusivities are described by a two-parameter semigroup
$\kappa_m = \varphi(m,M) \kappa$ and we are interested in the
pullback limit, for $m$ fixed, 
\begin{equation}\label{seq}
    {\cal K}(m;\kappa) = \lim_{M \to \infty} \varphi(m,M) \kappa.
\end{equation}
We then need to understand the behavior of
${\cal K}(m;\kappa)$ when
$\kappa \to 0$. Of practical importance is the fact that
the double asymptotics, $M \to \infty$ first, then $\kappa \to 0$
, can be replaced by a single asymptotics $\kappa \to 0$ using
\begin{equation}
    M(\kappa) ~~\mbox{such that}~~\sqrt{c_0} \epsilon_{M(\kappa)}^\beta \propto  \kappa.
\end{equation}
Indeed, $F_m(\kappa) \sim \kappa$ unless
$c_0 \epsilon_m^{2 \beta}/\kappa$ has the same order of magnitude
than $\kappa$.\\\\
Let
$z_m := \frac{\kappa_m}{\sqrt{c_0} \epsilon_m^\beta}$,
$z_M = \frac{\kappa}{\sqrt{c_0} \epsilon_M^\beta}$,
then 
\begin{equation}\label{seqz}
    z_{m-1} = G_m(z_m),~~{\rm with}~~ G_m(z) := g_m
    \left(z + \frac1z \right), g_m = \left(\frac{\epsilon_m}{\epsilon_{m-1}} \right)^\beta.
\end{equation}
In the geometric case, the scale separation is given by $\epsilon_m = \lceil \Lambda^m \rceil^{-1}$. Assuming, for simplicity, that $\Lambda \in \mathbb{N}$, it follows that $g_m = \Lambda^{-\beta}$. Therefore, Eq. (\ref{seqz}) becomes autonomous.
It is important to note that, in this case, the estimates presented in \cite{AV23} are no longer valid.
Still, it is very instructive to look at the behavior of the diffusivities sequence. A close inspection shows that
the diffusivities converge in the small $\kappa$ limit:
\begin{equation}
    {\cal K}_m := \lim_{\kappa \to 0} {\cal K}(m;\kappa)
    = \Lambda^{-\beta m} \sqrt{\frac{c_0}{\Lambda^\beta-1}}, \forall m \geq 0.
\end{equation}

Therefore, if one were able to establish a link between the renormalized sequence and the inviscid system in the geometric case as well, it would follow that the system is well-posed, yielding an atomic (Dirac) distribution, i.e. the absence of Eulerian spontaneous stochasticity.

\begin{figure}[h]
\centerline{\includegraphics[width=0.48\columnwidth]{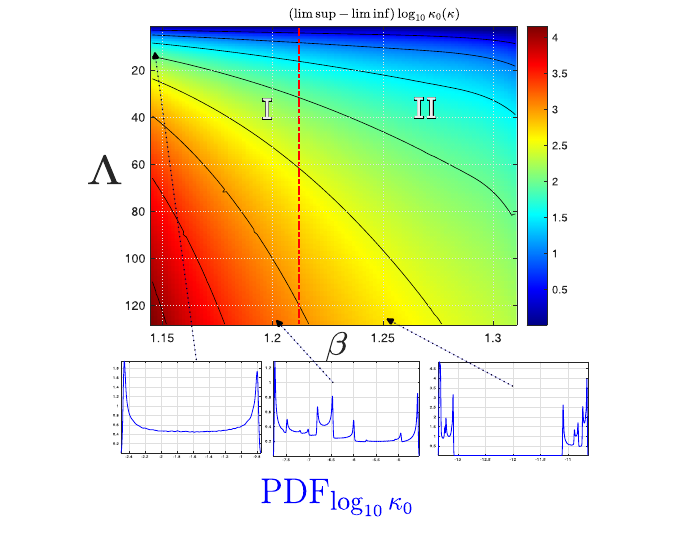}}
\centerline{\includegraphics[width=0.51\columnwidth]{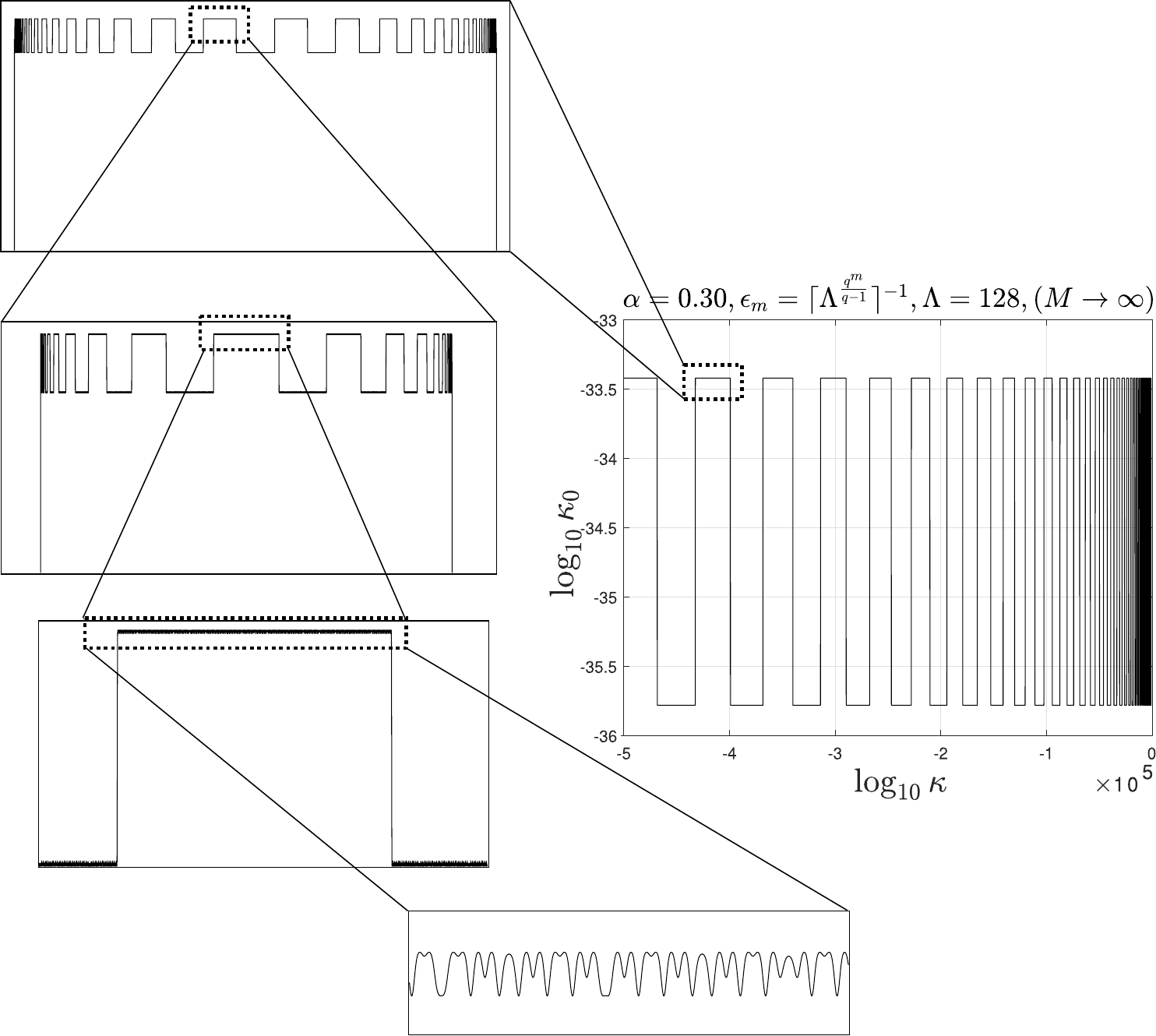}}
\caption{{Upper} panel: standard deviation of $\log_{10} \kappa_0$ as a function of $\beta$ and
$\Lambda$. Regime I corresponds to $\kappa_0$ distributions having a
simply-connected compact support,
for $1.14 \lessapprox \beta \lessapprox 1.21$, and regime II to a bimodal qualitative behavior of the $\kappa_0$ distribution. Three probability densities for the limit $\kappa \to 0$ are shown for $\beta=1.20$, $\beta=1.25$ ($\Lambda=128$) and $\beta = 1.143$ ($\Lambda=10$). {Lower} panel: behavior of $\log_{10} \kappa_0$ in $\log \kappa$ scale for $\beta=1.30$, $\Lambda=128$ with successive zooms revealing the fractal structure. Note that the frequency slows down suggesting a $\log \log$ scaling. Numerics are achieved using BigFloat precision.}
\label{fig:betaLBall}
\end{figure}

A radically different situation occurs when one considers an hypergeometric scale separation. The sequence (\ref{seqz}) becomes nonautonomous with
\begin{equation}
  G_m(z) = \Lambda^{-\beta q^{m-1}} \left( z + \frac{1}{z} \right).
\end{equation}
Figure \ref{fig:betaLBall} illustrates the behavior of $\limsup - \liminf > 0$ of the function $\kappa \mapsto \kappa_0(\kappa)$ as a function of the parameters $\beta$ and $\Lambda$ in the limit $\kappa \to 0$. The corresponding distributions are shown on a logarithmic scale. The lower panel presents the function itself for selected parameter values, highlighting its intricate fractal structure.

\begin{figure}[htbp]
\centerline{\includegraphics[width=0.75\columnwidth]{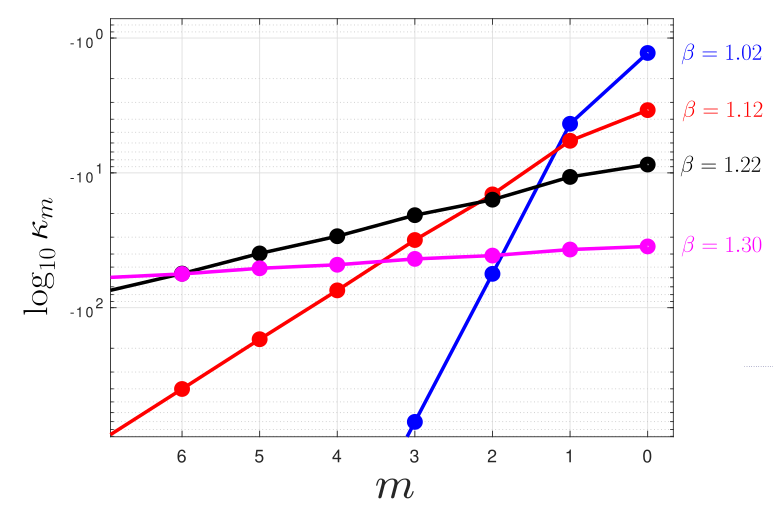}}
\vspace*{-0.3cm}
\caption{The diffusivity sequence $\kappa_m$ in loglog scale for four different values of $\beta$ using $\Lambda=128$. One can notice that as $\beta \to \frac43$, the slope tends to zero while the intercept diverges to minus infinity.}
\label{valkap}
\end{figure}

In general, $\kappa_0$ is not of order unity and depends significantly on $\beta$ (and $\Lambda$). It is $O(1)$ only in the limit $\beta \to 1^+$. As $\beta \to \frac43$, $\kappa_0$ becomes increasingly small -- dropping by several order of magnitude. Moreover, we numerically identify a critical value $\beta^\star \approx 1.1428$ above which the constraint admissible set ${\cal K}$ is not required to control the sequence. Figure
\ref{fig:betaLBall} uses $\beta$ values above this critical threshold $\beta^\star$. To meet the requirements outlined in \cite{AV24}, we fix $M$ and set $\kappa_M$ as $\theta \epsilon_M^{\frac{2\beta}{q+1}}$ for $\theta \in  (\frac12,2)$. We fix $\theta=1$. The sequence is shown in Fig. \ref{valkap} for different values of $\beta \in (1,\frac43)$.

\begin{figure}[htpb]
\centerline{\includegraphics[width=0.75\columnwidth]{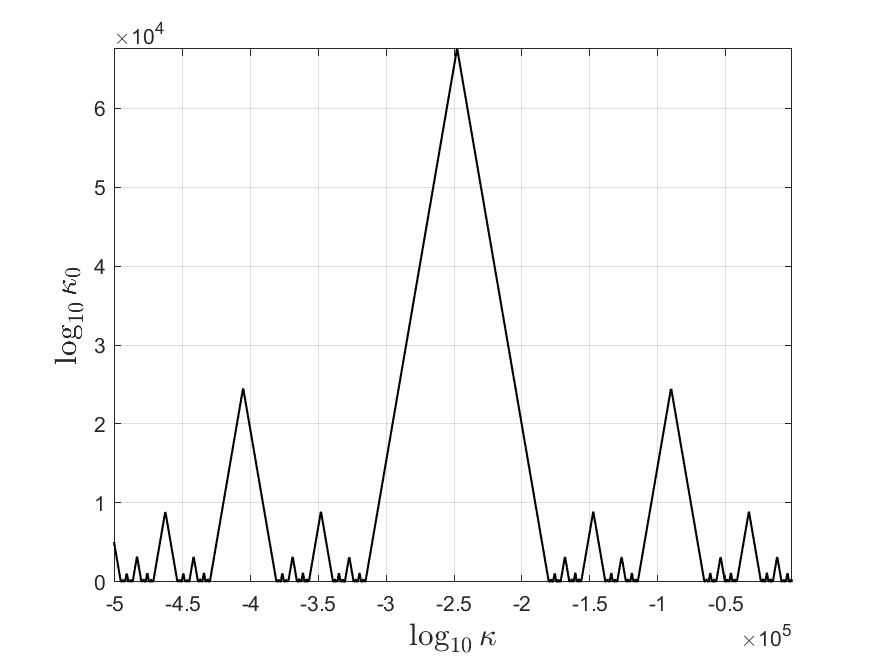}}
\vspace*{-0.3cm}
\caption{Values of $\kappa_0$ as a function of $\kappa_M = \kappa$ (in $\log_{10}$ scale) for $\beta = 1.10 < \beta^\star$ ($\Lambda = 128$). The renormalized diffusivity sequence becomes wild, with unbounded behavior unless $\kappa_M \in {\cal K}$ is imposed.}
\label{unboundk0}
\end{figure}

Only in the limit $\beta \to 1^+$, one expects $\kappa_0 = O(1)$. Note that the values of $\kappa_0$ depend on $\theta$ in all cases (not shown), indicating the absence of a selection principle at the level of $\kappa_0$. The selection occurs only at the level of the measures, which is the primary focus of this work. We also establish the existence of a critical value $\beta^\star \approx 1.1428$, above which the constraint set ${\cal K}$ is not required to control the sequence. For $\beta < \beta^\star$, however, it is necessary to enforce $\kappa_M \in {\cal K}$. This behavior is illustrated in Fig. \ref{unboundk0} for $\beta = 1.10$.

\section{Proof of Theorem \ref{M0M}} \label{Proof_M0M}
\subsection{Theorem \ref{M0M}: ${\cal M}= {\cal M}_0$}

As mentioned in the introduction, the proof is by successive steps.
\paragraph{${\cal M}_0$ is compact and convex and
$\mathcal{M}_0 = \overline{\operatorname{co}}(\mathcal{E})$.}
For the sake of consistency, we provide the proof which is classical.
\begin{proof}
It is convex since taking $\mu_1,\mu_2 \in {\cal P}({\cal S}_0)$ then for $\theta \in [0,1]$, $\mu({\cal S}_0)=(1-\theta) \mu_1({\cal S}_0) + \theta \mu_2({\cal S}_0) = 1$. Moreover, for all Borel sets $B \subset {\cal S}_0$, $\mu(B)\geq 0$ since
$\mu_1(B),\mu_2(B) \geq 0$ and thus $\mu \in {\cal M}_0$.

Due to the compacity of ${\cal S}_0$, one has immediate tightness: for all $\mu \in {\cal P}({\cal S}_0)$, $\mu({\cal S}_0)=1$ so that the probability measures in ${\cal P}({\cal S}_0)$ are tight. Prokhorov theorem then states that tightness of probability measures in a Polish space is equivalent to be relatively compact in the weak topology. It thus remains to show that the set is closed. Let $\mu_k \rightharpoonup \mu$ then since $\limsup_{k \to \infty} \mu_k({\cal S}_0) \leq \mu({\cal S}_0)$, then $\mu({\cal S}_0) \geq 1$. By weak convergence $\mu(H) = 1$ and $\mu(B) \geq 0$ for all Borel sets, therefore $\mu({\cal S}_0) \leq 1$ i.e. $\mu({\cal S}_0)=1$ and $\mu \in {\cal M}_0$.

${\cal M}_0$ is a compact convex subset of the locally convex space of signed measures and by virtue of Krein-Milman theorem it is the closed hull of its extreme points.
One must show that the extreme points of $\mathcal{M}_0$ are the Dirac measures. Let us take a Dirac measure $\delta_x$ for $x \in \mathcal{S}_0$, and assume $\delta_x = \theta \mu_1 + (1-\theta) \mu_2$ for $\theta \in (0,1)$, then $\delta_x(\{x \}) = 1$ and therefore $\mu_1(\{x\}) = \mu_2(\{x\}) = 1$ so that $\mu_1=\mu_2=\delta_x$. It is not possible to have other measures since it would decompose as a convex combination of Dirac measures. Therefore 
$\mathcal{M}_0 = \overline{\operatorname{co}}(\mathcal{E})$.
\end{proof}
\bigskip

\paragraph{$\mathcal{M}(\gamma) \subset {\cal M}_0$}
Let $\mu \in \mathcal{M}(\gamma)$, by definition one can find a subsequence $\tau_n \to \infty$ such that $\gamma_\# \mathrm{Leb}_{\tau_n} \rightharpoonup \mu$. Since $f \in \V$ converges uniformly to $f_0$ and $\gamma(\tau) = \Phi_t[f(\cdot,g(\tau))]x_0$, the set of accumulation points $G= \left\{ g \in H~:~\exists \tau_p \to \infty~\text{such that}~\gamma(\tau_p) \to g \right\} \subset {\cal S}_0$ and since ${\cal S}_0$ is closed, $\bar G
\subset {\cal S}_0$.
Let $N$ be some open set with $N \cap {\cal S}_0 = \emptyset$ and thus $N \cap \bar G = \emptyset$, then
$\mu(N) = \lim_{n \to \infty} \frac{1}{\tau_n} 
\int_0^{\tau_n} \mathbbm{1}_N (\gamma(s)) ds = 0$.
Therefore, $\mu(N)=0$ for all $N$ like above, i.e. the support of $\mu \subset {\cal S}_0$. 
\bigskip

\paragraph{$\mathcal{M}(\gamma)$ is compact}\label{Mcompact}
The space ${\cal M}(\gamma)$ is never nempty due to tightness of
the measures $\gamma_\# {\rm Leb}_\tau$ and Prokhorov theorem. We first show that it is 
compact in the weak topology. It suffices to show that it is closed and since it is in the compact set $\mathcal{M}_0$ it is compact as well. The proof is a classical diagonal argument. Let $\mu_k
\in {\cal M}(\gamma) \to \mu$.
By hypothesis, there is a sequence $\tau_{n,k} \to 0$ such that $
\lim_{n \to \infty} \frac{1}{\tau_{n,k}} \int_0^{\tau_{n,k}} F(\gamma(s)) ds = \langle \mu_k,F \rangle$. For each $n$, one can choose
$k=k(n)$ such that $\left| \frac{1}{\tau_{n,k}} \int_0^{\tau_{n,k}} F({\gamma(s)}) ds - \langle \mu_k,F\rangle \right| \leq \frac{1}{n}$.
The sequence $\tau_n' = \tau_{n,k(n)}$ is such that $\lim_{n \to \infty} \frac{1}{\tau_n'} \int_0^{\tau_n'} F({\gamma(s)}) ds = {\left\langle \mu, F \right \rangle }$, i.e. $\mu \in {\cal M}(\gamma)$.
\\

At this stage, it is interesting to note that $\mathcal{M}(\gamma)$ can be nonconvex. We give an example.
Let us assume that there exists a system such that $\gamma(s)=e^{i \log s}$.
We write the test functions like $F(z) = \sum_{k \geq 0} F_k z^k$. Let
$C(\epsilon) = \frac{1}{\epsilon} \int_0^\epsilon F(\gamma(s)) ds = \sum_{k \geq 0} G_k e^{i k \log \epsilon}$ with $G_k = \frac{F_k}{1+ik}$. Then we can exhibit subsequences $\epsilon_n^{(a)} \to 0$ of the form $\epsilon_n^{(a)} = \mathrm{exp}(a_n-2\pi n)$
with $a_n \to a$. It gives the family of measures $\langle \mu_a,F \rangle
= \sum_{k \geq 0} G_k e^{i k a}$. Consider two distinct measures $\mu_{a_1},\mu_{a_2}$ then one must find a sequence $\epsilon_n$ such that $C(\epsilon_n) \to \sum_{k \geq 0} G_k (\theta e^{i k a_1} + (1-\theta) e^{i k a_2})$. It gives the constraint: $\forall k \geq 0,
e^{i \log \epsilon_n} \to \left( \theta e^{i k a_1} + (1-\theta) e^{i k a_2} \right)^\frac1k$.
This is not possible unless $\epsilon_n$ depends on $k$.
Another consequence is that when ${\cal M}(\gamma)$ contains only Dirac measures as subsequential limits, then it cannot be convex by definition.
Otherwise, if it were convex, it would necessarily include convex combinations of Dirac measures that are not themselves Dirac.
\bigskip

\paragraph{$\mathcal{E} \subset \mathcal{M}$}\label{EinM}
\begin{proof}
Let $x \in \mathcal{S}_0$ we want to exhibit a curve denoted $\gamma_x$ such that $\gamma_x \in \Gamma$ and $\mathcal{M}(\gamma_x) = \{ \delta_x \}$. 
Consider the set of trajectories of the inviscid system 
$\left\{ \Phi_s[f_0(\cdot)] x_0, s \in [0,t] \right\}$. One can select a trajectory call $g$ in this set such that $g(t)=x$ (otherwise it would contradict $x \in S_0$). This trajectory is $C^1$ since $\dot g = f_0(g)$ and $f_0$ is continuous, therefore:
$$
\exists g \in C^1([0,t];H)~\text{such that}~\dot g = f_0(g),~g(0)=x_0,g(t) = x.
$$

We first use Stone-Weierstrass to approximate $\dot g$ by $C^\infty$ functions as the regularization parameter $\R$ goes to infinity. One obtains $g_\R \in C^\infty([0,t];H)$ such that $g_\R(0) = x_0$ and $\lim_{\R \to \infty} ||g_\R-g||_{C^1} = 0$. Moreover $g_\R(t) \to x$ as $\R \to \infty$. We then construct a vector field $F_\R = 
\dot g_\R$ in a small tubular neighborhood $\mathcal{T}_\R$ of $g$ of radius size $1/\R$. We use a smooth function cutoff $\theta_\R  \in C^\infty(H;[0,1])$ 
such that $\theta_\R = 0$ outside $\mathcal{T}_\R$ and $\theta_\R = 1$ inside say $\mathcal{T}_{2 \R}$. We have built a smooth vector field defined as
$$
f(\cdot,\frac{1}{\R}) := \theta_\R(\cdot) F_\R(\cdot) + (1-\theta_\R(\cdot)) f_0(\cdot),
$$
namely $f(\cdot,1/\R)$ is equal to $f_0$ outside the tube but connects smoothly to $F_\R$ inside. 

We now show that this vector field is in $V_0$. Inside the tube, one has $||f(x,1/\R)-f_0(x)|| = ||F_\R(x)-f_0(x)|| \leq ||F_\R(x) - f_0(g(x))|| + ||f_0(x)-f_0(g(x))||$. The first term goes to zero by the definition of $F_\R$ since $\dot g_\R \to \dot g = f_0(g)$ uniformly. The second term goes to zero uniformly by continuity of $f_0$ and the definition of the tubular neighborhood. Outside the tube $f(\cdot,1/\R) = f_0(\cdot)$. In order to conclude, one has by construction $||g_\R -g||_{C^0} \to 0$ and is the unique solution of $\dot x = f(x,1/\R), x(0)=x_0$.
\end{proof}
\bigskip

\paragraph{$\operatorname{co}(\mathcal{E}) \subset \mathcal{M}$}\label{coEM}
We aim to show that any finite convex combination of Dirac measures belongs to $\mathcal{M}$. We establish the stronger result:
\be \label{co_in_M_singleton}
\forall~\mu \in \operatorname{co}(\mathcal{E}),~\exists \gamma \in \Gamma \text{ such that } \mathcal{M}(\gamma) = \{\mu\}.
\de 
\begin{proof}
It suffices to consider convex combinations of two distinct Dirac masses. Let $\theta \in (0,1)$ and $x, y \in \mathcal{S}_0$ with $x \neq y$. We claim that
$$
\exists \gamma_\theta \in \Gamma \text{ such that } \mathcal{M}(\gamma_\theta) = \left\{ \theta \delta_x + (1-\theta) \delta_y \right\}.
$$

Let $f_x, f_y \in V_0$ be the regularizations constructed previously such that the inviscid limits of the associated trajectories converge to $x$ and $y$, respectively. Let us fix some $g \in {\cal A}_{\rm lg}$ and define $f_\theta$ by
\be \label{controlswitch}
\left\{
\begin{aligned}
f_\theta(\cdot, g(s)) & := a_\theta(s) f_x(\cdot, g(s)) + \left(1 - a_\theta(s)\right) f_y(\cdot, g(s)), \\
a_\theta(s) &= \frac{1}{2} + \frac{1}{2} \tanh\left( s \left( \sin {2\pi}{s} + c_\theta \right) \right), \quad c_\theta = \sin\left( \pi \left( \theta - \frac{1}{2} \right) \right).
\end{aligned}
\right.
\de 

This construction ensures that
$$
\lim_{\tau  \to \infty} \frac{1}{\tau} \int_0^\tau a_\theta(s) \, ds = \theta,
$$
and that $f_\theta$ approximates $f_x$ (resp. $f_y$) when $a_\theta(s) \approx 1$ (resp. $0$). Indeed, define $\gamma_\theta(s) := \phi_t[f_\theta(\cdot, g(s))] x_0$, we prove that
$$
\lim_{\tau \to \infty} \frac{1}{\tau} \int_0^\tau \delta_{\gamma_\theta(s)} \, ds
=
\lim_{\tau \to \infty} \frac{1}{\tau}  b 
   \int_0^\tau \delta_{\hat \gamma_\theta(s)} \, ds
$$
where
$$
\hat{\gamma}_\theta(s) := x \, \hat{a}_\theta(s) + y \left(1 - \hat{a}_\theta(s)\right), \quad
\hat{a}_\theta(s) := \frac{1}{2} + \frac{1}{2} \operatorname{sign} \left( \sin 2\pi s + c_\theta \right) \in \{0,1\}.
$$
It follows  from $\gamma_\theta(s) - \hat{\gamma}_\theta(s)\underset{s\to \infty}{=} o(1)$ and the dominated convergence theorem. Since $\hat{a}_\theta$ takes values in $\{0,1\}$, we obtain
$$
\lim_{\tau \to \infty} \frac{1}{\tau} \int_0^\tau \hat{a}_\theta(s) ds = \theta,
$$
and similarly for $1 - \hat{a}_\theta(s)$, yielding the result.
\\

It remains to verify that $f_\theta \in V_0$. Indeed,
$$
\| f_\theta - f_0 \|_\infty \leq \| a_\theta (f_x - f_0) + (1 - a_\theta)(f_y - f_0) \|_\infty 
\leq \| f_x - f_0 \|_\infty + \| f_y - f_0 \|_\infty,
$$
which goes to zero by assumption. Since $f_x,f_y \in \operatorname{Lip(H;H)}$ one has $f_\theta$ Lipschitz as well.

This construction extends to any finite convex combination of Dirac measures. Therefore,
\begin{equation} \label{coEMM0}
\operatorname{co}(\mathcal{E}) \subseteq \mathcal{M} \subseteq \mathcal{M}_0 = \overline{\operatorname{co}}(\mathcal{E}).
\end{equation}
\end{proof}
\paragraph{$\forall \mu \in {\cal M}_0$,
$\exists \gamma_\mu \in \Gamma$ such that ${\cal M}(\gamma_\mu) = \{\mu\}$}
We aim to show that $\mathcal{M}$ is closed, and that for any $\mu \in \mathcal{M}_0 = \overline{\mathcal{M}}$, there exists a regularized curve $\gamma \in \Gamma$ such that
$$
\mathcal{M}(\gamma) = \{ \mu \}.
$$
\begin{proof}
The strategy is to renormalize a sequence $\mu_n \to \mu$ and glue together the corresponding curves $\gamma_n$ in a controlled way. This is feasible because no bounded variation constraint is imposed—only the asymptotic behavior via Birkhoff averages is relevant.

Let $\mu_n \in \mathcal{M}$ be such that $\mu_n \rightharpoonup \mu$. By taking the closure of 
Eq.~\eqref{coEMM0}), one has $\overline{\cal M} = 
\overline{\operatorname{co}}(\mathcal{E}) $, namely
one can choose $\mu_n \in \operatorname{co}(\mathcal{E}) \subset {\cal M}$. Therefore from \eqref{co_in_M_singleton},  there exists a curve $\gamma_n$ such that
$$
\mathcal{M}(\gamma_n) = \{ \mu_n \}.
$$

\smallskip

Let $(\R_n^-)$, $(\R_n)$, and $(\lambda_n)$ be increasing sequences with $\R_n^-, \R_n, \lambda_n \to \infty$, $\R_0 = \R_0^- = 0$, and $\R_n^- < \R_n$, where $|\R_n - \R_n^-| \ll 1$. Fix $g \in \alg$ and set $S := g(s)$.

\begin{figure}[htbp] \centering\includegraphics[scale=0.5]{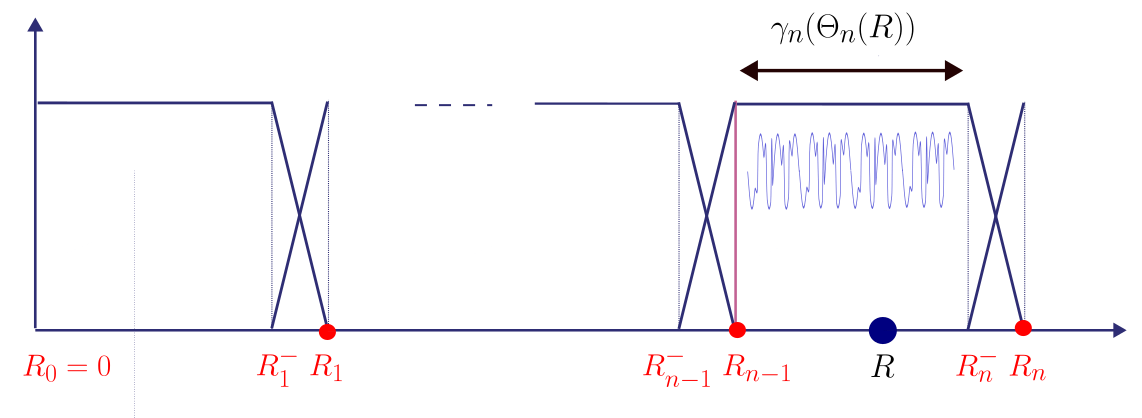}
\caption{A sketch view of the rescaling and concatenation of the sequence $\mu_n$.}
\label{gamu}
\end{figure}

Introduce a smooth partition of unity $(\chi_n)_{n \geq 1}$ with $0 \leq \chi_n \leq 1$, such that (see Fig. \ref{gamu}):
$$
\operatorname{supp} \chi_n = U_n := [\R_{n-1}^-, \R_n], \qquad
\chi_n(S) = 1 \text{ on } J_n := [\R_{n-1}, \R_n^-],
\qquad \sum_{n \geq 1} \chi_n(S) = 1.
$$

Define the vector field and corresponding regularization curve:
$$
f(\cdot, S) := \sum_{n \geq 1} \chi_n(S) f_n\left( \cdot, \Theta_n(S) \right),
\quad \Theta_n(S) := \lambda_n \frac{S - \R_{n-1}}{\R_n^- - \R_{n-1}},
\quad \gamma(S) := \phi_t[f(\cdot, S)] x_0.
$$
In particular, for $S \in J_n$, we recover:
$$
\gamma(S) = \gamma_n(\Theta_n(S)).
$$

Because $(\chi_n)$ is a partition of unity and each $f_n \in \V$, we have $f \in \V$ since $f_n \in \operatorname{Lip}$, and moreover
$$
\lim_{S \to \infty} \| f(\cdot, S) - f_0(\cdot) \|_\infty
\leq \lim_{S \to \infty} \sum_{n \geq 1} \chi_n(S)
\| f_n(\cdot, \Theta_n(S)) - f_0(\cdot) \|_\infty = 0.
$$
\\\\
Fix $\R \in (\R_{n-1}, \R_n^-]$. Define:
$$
A := \frac{1}{\R} \int_0^R F(\gamma(S)) \, dS = A_0 + A_1 + A_2,
$$
with
$$
\begin{aligned}
A_0 &= \frac{1}{\R} \sum_{k = 1}^{n-1} \int_{J_k}
F(\gamma_k(\Theta_k(S))) \, dS 
= \sum_{k=1}^{n-1} \frac{\R_k^- - \R_{k-1}}{\R}  \frac{1}{\lambda_k}
\int_0^{\lambda_k} F(\gamma_k(S)) \, dS, \\
A_1 &= \frac{1}{\R} \sum_{k = 1}^{n-1} \int_{\R_k^-}^{\R_k}
F(\gamma(S)) \, dS, \\
A_2 &= \frac{1}{\R} \int_{\R_{n-1}}^{\R} F(\gamma_n(\Theta_n(S))) dS = \frac{\R_n^- - \R_{n-1}}{\lambda_n R} \int_0^{\Theta_n(R)} F(\gamma_n(S)) dS
= \frac{\R-\R_{n-1}}{\R} \frac{1}{\Theta_n(\R)}\int_0^{\Theta_n(\R)} F(\gamma_n(S)) dS.
\end{aligned}
$$
Define the error term
$$
\epsilon_k(l) := \frac{1}{l} \int_0^l F(\gamma_k(S)) \, dS - \langle \mu, F \rangle,~\text{and}~\delta_k := \R_k-\R_k^- \ll 1.
$$
Since $\mu_k \rightharpoonup \mu$ and $\mathcal{M}(\gamma_k) = \{ \mu_k \}$, we have $\epsilon_k(l) \to 0$ as $l \to \infty$.
We obtain
\be  \label{A0A2}
A_0 + A_2 - \langle \mu, F \rangle = 
\frac{\R-\R_{n-1}}{\R} \epsilon_n(\Theta_n(\R)) 
-\frac{1}{\R} \langle \mu,F \rangle \sum_{k=1}^{n-1} \delta_k 
+ \frac{1}{\R}\sum_{k=1}^{n-1} (\R_k^- - \R_{k-1}) \epsilon_k(\lambda_k).
\de 
We end up with
$$
A - \langle \mu,F \rangle = \frac{1}{\R} 
\left\{ (\R_n^--\R_{n-1}) \theta_R \epsilon_n(\lambda_n \theta_R)  +
 \sum_{k=1}^{n-1} (\R_k^- - \R_{k-1}) \epsilon_k(\lambda_k) -\langle \mu,F \rangle
\sum_{k=1}^{n-1}\delta_k + \sum_{k=1}^{n-1} 
\int_0^{\delta_k} F(\gamma(S+\R_k^-) dS \right\},
$$
where  $\theta_\R = \frac{\R-\R_{n-1}}{\R_n^--\R_{n-1}} \in (0,1]$.
Thus, we have the estimate:
\begin{equation} \label{upest}
\left| \frac{1}{\R} \int_0^\R F(\gamma(S)) \, dS - \langle \mu, F \rangle \right|
\leq   \frac{1}{\R} \left\{ 
\left(C + |\langle \mu, F \rangle|\right) \sum_{k = 1}^{n-1} \delta_k
+  \sum_{k = 1}^{n-1} |\epsilon_k(\lambda_k)| (\R_k - \R_{k-1})
+ (\R_n - \R_{n-1}) \theta_R |\epsilon_n(\lambda_n \theta_R)|  \right\}
\end{equation}
The terms involving $\delta_k$ are easily controlled by choosing $\delta_k$ small enough such that $\sum_{k \geq 1} \delta_k \leq M$. Similarly, since we are free to choose $\lambda_k$ as large as we want, the terms
$\epsilon_k$ can be made small and we choose $\lambda_k$ such that $\sum_{k \geq 1} |\epsilon_k(\lambda_k)| (\R_k - \R_{k-1}) \leq M$. The third term is also bounded uniformly in $\R$. As a consequence
$$
\left| \frac{1}{\R} \int_0^R F(\gamma(S)) \, dS - \langle \mu, F \rangle \right| \leq \frac{C}{\R} \to 0.
$$
Finally, if $\R \in [\R_n^-, \R_n]$, then we set $A_2 = 0$ and the residual term is absorbed into $A_1$, the bracket in \eqref{A0A2} becomes
$$
-\frac{1}{\R} \sum_{k = 1}^{n-1} \delta_k + \frac{\R_n^-}{\R} - 1,
$$
which leads to a similar bound using $\R - \R_n^- \leq \delta_n$ with $\sum_{k=1}^n \delta_k$ instead. The result follows.
The direct consequence is that ${\cal M}$ is closed for the weak topology.
\end{proof}
Theorem is proven by taking the closure of (\ref{coEMM0}) with $\overline{\cal M} = {\cal M}$:
\be 
{\cal M} = {\cal M}_0 = \overline{\rm co}({\cal E}).
\de

\section{$\SP$ and the RG formalism}\label{RG_up}

We provide here mathematical proofs of the
results described in  Section~\ref{RGSP}. The goal is to connect renormalization--group (RG) dynamics with the emergence of statistical attractors in both the autonomous (semigroup) and non-autonomous (general) regularization settings, and to reinterpret spontaneous stochasticity (\(\SP\)) in light of these developments.

\subsection{Autonomous Regularization: The RG Framework}
\label{RG_auto}

\begin{proof}
First, we show that $\omega(\mu)$ is non-empty and compact. 
Because $K$ is compact and absorbing, ${\rm supp}\,({\cal R}_\tau)_\#\mu\subset K$ for every $\tau$, whence ${\rm supp}\,\mu_\R(\mu)\subset K$. Hence the family $(\mu_\R(\mu))_{\R\ge0}$ is tight. By Prokhorov's theorem it is relatively compact in ${\cal P}(K)$, so any sequence $\mu_{\R_n}(\mu)$ admits a convergent subsequence, say $\mu_{\R_{n_k}}(\mu) \rightharpoonup\mu^\star$. Consequently $\mu^\star\in\bigcap_{\R}\overline{\{\mu_{\R'}(\mu)~:~\R'\ge\R\}}$, proving that $\omega(\mu)$ is non-empty. The intersection of closed subsets of the compact space ${\cal P}(K)$ is compact, so $\omega(\mu)$ is compact as well.

We next establish forward invariance. An equivalent description of the $\omega$-limit set is
$$
\omega(\mu)=\bigl\{\eta \mid\exists\,\R_n\to\infty,\ \mu_{\R_n}(\mu)\rightharpoonup\eta \bigr\}.
$$
Fix $\tau>0$ and $\eta \in\omega(\mu)$. Then, there exists $\R_n \to \infty$ such that $\eta$ is the limit of $\mu_{\R_n}(\mu)$ and
$$
({\cal R}_\tau)_\#\eta
=({\cal R}_\tau)_\#\lim_{n\to\infty} \mu_{\R_n}
=\lim_{n\to\infty}\frac{1}{\R_n}\int_0^{\R_n}({\cal R}_{\tau+s})_\#\mu\,ds
=\lim_{n\to\infty}\left( \frac{1}{\R_n}\int_0^{\R_n}({\cal R}_s)_\#\mu\,ds+E_n \right),
$$
where
$$
E_n:=\frac{1}{\R_n}\int_{\R_n}^{\R_n+\tau}({\cal R}_s)_\#\mu\,ds-\frac{1}{\R_n}\int_0^\tau({\cal R}_s)_\#\mu\,ds.
$$
For every $F \in C_b(H;\mathbb{R})$,
$
\bigl|\langle E_n,F\rangle\bigr|\le\frac{C}{\R_n},
$
with $C$ independent of $F$, $\mu$, and $n$, so $E_n\to0$ as $n\to\infty$. Therefore, one has $({\cal R}_\tau)_\# \eta = \eta$ and 
forward invariance holds and ${\cal M}_{\cal R}^\# \subset {\rm Inv}({\cal R})$. The converse is trivial: If $\eta \in {\rm Inv}({\cal R})$ then $\mu_\R(\nu) = \nu$ giving $\omega(\eta) = \{ \eta \}$.
The set $\operatorname{Inv}({\cal R})$ is compact since is closed and is a subset 
of the compact set ${\cal P}(K)$ with $K$ compact. It is also convex using the definition of invariance.
\\\\
The statement (ii) mostly follows from the definitions. Let us justify that
$\operatorname{Inv}({\cal R}) = \overline{\operatorname{co}} ({\cal E}_{\cal R})$. 
\\

Ergodic invariant measures are extreme points: 
Let $\mu$ be $\mathcal{R}_t$-invariant and ergodic. Suppose
$\mu=\theta\mu_1+(1-\theta)\mu_2$ with $\mu_i\in\operatorname{Inv}(\mathcal{R})$,
$\mu_1\neq\mu_2$, and $\theta\in(0,1)$. Notice that $\mu_i \ll \mu$ since for a Borel set such that $\mu(B) = 0$, one has $\theta \mu_1(B) + (1-\theta) \mu_2(B) = 0$, it implies $\mu_i(B)=0$.
Hence there exist
$f_i=\frac{d\mu_i}{d\mu}\ge 0$ with $\int f_i\,d\mu=1$ and
$1=\theta f_1+(1-\theta)f_2$ $\mu$-a.e.
For any Borel sets $A$ and $t \geq 0$,
$
\int_A f_i\,d\mu=\mu_i(A)=\mu_i(\mathcal{R}_t^{-1}A)=\int_{\mathcal{R}_t^{-1}A} f_i\,d\mu
=\int_A f_i\circ\mathcal{R}_t\,d\mu,
$
using invariance of $\mu_i$ and $\mu$. Therefore, 
$f_i\circ\mathcal{R}_t=f_i$ $\mu$-a.e.
By ergodicity of $\mu$, every $\mathcal{R}_t$-invariant $L^1(\mu)$ function is
$\mu$-a.e. constant, so $f_i\equiv c_i$. From $\int f_i\,d\mu=1$ we get $c_i=1$,
and then $1=\theta f_1+(1-\theta)f_2$ forces $f_1=f_2=1$. Thus $\mu_1=\mu_2=\mu$,
contradicting $\mu_1\neq\mu_2$. Therefore $\mu$ is an extreme point of
$\operatorname{Inv}(\mathcal{R})$.
\\

Extreme points are ergodic invariant measures: we show the contrapositive statement, assume $\mu$ is not ergodic, then there is a Borel set $B_\star$ such that $0< \mu(B_\star) < 1$, then one builds two invariant measures 
$\mu_1(B) = \mu(B_\star \cap B)/\mu(B_\star)$, $\mu_2 = \mu(B_\star^c \cap B)/\mu(B_\star^c)$. Therefore $\mu = \mu(B_\star) \mu_1 + \mu(B_\star^c) \mu_2
= \mu(B_\star) \mu_1 + (1-\mu(B_\star) \mu_2$ and $\mu$ cannot be extreme.
\\

To conclude, one uses the fact that $\operatorname{Inv}({\cal R})$ is compact and convex and moreover, the extreme points are ${\cal E}_{\cal R}$. Krein-Milman theorem then applies. 
\end{proof}
\begin{remark}
\leavevmode
\begin{enumerate}
    \item 
The compact absorbing set can be taken as $K=B_\rho$
for a closed ball of large enough radius $\rho$. It thus contains every RG orbits. The compacity is required to establish that $\omega(\mu)$ is indeed non-empty.
\item 
In the infinite-dimensional setting when considering flow maps:
$x_0 \mapsto \Phi_t[f(\cdot,\tau)]x_0$, it is more difficult since $B_\rho$ is no more compact. One needs additional hypotheses such as equicontinuity. Without loss of generality, assume that for a fixed initial measure $\mu\in{\cal P}(B_\rho)$ for some large $\rho$, i.e.\ ${\rm supp}\,\mu\subset B_\rho$. Define
$$
{\cal F}_\mu:=\{{\cal R}_\tau\phi~:~\phi\in{\rm supp}\,\mu\}.
$$
We require ${\cal F}_\mu$ to be uniformly equicontinuous; in that case, by the Arzel\`a--Ascoli theorem, it is precompact, and we set $K_\mu=\overline{{\cal F}_\mu}$. This set is a compact, absorbing set with ${\cal R}_\tau K_\mu \subset K_\mu$, ${\rm supp}~\mu \subset K_\mu$.
However, we again emphasize that our settings propose a much simpler alternative by freezing $x_0$ and just focusing on states in $H$ not on genuine flow maps (see Theorem \ref{CN}).
\item 
The link with \eqref{P_0} is implicit in the definition of the RG flow; see Definition \ref{RGflow}. In particular, the invariant measures have their support in ${\cal S}_0 \subset K$.
\end{enumerate}
\end{remark}

\subsection{General Non-Autonomous Regularizations: The Bebutov Flow}\label{BEBU_auto}
Since the objects to investigate are regularization curves $\gamma \in \Gamma$ and since one is only interested in their marginal statistical behavior in the inviscid limit, one needs to restrict the set of observables through the lens of the evaluation map 
$\Psi$, namely to consider the class of observables:
$$
\mathcal{O}_\Psi := \left\{ F \circ \Psi \;:\; F \in C_b(H;\mathbb{R}) \right\},
$$
i.e., observables that depend only on the evaluation $\gamma(0)$ (or any $\gamma(\tau_0)$).
It implies to consider a coarser topology which makes $\Psi_\#$ continuous. The following general lemma will be useful:
\begin{lemma}\label{topolemma}
 Let $X$ and $Y$ be topological spaces and $Y$ a Polish space, $\Psi:X \to Y$ and its pushforward
 $\Psi_\#: {\cal P}(X) \to {\cal P}(Y)$, with the standard weak topology. We define the $\Psi$-topology, the coarsest topology that makes $\Psi_\#$ continuous:
 $$
 \nu_n \overset{\scriptscriptstyle \Psi}{\rightharpoonup} \nu \Longleftrightarrow \Psi_\# \nu_n \rightharpoonup \Psi_\# \nu.  
 $$
For $G \subset {\cal P}(X)$, the following hold:
\begin{enumerate}
\item $G$ is $\Psi$-closed $\Longleftrightarrow$ there exist $ C \subset {\cal P}(Y)$ closed such that $G = \Psi_\#^{-1} C$.
\item $G$ is $\Psi$-compact $\Longleftrightarrow$ $\Psi_\# G$ is compact.
\end{enumerate}

\end{lemma}
\begin{proof} 1. $\Rightarrow$: 
Let $C = \overline{\Psi_\# G}$, one has $\Psi_\# G \subset \overline{\Psi_\# G}$, giving 
$G \subset \Psi_\#^{-1}(C)$. Let $\nu \in \Psi_\#^{-1}(C)$, i.e. $\Psi_\# \nu \in C=\overline{\Psi_\# G}$, one has therefore a sequence 
$\Psi_\# \nu_n \rightharpoonup \Psi_\# \nu$, i.e. 
$\nu_n \overset{\scriptscriptstyle \Psi}{\rightharpoonup} 
\nu$ and since $G$ is $\Psi$-closed $\nu \in G$ and
$\Psi_\#^{-1}(C) \subset G$.
1. $\Leftarrow$: since $G = \Psi_\#^{-1}(C)$ with $C$ closed and $\Psi_\#$ is continuous then its preimage is closed in the $\Psi$-topology. 2. $\Rightarrow$: since $\Psi_\#$ is continuous for the $\Psi$-topology, then the image by $\Psi_\#$ of a compact set is compact.
2. $\Leftarrow$: Let $g_n$ be a sequence in $G$, we need to exhibit a subsequence such that it converges to some $g \in G$. Since, $\Psi_\# G$ is compact, the sequence $\Psi_\# g_n$ has a subsequence $\Psi_\# g_{n_k} \rightharpoonup y \in \Psi_\# G$. Let $x \in {\cal P}(X)$ such that $\Psi_\# x = y$, i.e. $x \in \Psi_\#^{-1}\{y\}$ and since 
$\Psi_\#^{-1}\{y \} \cap G \neq \emptyset$ one can choose $x \in G$.
\end{proof}

\begin{proof}
The proof uses Lemma \ref{topolemma} 2. with $X=\Gamma,Y=H$ where $Y$ is Polish. Moreover, $\Psi(\Gamma)$ closed means that ${\rm Im}(\Psi_\#)$ is closed. In term of curves $\gamma \in \Gamma$, it translated to $\cup_{\gamma \in \Gamma} \{ \gamma(0) \}$ is closed, where $\gamma(0)$ is describing some highly regularized version of the inviscid problem $({\cal P}_0)$. This is therefore a very weak constraint on the set $\Gamma$.

\smallskip
\emph{(i)} Let $N_\R:= \{\nu_{\R'}(\nu)~:~\R' \geq \R \} \subset{\cal P}(\Gamma)$. Then, one has for $\rho$ large enough
$$
\Psi_\# N_R \subset {\cal P}(B_\rho),
$$
where $B_\rho$ is the closed ball of $H$ of radius $\rho$ which is compact. Thus by Prokhorov's Theorem, $\overline{\Psi_\# N_R}$ is compact. Moreover, $\Psi_\# \omega(\nu) = \bigcap_{\R > 0} 
\Psi_\# \overline{N_R}^\Psi = \bigcap_{\R > 0} 
\overline{\Psi_\# N_R} \cap {\rm Im}(\Psi_\#) = 
\bigcap_{\R > 0} 
\overline{ \Psi_\# N_R}$ since ${\rm Im}(\Psi_\#)$ is closed.
The intersection of compact sets is also compact, and thus $\Psi_\# \omega(\nu) $ is compact. By applying Lemma \ref{topolemma} 2. one concludes that $\omega(\nu)= \mathscr{M}(\nu)$ is $\Psi$-compact. Note also
that since $\Psi_\# \omega(\delta_\gamma) = {\cal M}(\gamma) \neq \emptyset$ then $\mathscr{M}(\nu)$ is non-empty.
For the forward invariance, we use the alternative expression 
$$
\mathscr{M}(\nu) = \omega(\nu) = \left\{ \eta~:~\exists \R_n \to \infty,  \nu_{\R_n} (\nu)\rightharpoonup \eta ~\mbox{in $\Psi$-topology}\right\}.
$$
Let $\eta \in \mathscr{M}(\nu)$,  
one has some $\R_n \to \infty$ such that $\eta$ is the limit of $\nu_{\R_n}(\nu)$ and then
$$
(\mathscr{S}_\tau)_\# \eta = (\mathscr{S}_\tau)_\# \lim_{n \to \infty} \nu_{\R_n}(\nu) = \lim_{n \to \infty}
\nu_{\R_n}(\nu) + \R_n^{-1} \left(
\int_{\R_n}^{\R_n+\tau} (\mathscr{S}_\tau')_\# \nu d\tau' - 
\int_0^\tau (\mathscr{S}_\tau')_\# \nu d\tau'
\right).
$$
The remainder converges (weakly) to zero in the $\Psi$-topology implying that $(\mathscr{S}_\tau)_\# \eta = \eta$.
As a consequence, one has $(\mathscr{S}_\tau)_\# \mathscr{M}(\nu) \subset \mathscr{M}(\nu)$.
One therefore concludes (i), i.e. $\mathscr{M}(\nu)$ is a statistical attractor.

\smallskip
\emph{(ii)} We use the fact that $\Psi_\#$ is continuous, 
and the alternative expression: 
$\mathscr{M}(\nu) = \omega(\nu) = \left\{ \eta~:~\exists \R_n \to \infty,  \nu_{\R_n} (\nu)\rightharpoonup \eta \right\}$. In such a case, 
$\Psi_\# \omega(\nu) \subset \omega(\Psi_\# \nu) \subset {\cal M}_0$. For $\nu = \delta_\gamma$, using (\ref{Bebuid}), 
one
has $\Psi_\# \omega(\delta_\gamma) = \omega(\Psi_\# \delta_\gamma) = {\cal M}(\gamma)$.

\smallskip
\emph{(iii)}  
From (ii) one has $\Psi_\#\mathscr M\subset\mathcal M_0$.
Conversely, Theorem~\ref{M0M} provides for every
$\mu\in\mathcal M_0$ a curve $\gamma_\mu$ with
$\mu\in\Psi_\#\mathscr M(\delta_{\gamma_\mu})\subset\Psi_\#\mathscr M$,
so equality $\Psi_\# \mathscr{M} = {\cal M}_0$ holds.
Step (i) also shows that $\mathscr{M} \subset {\rm Inv}(\mathscr{S})$. Conversely, if $\eta \in {\rm Inv}(\mathscr{S})$, then $\nu_\R(\eta) = \eta$, i.e. $\omega(\eta) = \{ \eta \}$. Since
${\cal M}_0$ is compact, from Lemma \ref{topolemma}, $\mathscr{M}$
is $\Psi$-compact (non-empty) and it is also forward invariant, therefore it is a statistical attractor.
\end{proof}

\section{Examples}\label{Examples}
We provide various simple examples exhibiting $\SP$ in 
Sections \ref{Ex1}--\ref{Ex4}. We also extend the discussion on RG flows Section \ref{RGSP} to systems having an isolated singularity in Section \ref{RGreg}.
\subsection{Example 1}\label{Ex1}
We consider the well-known example $\dot x = f_0(x), x(0) = x_0$ with $f_0(x) = x^\frac13$. The function $f_0$ has a non-Lipschitz singularity at 0.
Following \cite{Drivas21}, we consider the regularized system 
$\dot x = f(x,\epsilon)$, and $f(\cdot,\epsilon)$ is defined as
\begin{equation}\label{toy1_3}
({\cal P}_\epsilon):~\left\{
\begin{array}{lllcccc}
\dot x & = & f_0(x) & |x| & > & \epsilon \\
\dot x & = & \epsilon^\frac13 G(\frac{x}{\epsilon},\epsilon) & |x| & \leq & \epsilon
\end{array}\right.,
\end{equation}
where
$G(x,\epsilon) =\xi(x) f_0(x) + (1-\xi(x)) \frac{\omega(\epsilon)+x}{2}$ 
with $\xi(x) = 3x^2 - 2 |x|^3$.
Since obviously, $({\cal P}_\epsilon)$ is well-posed (Lipschitz),
it guarantees that $\gamma: s \mapsto \Phi_t[f(\cdot,s)] x_0$ is well-defined as a function. The regularized system is
well-posed. We will use two different regularizations denoted $\gamma_1$ and
$\gamma_2$ with resp.
$\omega_1(\epsilon) = \sin \frac{1}{\epsilon}$ and $\omega_2 = 0.5 + \sin (\frac{1}{\epsilon}+{\rm cst})$.
Note that these deterministic functions rapidly change their sign when  $\epsilon$ becomes small. They introduce a source of uncertainty that vanishes in the limit as $\epsilon \to 0$,
playing a role similar to that of a random variable.

\begin{figure}[htbp]
\centerline{\includegraphics[scale=0.8]{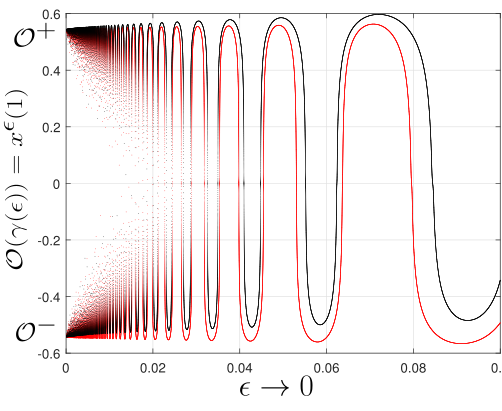}}
\caption{$\gamma(\epsilon) = x^\epsilon(1)$ solution of (\ref{toy1_3}) for $\epsilon$ uniformally distributed in $[10^{-4},10^{-1}]$: in red $\gamma_1$ and in black $\gamma_2$. Both measures concentrate on $\pm a$ with $a \approx 0.544$.}
\label{Toyess}
\end{figure}

Considering Definition \ref{LSPdef}, one uses the observable ${\cal O}(x)=x$, then defines
$\gamma_{1,2}(\epsilon) =  x^\epsilon(1)$ for the two functions $\omega_{1,2}$. 
This is a continuous function due to well-posedness. The result is shown in Fig.~\ref{Toyess}.

For the two regularizations  $\gamma_1$ and $\gamma_2$, the pushforward measures are not equal but their support is the same: the first perturbation yields 
$\frac{\delta_{+a} + \delta_{-a}}{2}$
whereas the second perturbation is $\theta \delta_{+a} + (1-\theta) \delta_{-a}$ with $\theta \approx 0.7$.
To summarise:\hspace*{-0,2cm}
\begin{itemize}
\item[$\bullet$] $({\cal P}_\epsilon)$ is strongly $\SP$ for 
$\gamma_1,\gamma_2$ and regularization by additive noise. One must necessarily have
$x_0=0$ for any choice $t > 0$.
\item[$\bullet$] 
When taking $\omega(\epsilon) = {\rm cst}$, a solution is selected in the limit.
\item[$\bullet$] $({\cal P}_\epsilon): dX_t = {\rm sign}(\sin X_t) |\sin X_t|^\alpha~dt + 
\sqrt{2\epsilon} dW_t, x(0) = x_0 ~\alpha \in (0,1)$ is strongly $\SP$ for $x_0 = 2k\pi, k \in \mathbb{Z}$. Theorem \ref{CN} is satisfied with Dini derivatives equal to $+\infty$ only at $x_0 = 2k\pi, k \in \mathbb{Z}$ so that $\Gamma^\star = (2k\pi)_{k \in \mathbb{Z}}$. Moreover ${\cal W}^-(\Gamma^\star) = \Gamma^\star$. 
\end{itemize}

\begin{figure}[htbp]
\centerline{\includegraphics[width=\linewidth]{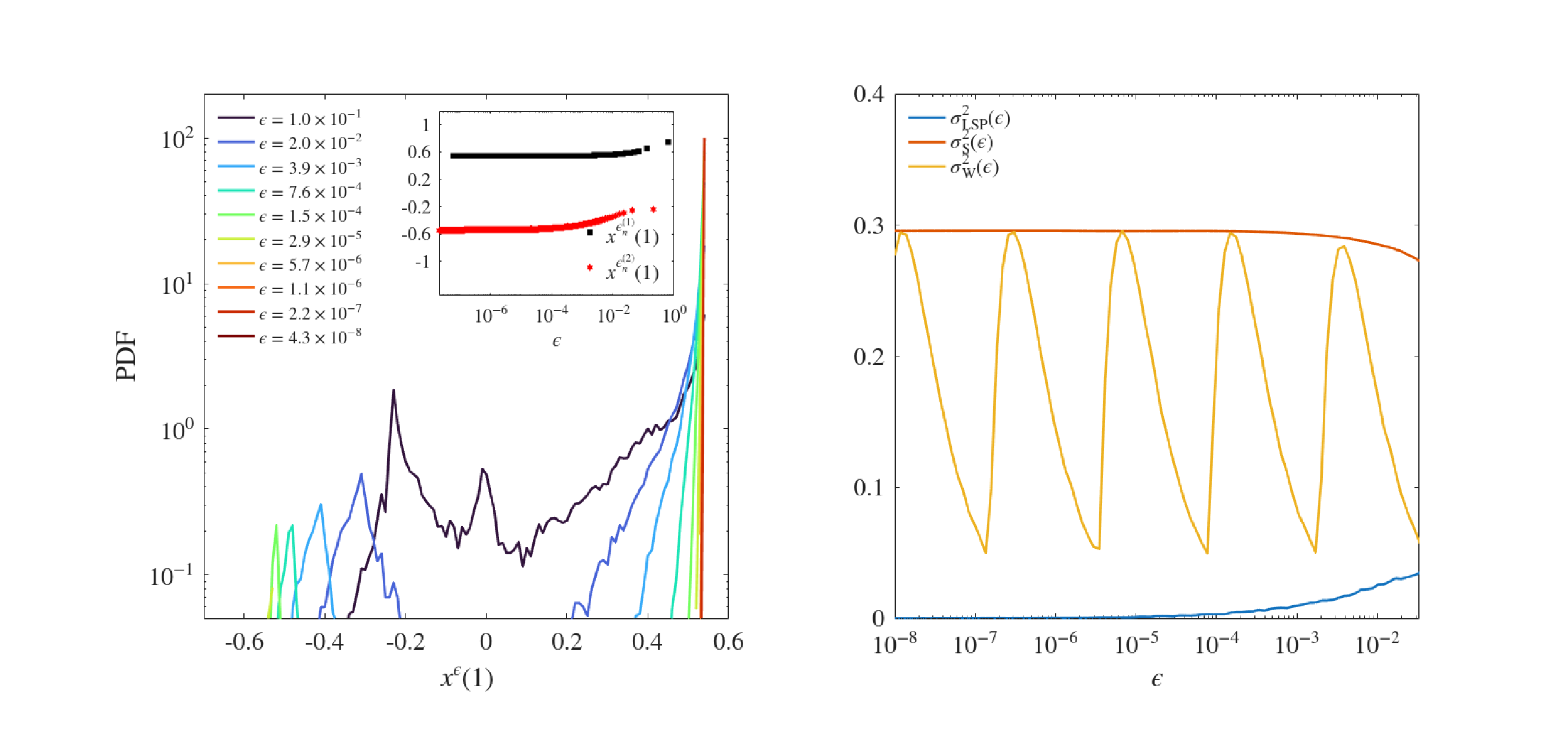}}
\caption{ 
Simulations of system~\eqref{toy1_3} for different choices of the function $\omega$. Specifically, we consider $\omega_1(\epsilon) = \sin \tfrac{1}{\epsilon}$, $\omega_2(\epsilon) = \sin \ln \tfrac{1}{\epsilon}$, and $\omega_3(\epsilon) = 1 - \epsilon + \sin \tfrac{1}{\epsilon}$ to illustrate, respectively, Strong-$\SP$, Weak-$\SP$, and $\delta$-$LSP$.
Left: Probability density of $x^\epsilon(1)$ in the $\omega_3$ case, for various values of $\epsilon$, where the regularization parameter is uniformly distributed over $[0,\epsilon]$. We observe clear convergence toward a Dirac mass. The inset displays two subsequences $x^{{\epsilon_n}^{(1,2)}}$ converging toward distinct values, thereby illustrating $LSP$. We used $\epsilon_n^{-1} = \pi n / 2$, with subsequences corresponding to $\mathrm{mod}(n,4) = 1$ and $\mathrm{mod}(n,4) = 3$.
Right: Weak average $\langle \mu_\epsilon, F \rangle$ for $F(x) =x^2-\bar x^2 
$, with $\bar x=\int x \mu(dx)$, that is, the variance of $x^\epsilon(1)$, plotted over the support of $\mu_\epsilon$. For $\omega_1$, the orange curve exhibits convergence toward a finite value, consistent with Strong-$\SP$. In contrast, the yellow curve for $\omega_2$ displays strong oscillations, compatible only with Weak-$\SP$. Finally, the blue curve corresponding to $\omega_3$ converges to zero, as expected when $\mu_\epsilon \rightharpoonup \delta_a$; in this case, $\omega_3$ corresponds to $\delta$-$LSP$ and does not yield a genuine selection principle (see inset, right panel).}
\label{WSSP}
\end{figure}

We finally illustrates the three manifestations of $LSP$ in Fig. \ref{WSSP}, namely $\delta$-$LSP$, Weak-$\SP$ and Strong-$\SP$ (see Fig. \ref{tricho}).
\subsection{Example 2}\label{exAmbrosio}
\begin{figure}[htbp]
\centerline{\includegraphics[scale=0.8]{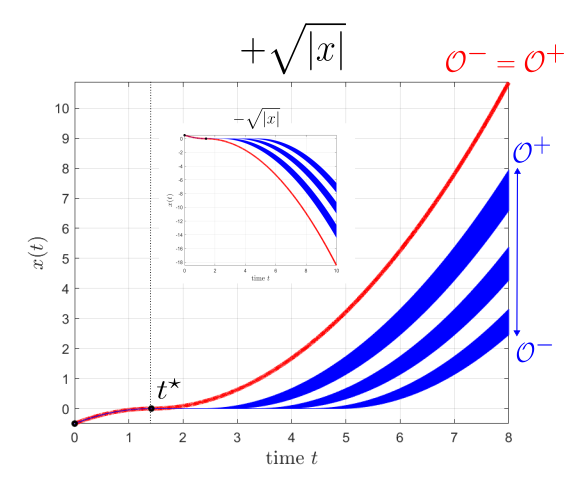}}
\caption{Solutions selected in the limit $\epsilon \to 0$ for the system 
$\dot x = \sqrt{|x|}, x(0)=-c^2$ for two different regularizations, in red by additive noise: the system is not $\SP$ ($\gamma^+=\gamma^-$), in blue using the regularization (\ref{regamb}) with $T \in [1,\frac32] \cup [2,\frac52]
\cup [3,\frac72]$ uniformly distributed, the system becomes $\SP$ for all $t > 2|c|+1$ ($\gamma^+ > \gamma^-$). The black dots are for $c=1/\sqrt{2}$, $(t_0,x_0)=(0,-0.5)$ and $(t^\star,x^\star)=(\sqrt{2},0)$. The inset shows the case with $-\sqrt{|x|}$ and
$x_0 = +c^2$ (minus the regularization (\ref{regamb})).}
\label{Ambrosio_ex}
\end{figure}

We discuss again a 1-D example which deserves some important comments. It is inspired
by \cite{ambrosio2004transport,Flandoli2009}. The inviscid system is the simple $\dot x = \sqrt{|x|}$ with
initial condition $x(0)=-c^2 < 0$. It has an infinity of solutions: the pre-blowup
life is until time $t^\star = 2|c|$ where it reaches the H\"older singularity 0. Then the solution can
wait an arbitrary long time $T\geq 0$ at 0 until it takes off: $x(t) = \frac14 (t-T-2|c|)^2$.
We consider two different regularizations, one is adding some additive noise 
$dx = \sqrt{|x|} dt + \sqrt{\epsilon} dW_t, x(0)=-c^2$, the
other takes the form
\begin{equation} \label{regamb}
f(x,\epsilon)= \mathbbm{1}_{(-\infty,-\epsilon^2]}(x) \sqrt{|x|} + 
\mathbbm{1}_{[-\epsilon^2,\epsilon T-\epsilon^2]}(x) \epsilon + \mathbbm{1}_{]\epsilon T - \epsilon^2,+\infty)}(x) \sqrt{x - \epsilon T + 2 \epsilon^2}, 
\end{equation}
where $\mathbbm{1}_I(x)$ is the characteristic function equals to 1 if $x \in I$ and 0 else.
As discussed in \cite{ambrosio2004transport}, this regularization selects a solution that remains at 0 for a duration of time $T$. Consequently, the regularized system can be fully controlled to match any target distribution. In Fig.~\ref{Ambrosio_ex}, we deliberately randomize $T$ so that the system becomes (strongly) $\SP$ and selects three equally likely, disjoint intervals! Alternatively, one could carefully design some rapidly oscillating function $T = T(\epsilon)$ inspired by the technique (\ref{controlswitch}), as in the previous example which would give the same result (not shown).

There is in fact no need to randomize $f(\cdot,\epsilon)$.
The system is not $\SP$ when regularized by classical additive noise.
This example once again demonstrates that one cannot expect $\SP$ to have large universality classes. This system (or any other) exemplifies Theorem~\ref{M0M}: it is in fact possible to design a regularization of $\sqrt{|x|}$ that converges to any target probability distribution in ${\cal M}_0$ with support ${\cal S}_0 = [0,\frac14 (t-2|c|)^2], t \geq 2|c|$ (see Eq. (\ref{M0})) as $\epsilon \to 0$.

\subsection{Example 3}\label{Ex3}
We consider a system with an isolated H\"older singularity at the origin. 
%Isolated H\"older singularities are believed to play an important role in shell models of turbulence \cite{Dombre1998,Maily2012}. 
A detailed measure-theoretic analysis can be found in \cite{Drivas21,Drivas24}. These systems take the form
$$
\dot x = |x|^\alpha F(y),~y = \frac{x}{|x|} \in \mathbb{S}^{n-1},
F:\mathbb{S}^{n-1} \to \mathbb{R}^n.
$$ 
Using the convention of \cite{Drivas21}, one can write $F$ as a decomposition into
radial and tangential components: $F(y) = F_r(y) y + F_s(y)$, with
$F_r: \mathbb{S}^{n-1} \to \mathbb{R}$ and $F_s: \mathbb{S}^{n-1} \to T\mathbb{S}^{n-1}$.
We note that the Dini directional derivatives along some vector $v \in \mathbb{S}^{n-1}$ at the singularity $x=0$, reduce in this case to 
$$
\Lambda^+_z(0,v) = \limsup_{t \to 0^+} t^{\alpha-1} F_r(v) \in \{-\infty,0,+\infty\}.
$$
In order for Theorem \ref{CN} to apply, one must necessarily have directions $v$ for which
$F_r(v) > 0$. This property is indeed the same than the {\it defocusing}
property considered
in \cite{Drivas24} (see (b), p.1859). In general, such singularities can be seen as generalized 'saddles', including purely unstable critical points, and are distinguished by having both nontrivial stable and unstable sets.

\begin{figure}[htbp]
\centerline{\includegraphics[scale=0.8]{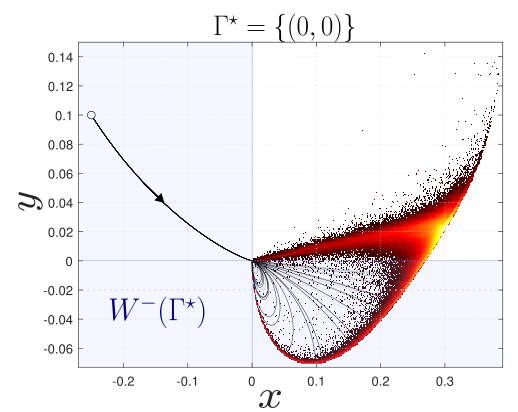}}
\caption{Distribution of the solutions of (\ref{Ex2eq}) at $t=1.5$ starting from the initial condition
$x_0=(-0.25,0.1)$ (log-scaled colormap). The system has been regularized by additive noise of amplitude
$\sqrt{\epsilon}$ with $\epsilon \to 0$. The solution for this initial condition reach the singularity at finite blowup time $t^\star 
\approx 0.875$. The region in light blue color corresponds to the stable set
associated with initial conditions giving $\SP$.}
\label{figex2}
\end{figure}

Note that we also allow the initial condition to be on the singularity
like in 1-D but the stable set ${\cal W}^-(0)$ in fact provides a detailed characterization of all "pre-blowup" initial conditions. 
The example presented here is a slight modification of the 2 d.o.f. example studied in \cite{Drivas21}, but, unlike the original, it does exhibit strong $\SP$.
It takes the form
\be \label{Ex2eq}
F(y) = y(y_1+y_2) - y^\perp y_1y_2^2,~y^\perp = (-y_2,y_1).
\de 
This example is not trivial and has in particular homoclinic orbits with overlapping stable and unstable sets. Indeed, using the conventions of \cite{Drivas21,Drivas24}, the
renormalized system $\dot y = (F(y) \cdot y^\perp) y^\perp$ on $\mathbb{S}^1$ has a well-defined focusing attractor at $\varphi= \frac{3\pi}{2}$ (it guarantees finite-time blowup), but the defocusing attractor at $\varphi=0$ in 
$\mathbb{S}^1$ is non-hyperbolic and one-sided (and has a physical measure $\delta_{\varphi=0}$). The distribution at some fixed time of the solutions, for a single pre-blowup initial condition $x_0
\in {\cal W}^-(0)$, is shown in Fig. \ref{figex2}.

\subsection{Example 4}\label{Ex4}
\begin{figure}[htbp]
\centerline{\includegraphics[scale=0.8]{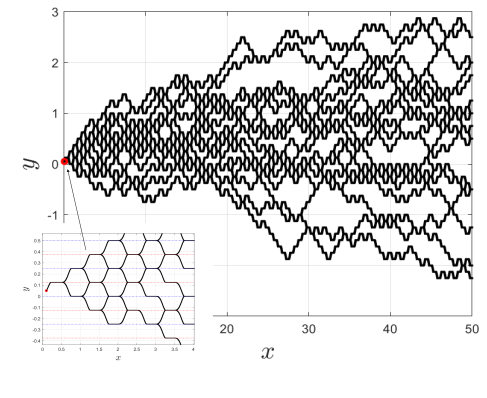}}
\caption{System (\ref{ex3}) for  $t \in [0,50]$, $\mu=1$ and $\sigma=4$ starting from the initial condition 
$x_0 = (0.1,0.05)$ (the system has been regularized by additive noise of amplitude
$\sqrt{\epsilon}$, $\epsilon \to 0$ integrated for 20 different realizations).
The blowup time is $t^\star \approx 0.5$ (see zoom: the levels $x_2 = Y_{p,1},Y_{p,2}$ of $\Gamma^\star$ are also shown in red and blue dash-dotted lines).}
\label{figex3}
\end{figure}

We now discuss a simple example which has $\SP$ although there is no critical points. It is in fact a 1-D nonautonomous system translated into an autonomous one. Let us define $s_{\alpha,\sigma}(x_2) = 
{\rm sign}(\sin 2 \pi \sigma x_2) 
|\sin 2 \pi \sigma x_2|^\alpha$ and for $\alpha \in (0,1)$, $\mu \in \mathbb{R}$, 
\be \label{ex3}
\begin{array}{lll}
\dot x_1 & = & \mu,  \\
\dot x_2 & = & s_{\alpha,\sigma}(x_2) \sin(2\pi x_1)
\end{array}
\de
then by an abuse of notation, we can write 
\begin{equation}
\Lambda^+_z(x,v) = 
(s_{\alpha,\sigma}(x_2) \cos 2\pi x_1) 2 \pi v_1 + (\sigma \alpha s_{\alpha-1,\sigma}(x_2) \cos(2\pi \sigma x_2) \sin 2\pi x_1)) 2 \pi v_2,
~|v|=1.
\end{equation}
Therefore, one can easily identify a nonempty singular set $\Gamma^\star$ corresponding
to all $x=(x_1,x_2)$ such that $\sin 2\pi \sigma x_2=0$, i.e. $x_2 = \frac{j}{2\sigma},
j \in \mathbb{Z}$ such that $(-1)^j \sin (2 \pi x_1) > 0$ or equivalently
introducing the two open intervals $I_{k,1} = (2k \pi,(2k +1)\pi)$, $I_{k,2} = 
I_{k,1} - \pi$ and $Y_{p,1} = \frac{p}{\sigma}$, $Y_{p,2} = Y_{p,1} + \frac{1}{2\sigma}$:
$$
\Gamma^\star = \bigcup_{j,p \in \mathbb{Z}^2} I_{j,1} \times \{Y_{p,1} \}
\cup I_{j,2} \times \{Y_{p,2} \}.
$$
For all $x \in \Gamma^\star$, one has $\Lambda^+_z(x,v) = +\infty$ provided $v_2 \neq 0$.
The stable set ${\cal W}^-(\Gamma^\star)$ is more difficult to characterize and depends crucially on $\mu$, but it includes
$\Gamma^\star$ by definition. We claim that, at least for the parameters chosen in Fig.~\ref{figex3},

${\cal W}^-(\Gamma^\star) = \mathbb{R}^2$,
so that all initial conditions meet the singular set in finite time. 
This system is obviously $\SP$, leading to the formation of space-filling beehive-like patterns. The hexagonal cells are constrained by the set $\Gamma^\star$ in the $x_2$-direction and by $\mu$ in the 
$x_1$
direction.
These patterns result from a periodic alternation: pairs of trajectories merge into one or split into two, dictated by the 
$x_1$  
position and driven by alternating phases of contraction and expansion.
The number of trajectories reaching the axis $x_1 = t$ with initial condition at $t_0$ scales like
 $2^{\lfloor \lambda(\sigma,\mu) (t-t_0) \rfloor}$. We do not show the distribution at fixed time, it can be guessed from Fig.~\ref{figex3} as simply corresponding to a sum of Dirac terms with complicated weights.
\subsection{RG regularization}\label{RGreg}
This section presents a formal discussion of $\SP$ for a specific class of systems whose inviscid vector field possesses a non-Lipschitz singularity at the origin.  
It illustrates how the general framework developed in Section~\ref{RGSP} and Theorem~\ref{RGattractors}, as well as the results of Section~\ref{allprob}, admit concrete applications.  
In principle, the approach extends to singular sets as well; see Theorem~\ref{CN}.  
In all cases, the central idea is to regularize the vector field within a controlled neighborhood of the singular set or singularities of the inviscid system, with this neighborhood shrinking to zero in the inviscid limit, here corresponding to $\delta \to \infty$.

\paragraph{A general discussion}
We assume that the inviscid system with vector field $f_0 \in C_b(\mathbb{R}^d)$ is Lipschitz on $\mathbb{R}^d \setminus \{0\}$ and satisfies $f_0(0) = 0$, with an isolated (expanding) Lipschitz singularity at the origin. Typical examples are systems of the form 
${\displaystyle f_0(x) = |x|^\alpha F_0\left(\frac{x}{|x|} \right),~\alpha \in (0,1)}$, see \cite{Drivas21,Drivas24}.

For each $\delta > 0$ and $\tau \geq 0$, we define $f$ in~\eqref{RG_Phi2} by
\be \label{reguiso}
f(\Phi,\tau,y)=
\begin{cases}
h(\Phi,\tau,y), & |\Phi|\le \delta(\tau),\\
f_0(\Phi),      & |\Phi|\ge \delta(\tau),
\end{cases}
\qquad
\lim_{\tau\to\infty}\delta(\tau)=0,
\de 
where, for every $\tau,y$, the map $\Phi \mapsto h(\Phi,\tau,y)$ is Lipschitz and $h$ is such that
$f(\cdot,\tau,y)$ is continuous across the boundary of the ball $B_{\delta(\tau)}$.
To avoid a discussion on $x_0$ and the time $t$ at which we should observe the system; see Theorem \ref{CN}, we set
$$
x_0=0~\text{and}~t = 1.
$$
The key quantity is the \emph{last exit time} from $B_{\delta(\tau)}$, rather than the detailed flow inside.  Let $\Phi_{\tau,y}^{(h)}(s)$ be the solution of
\be 
\dot\Phi = h(\Phi,\tau,y),\quad \Phi(0)=0.
\de 
Define
\be  
t^\star(\tau,y)
:=\sup\{\,s\in[0,1]:\|\Phi_{\tau,y}^{(h)}(s)\|\ge\delta(\tau)\},
\quad
\Phi^\star(\tau,y)
:=\Phi_{\tau,y}^{(h)}\bigl(t^\star(\tau,y)\bigr).
\de 
We also denote by $\Phi_{\tau,y}^{(f_0)}(s)$ the solution of 
\be \label{reguP0sol}
\dot\Phi = f_0(\Phi),\quad \Phi\bigl(t^\star(\tau,y)\bigr)=\Phi^\star(\tau,y).
\de 
Then from \eqref{RG_system} we have
\be \label{RGconstr}
\mathcal{R}_\tau y =
\begin{cases}
\Phi_{\tau,y}^{(h)}(1), & t^\star(\tau,y)\ge1,\\
\Phi_{\tau,y}^{(f_0)}(1), & t^\star(\tau,y)\le1.
\end{cases}
\de 
The problem reduces to understand \eqref{RGconstr} in the limit $\tau \to \infty$. In full generality, one does not expect $t^\star(\tau,y)$ to have a limit as $\tau \to \infty$. We first assume that the solution stays in $B_\delta$ for all $\tau$: 
\be \label{trap}
\forall \tau \geq 0, t^\star(\tau,y) \geq 1.
\de 
Since $\| \Phi^\star \| = \delta(\tau) \to 0$, it gives the constraint that 
${\cal R}_\tau$ must have an attractor (together with its basin of attraction) which is $\{0 \}$. Otherwise, one cannot impose the regularization coming from ${\cal R}_\tau$.

The interesting case is when \eqref{trap} is not satisfied. Assume that, we now have an upper bound:
\be \label{notrap}
\forall \tau \geq 0, t^\star(\tau,y) \leq 1.
\de 
Let ${\cal L}_y$ denote the set of accumulation points as 
$\tau \to \infty$, solutions of \eqref{reguP0sol} at time $t=1$:
\be 
{\cal L}_y:= \bigcap_{T \geq 0} \overline{\left\{ \Phi^{(f_0)}_{\tau,y}(1) ~:~\tau \geq T
\right\}} \subset {\cal S}_0,
\de 
where ${\cal S}_0$ is defined in \eqref{S0}.
We claim without proving that the $\omega$-limit set
of the semigroup ${\cal R}_\tau$:
$\omega(y):= \bigcap_{T \geq 0} \overline{\{ {\cal R}_\tau y~:~\tau \geq T \}} $ satisfies
\be 
\omega(y) = {\cal L}_y \subset {\cal S}_0,
\de 
as a direct consequence of the definitions and corresponds to the asymptotics formulation of \eqref{RGconstr}.
This set equality is rather useful: one does not need to look at the precise form of regularizations \eqref{reguiso} but rather at the limit set of the semigroup generated by $g$. The only important constraint is that this limit set must have support in ${\cal S}_0$. Note that $\omega(y)$ in state space is not the same than $\omega(\delta_y)$ in measure space; see Theorem \ref{RGattractors}. By definition, the limit set $\omega(\delta_y)$ contains invariant measures with support in $\omega(y)$. Obviously, one has
\begin{center}
    if $\omega(\delta_y) = \{ \mu \}$ non-Dirac then \eqref{reguiso} is Strong-$\SP$ for the RG regularization associated with $y$.
\end{center}

There is an interesting third possibility where the exit time neither satisfies \eqref{trap} nor \eqref{notrap}. As a simple example, one can think of regularizations having fast oscillations in $\tau$, say
$h(\Phi,\tau,y) = \sin \pi \tau$ and thus one must solve $(\sin \pi \tau) t^\star(\tau) = \delta(\tau)$. For $\tau \in \mathbb{N}$, the solution stays at the origin and $t^\star = +\infty$.
\\\\
Independently, the notion of \emph{extremal} solutions, often cited as being the solutions selected in inviscid limits has a net and simple definition in our context:
\begin{definition}[Extremal solutions]
An extremal solution of \eqref{P_0} is such that  $\Phi_t[f_0(\cdot)] x_0 \in \partial {\cal S}_0$.
\end{definition}
The mechanical consequence of this definition is that all regularizations for which the exit time tends to zero as $\tau \to \infty$ select states lying on the boundary of ${\cal S}_0$:
\be 
\bigcup_{y \in {\cal Y}_0} \omega(y) \subset \partial {\cal S}_0, \quad
{\cal Y}_0 := \{ y : \lim_{\tau \to \infty} t^\star(\tau,y) = 0 \}.
\de 
It is worth noting the SDE case for regularizations satisfying
$
dX_s = f_0(X_s)\, ds + \frac{1}{\sqrt{\tau}}\, dW_s, \quad X_0 = x_0.
$
In this case, the exit time is expected to vanish in the limit, leading to the selection of extremal solutions. We do not discuss this situation further.

\paragraph{A simple example}

We consider the same example than in Section \ref{Ex1}, namely $f_0(x) = x^\frac13, x_0=0$ and $t=1$ giving
${\cal S}_0  = \left[ -\left(\frac23 \right)^\frac32,
+\left(\frac23 \right)^\frac32 \right]$. Our aim is to build an explicit system \eqref{RG_system}. We choose some monotonic decreasing $g$ such that ${\cal R}_\tau y$ converges to $\chi$ such that it crosses the real axis at one point $\chi$:
$$
g(\phi) = \chi - \phi,~{\cal R}_\tau y = \chi + (y-\chi) e^{-\tau},
$$
For simplicity, we do not write the dependencies on $y,\tau$ unless required.
Let $t^\star$ be the \emph{exit time} from the ball $B_\delta$ for the system
$\dot \Phi = h(\Phi), \Phi(0)=0$. Note that, as soon as the solution leaves $B_\delta$, it cannot return back.
Let us assume that \eqref{notrap} holds. The solution is therefore $\Phi(t) = \pm \left( \delta^\frac23 + \frac23 (1-t^\star) 
 \right)^\frac32 $ where the sign depends on whether the solution leaves to the left or right. Let us focus on the case where the solution leave $B_\delta$ to the right. Then one must have
 $
 \left( \delta^\frac23 + \frac23 (1-t^\star) 
 \right)^\frac32  = {\cal R}_\tau y
 $; see \eqref{RG_system},\eqref{RGconstr}. We observe that
 $\lim_{\tau \to \infty} {\cal R}_\tau y = \chi$ and thus in light of the constraint, the corresponding regularization must be such that $\lim_{\tau \to \infty} t^\star(\tau,y)$ exist, call this limit $T^\star$. It gives the relation
 \be \label{T_star}
 T^\star = 1 - \frac32 \chi^\frac23.
 \de 
 In fact, one can exhibit some $h$ such that $T^\star$ takes any values depending on $\delta$. Consider for instance $h(\Phi,\tau,y) = \Phi + \sigma(\tau)$ where
 $\sigma > 0$. From the previous remark, the limit  $\lim_{\tau \to \infty} \sigma(\tau)$ must exist. It gives the solution 
 $\Phi^{(h)}_{\tau,y}(s) = \sigma(\tau) (e^s-1)$ namely
 $$
 t^\star(\tau) = \log \left( 1 + \frac{\delta(\tau)}{\sigma(\tau)} \right).
 $$
 One is free to choose the ratio as we want, e.g. $\sigma = \frac{\delta}{e^{T^\star}-1}$, in which case $\lim_{\tau \to 0} t^\star(\tau) = T^\star$ with
 the constraint \eqref{T_star}.
 One has thus distinguished for each state in ${\cal S}_0$ a simple dynamical system 
 ${\cal R}_\tau$ which selects $\chi \in {\cal S}_0=\left[ -\left(\frac23 \right)^\frac32,
+\left(\frac23 \right)^\frac32 \right]$.
These regularizations, by construction, cannot display $\SP$, the obstruction being purely one-dimensional.  
In dimension one, dissipative dynamical systems have attractors reduced to single points, and mixing or ergodic dynamics with continuous $\tau$ simply do not occur; at the very least, one must resort to \emph{discrete} dynamics.  
RG regularizations are therefore confined to visiting only the extreme points of $\mathcal{M}_0$, namely $\mathcal{E} = \{\delta_x\}_{x \in \mathcal{S}_0}$ (see Theorem~\ref{M0M}), without ever reaching the convex combinations required for $\SP$.  
Example~\ref{Ex1} shows that $\SP$ can nonetheless arise but for more general (non-autonomous) regularizations lying outside the RG framework.
\\

The notion of universality in this context depends sensitively on the choice of RG--flow regularizations, since it is tied to the basins of attraction 
of the dynamical system \eqref{RG_Phi2}. As an example, for the same inviscid system we produce a regularization with $n$ distinct classes, not all of equal size. A different choice of RG flow can yield a \emph{totally} different scenario.
Assume that the RG semigroup generator $g$ crosses the real axis at several points within the interval $\left[ -\left(\frac{2}{3}\right)^{\frac{3}{2}}, +\left(\frac{2}{3}\right)^{\frac{3}{2}} \right]$. Suppose $g$ has $n$ roots $\rho_k$ separated by $n-1$ saddle points $\sigma_k$:
$$
-\left(\frac{2}{3}\right)^{\frac{3}{2}} \leq \rho_1 < \sigma_1 < \rho_2 < \sigma_2 < \cdots < \sigma_{n-1} < \rho_n \leq +\left(\frac{2}{3}\right)^{\frac{3}{2}}.
$$
The corresponding RG flow then has $n$ basins of attraction, denoted ${\cal B}_k$, such that all regularizations with parameter $y \in {\cal B}_k$ select the same state $\rho_k$ in the inviscid limit. These basins are
$$
{\cal B}_k = (\sigma_{k-1}, \sigma_k), \quad k=1,\dots,n,
$$
with the convention $\sigma_0 = -\infty$ and $\sigma_n = +\infty$. In this setting one typically obtains a Morse–Smale system satisfying Corollary~\ref{coroDS}. In polar coordinates this yields Strong-$\SP$ regularizations with measures supported on limit cycles, though not in one dimension, where the dynamics can only be gradient.

The extreme version is when $n \to \infty$, i.e. when $g|_{{\cal S}_0}=0$ and $g(\phi) > 0, \phi < -(2/3)^{3/2}$ and $g(\phi) < 0,~\phi > (2/3)^{3/2}$
so that ${\cal S}_0$ is attracting all trajectories for $y \notin {\cal S}_0$. One has ${\cal R}_\tau$ having ${\cal S}_0$ as a global attractor, and any $\mu \in {\cal P}({\cal S}_0)$ is an invariant (non-physical) measure for ${\cal R}_\tau$. In view of Theorem  \ref{RGattractors}, one has in this case:
${\cal E}_{\cal R} = \{ \delta_x \}_{x \in {\cal S}_0} = {\cal M}_{\cal R} \subset {\cal M}_{\cal R}^\# = \operatorname{Inv}({\cal R}) = {\cal P}({\cal S}_0)$. 
This dynamical system is not hyperbolic, and is strongly degenerated.
Such example can indeed be generalized to any system by imposing 
$g|_{{\cal S}_0} = 0$, to produce an alternative simple proof that ${\cal E}_{\cal R} = {\cal E} \subset {\cal M}_0$; see \ref{EinM}.

\section*{Acknowledgment} \label{Ack}
We would like to thank S.Thalabard and J.Bec for stimulating discussions on the subject and the IDEX summer school "100 years of cascades", from which this project has started. We also gratefully acknowledge the Calisto team at INRIA for their warm hospitality and continuous support throughout the project.
\bibliography{REFESS.bib}

\begin{thebibliography}{10}
\providecommand{\url}[1]{\texttt{#1}}
\providecommand{\urlprefix}{URL }
\expandafter\ifx\csname urlstyle\endcsname\relax
  \providecommand{\doi}[1]{doi:\discretionary{}{}{}#1}\else
  \providecommand{\doi}{doi:\discretionary{}{}{}\begingroup
  \urlstyle{rm}\Url}\fi

\bibitem{Agarwal_Lak}
R.~Agarwal and V.~Lakshmikantham, \textit{Uniqueness and Nonuniqueness Criteria
  for Ordinary Differential Equations}, \textit{Series in Real Analysis},
  vol.~6, World Scientific, Singapore, 1993.

\bibitem{alberti2019loss}
G.~Alberti, G.~Crippa, and A.~L. Mazzucato, \textit{Loss of regularity for the
  continuity equation with non-lipschitz velocity field}, Annals of PDE
  \textbf{5} (2019), no.~1, 9.

\bibitem{ambrosio2004transport}
L.~Ambrosio, \textit{{Transport equation and Cauchy problem for BV vector
  fields}}, Inventiones mathematicae \textbf{158} (2004), no.~2, 227--260.

\bibitem{AV23}
S.~Armstrong and V.~Vicol, \textit{Anomalous diffusion by fractal
  homogenization}, arXiv preprint arXiv:2305.05048  (2023).

\bibitem{AV24}
S.~Armstrong and V.~Vicol, \textit{{Anomalous Diffusion by Fractal
  Homogenization}}, Annals of PDE \textbf{11} (2025), 2.

\bibitem{AAV25}
S.~Armstrong and V.~Vicol, \textit{Anomalous diffusion via iterative
  quantitative homogenization: an overview of the main ideas}, arXiv preprint
  arXiv:2503.11744  (2025).

\bibitem{arnold1999bifurcation}
V.~I. Arnold, V.~S. Afrajmovich, Y.~S. Il'yashenko, and L.~P. Shil'nikov,
  \textit{Bifurcation Theory and Catastrophe Theory}, \textit{Encyclopaedia of
  Mathematical Sciences}, vol.~5, Springer-Verlag, Berlin, Heidelberg, 1999.

\bibitem{Bahouri2011}
H.~Bahouri, J.-Y. Chemin, and R.~Danchin, \textit{Fourier Analysis and
  Nonlinear Partial Differential Equations}, \textit{Grundlehren der
  mathematischen Wissenschaften}, vol. 343, Springer, 2011,
  \doi{10.1007/978-3-642-16830-7}.

\bibitem{Ball1997}
J.~Ball, \textit{{Continuity Properties and Global Attractors of Generalized
  Semiflows and the Navier-Stokes Equations}}, J. of Nonlinear Sci. \textbf{7}
  (1997), 475--502.

\bibitem{Gawedzki98}
D.~Bernard, K.~Gaw\c{e}dzki, and A.~Kupiainen, \textit{{Slow modes in passive
  advection}}, J. Stat. Phys. \textbf{90} (1998), no.~3, 519--569.

\bibitem{biferale2018rayleigh}
L.~Biferale, G.~Boffetta, A.~Mailybaev, and A.~Scagliarini,
  \textit{Rayleigh-taylor turbulence with singular nonuniform initial
  conditions}, Physical Review Fluids \textbf{3} (2018), no.~9, 092601.

\bibitem{BEC}
R.~Bitane, H.~Homann, and J.~Bec, \textit{Time scales of turbulent relative
  dispersion}, Phys. Rev. E \textbf{86} (2012), no.~4, 045302.

\bibitem{Borelli}
V.~Borrelli, S.~Jabrane, F.~Lazarus, and B.~Thibert, \textit{Flat tori in
  three-dimensional space and convex integration}, Proceedings of the National
  Academy of Sciences \textbf{109} (2012), no.~19, 7218--7223.

\bibitem{Brenier1989}
Y.~Brenier, \textit{The least action principle and the related concept of
  generalized flows for incompressible perfect fluids}, Journal of the American
  Mathematical Society \textbf{2} (1989), no.~2, 225--255.

\bibitem{bruckner1978}
A.~Bruckner, \textit{Differentiation of Real Functions}, \textit{Lecture Notes
  in Mathematics}, vol. 659, Springer Berlin, Heidelberg, 1978,
  \doi{10.1007/BFb0069821}.

\bibitem{BuckVic}
T.~Buckmaster and V.~Vicol, \textit{{Nonuniqueness of Weak Solutions to the
  Navier-Stokes Equation}}, Annals of Mathematics \textbf{189} (2019), no.~1,
  101--144.

\bibitem{ReviewConvex}
T.~Buckmaster and V.~Vicol, \textit{Convex integration and phenomenologies in
  turbulence}, EMS Surveys in Mathematical Sciences \textbf{6} (2020), no.~1,
  173--263.

\bibitem{Admi_Onsager}
T.~Buckmaster, C.~De~Lellis, L.~Székelyhidi~Jr, and V.~Vicol,
  \textit{Onsager’s conjecture for admissible weak solutions}, Communications
  on Pure and Applied Mathematics \textbf{72} (2018), no.~2, 229--274.

\bibitem{burczak2023anomalous}
J.~Burczak, L.~Sz{\'e}kelyhidi~Jr, and B.~Wu, \textit{{Anomalous dissipation
  and Euler flows}}, arXiv preprint arXiv:2310.02934  (2023).

\bibitem{Ciro}
C.~Campolina and A.~Mailybaev, \textit{{Chaotic Blowup in the 3D Incompressible
  Euler Equations on a Logarithmic Lattice}}, Phys. Rev. Lett. \textbf{121}
  (2018), no.~6, 064501.

\bibitem{Campolina_2021}
C.~S. Campolina and A.~A. Mailybaev, \textit{Fluid dynamics on logarithmic
  lattices}, Nonlinearity \textbf{34} (2021), no.~7, 4684.

\bibitem{James2003}
T.~Caraballo, P.~Marín-Rubio, and J.~Robinson, \textit{A comparison between
  two theories for multi-valued semiflows and their asymptotic behaviour},
  Set-Valued Analysis \textbf{11} (2003), no.~3, 297--322.

\bibitem{Cardy2008}
J.~Cardy, G.~Falkovich, and K.~Gaw\c{e}dzki, \textit{Non-equilibrium
  statistical mechanics and turbulence}, S.~Nazarenko and O.~V. Zaboronski
  (eds.), \textit{Non-Equilibrium Statistical Mechanics and Turbulence},
  Cambridge University Press, 2008. 1--161, \doi{10.1017/CBO9780511755756}.

\bibitem{Chaves2003}
M.~Chaves, K.~Gawedzki, P.~Horvai, A.~Kupiainen, and M.~Vergassola,
  \textit{{Lagrangian Dispersion in Gaussian Self-Similar Velocity Ensembles}},
  J. Stat. Phys. \textbf{113} (2003), 643--692.

\bibitem{ChowHwa1995}
C.~C. Chow and T.~Hwa, \textit{Defect-mediated stability: An effective
  hydrodynamic theory of spatio-temporal chaos}, Physica D: Nonlinear Phenomena
  \textbf{84} (1995), no. 3-4, 494--521.

\bibitem{colombo}
M.~Colombo, G.~Crippa, and M.~Sorella, \textit{{Anomalous Dissipation and Lack
  of Selection in the Obukhov--Corrsin Theory of Scalar Turbulence}}, Annals of
  PDE \textbf{9} (2023), no.~21.

\bibitem{CET94}
P.~Constantin, W.~E, and E.~Titi, \textit{{Onsager’s Conjecture on the Energy
  Conservation for Solutions of Euler’s Equation}}, Comm. in Math. Phys.
  \textbf{165} (1994), no.~1, 207--209.

\bibitem{corrsin1951}
S.~Corrsin, \textit{On the spectrum of isotropic temperature fluctuations in an
  isotropic turbulence}, Journal of Applied Physics \textbf{22} (1951), no.~4,
  469--473.

\bibitem{crisanti1993intermittency}
A.~Crisanti, M.~Jensen, A.~Vulpiani, and G.~Paladin, \textit{Intermittency and
  predictability in turbulence}, Physical review letters \textbf{70} (1993),
  no.~2, 166.

\bibitem{hprinciple}
C.~De~Lellis and L.~Székelyhidi~Jr, \textit{The h-principle and the equations
  of fluid dynamics}, Annals of Mathematics \textbf{193} (2013), no.~3,
  1013--1060.

\bibitem{diperna1989ordinary}
R.~DiPerna and P.-L. Lions, \textit{Ordinary differential equations, transport
  theory and sobolev spaces}, Inventiones mathematicae \textbf{98} (1989),
  no.~3, 511--547.

\bibitem{DiPerna1987}
R.~DiPerna and A.~Majda, \textit{Oscillations and concentrations in weak
  solutions of the incompressible fluid equations}, Communications in
  Mathematical Physics \textbf{108} (1987), 667--689.

\bibitem{Drivas17}
T.~Drivas and G.~Eyink, \textit{{A Lagrangian fluctuation–dissipation
  relation for scalar turbulence}}, J. Fluid Mech. \textbf{829} (2017),
  153--189.

\bibitem{Drivas21}
T.~Drivas and A.~Mailybaev, \textit{{'Life after death' in ordinary
  differential equations with a non-Lipschitz singularity}}, Nonlinearity
  \textbf{34} (2021), 2296.

\bibitem{Drivas24}
T.~Drivas, A.~Mailybaev, and A.~Raibekas, \textit{{Statistical determinism in
  non-Lipschitz dynamical systems}}, Ergodic Theo. and Dyn. Syst. \textbf{44}
  (2024), 1856--1884.

\bibitem{Drivas_Elgindi}
T.~Drivas, T.~Elgindi, G.~Iyer, and I.-J. Jeong, \textit{{Anomalous dissipation
  in passive scalar transport}}, Arch. Ration. Mech. Anal. \textbf{243} (2022),
  no. 243(3), 1151--1180.

\bibitem{EVDE}
W.~E and E.~Vanden~Eijnden, \textit{Generalized flows, intrinsic stochasticity,
  and turbulent transport}, Proc. Natl. Acad. Sci. \textbf{97} (2000),
  8200--8205.

\bibitem{E2003}
W.~E and E.~Vanden-Eijnden, \textit{A note on generalized flows}, Physica D:
  Nonlinear Phenomena \textbf{183} (2003), 159--174.

\bibitem{ElgLiss24}
T.~M. Elgindi and K.~Liss, \textit{Norm growth, non-uniqueness, and anomalous
  dissipation in passive scalars}, Archive for Rational Mechanics and Analysis
  \textbf{248} (2024), no.~6, 120.

\bibitem{Eyink_Bandak20}
G.~Eyink and D.~Bandak, \textit{Renormalization group approach to spontaneous
  stochasticity}, Phys. Rev. Res. \textbf{2} (2020), 043161.

\bibitem{Falkovich2001}
G.~Falkovich, K.~Gawedzki, and M.~Vergassola, \textit{Particles and fields in
  fluid turbulence}, Reviews of Modern Physics \textbf{73} (2001), 913--975.

\bibitem{Feigenbaum1976}
M.~Feigenbaum, \textit{Universality in complex discrete dynamics}, Los Alamos
  Theoretical Division Annual Report  (1976), 98--102.

\bibitem{Flandoli2009}
F.~Flandoli, \textit{Remarks on uniqueness and strong solutions to
  deterministic and stochastic differential equations}, Metrika \textbf{69}
  (2009), 101--123.

\bibitem{Foias1972}
C.~Foias, \textit{{Statistical Study of Navier-Stokes Equations I}}, Rendiconti
  del Seminario Matematico della Università di Padova \textbf{48} (1972),
  219--348.

\bibitem{Foias1976}
C.~Foias and G.~Prodi, \textit{{Sur les solutions statistiques des \'equations
  de Navier–Stokes}}, Annali di Matematica Pura ed Applicata \textbf{111}
  (1976), no.~4, 307--330.

\bibitem{Foias2013}
C.~Foias, R.~M. Rosa, and R.~Temam, \textit{{Properties of Time-Dependent
  Statistical Solutions of the Three-Dimensional Navier-Stokes Equations}},
  Annales de l'Institut Fourier \textbf{63} (2013), no.~6, 2515--2573.

\bibitem{FrischBec1}
U.~Frisch, T.~Matsumoto, and J.~Bec, \textit{{Singularities of Euler Flow? Not
  Out of the Blue!}}, J. Stat. Phys. \textbf{113} (2003), no. 5-6, 761--781.

\bibitem{Gilson1998}
J.~Gilson, I.~Daumont, and T.~Dombre, \textit{A two-fluid picture of
  intermittency in shell-models of turbulence}, \textit{Advances in Turbulence
  VII}, Springer, 1998, 219--222.

\bibitem{giorgi1992dini}
G.~Giorgi and S.~Koml{\'o}si, \textit{{D}ini derivatives in optimization —
  part i}, Rivista di matematica per le scienze economiche e sociali
  \textbf{15} (1992), 3--30.

\bibitem{Gromov}
M.~Gromov, \textit{Local and Global in Geometry}, Balzan Prize, 1999.

\bibitem{Hartman2002}
P.~Hartman, \textit{Ordinary Differential Equations}, \textit{Classics in
  Applied Mathematics}, vol.~38, 2nd edn., SIAM, 2002,
  \doi{10.1137/1.9780898719222}.

\bibitem{Titi2023}
L.~Huysmans and E.~Titi, \textit{{Non-Uniqueness and Inadmissibility of the
  Vanishing Viscosity Limit for the Passive Scalar Transport Equation}},
  arXiv:2307.00809  (2023).

\bibitem{Isett}
P.~Isett, \textit{A proof of onsager’s conjecture}, Annals of Mathematics
  \textbf{188} (2018), no.~3, 871--963.

\bibitem{Jullien2000}
M.-C. Jullien, P.~Castiglione, and P.~Tabeling, \textit{Experimental
  observation of batchelor dispersion of passive tracers}, Phys. Rev. Lett.
  \textbf{85} (2000), 3636--3639.

\bibitem{KARABACAK_ASHWIN_2011}
O.~Karabacak and P.~Ashwin, \textit{On statistical attractors and the
  convergence of time averages}, Mathematical Proceedings of the Cambridge
  Philosophical Society \textbf{150} (2011), no.~2, 353--365.

\bibitem{Kolmogorov1941}
A.~Kolmogorov, \textit{Dissipation of energy in locally isotropic turbulence},
  Doklady Akademii Nauk SSSR \textbf{31} (1941), 538--540.

\bibitem{Kraichnan}
R.~Kraichnan, \textit{{Small‐Scale Structure of a Scalar Field Convected by
  Turbulence}}, Physics of Fluids \textbf{11} (1968), 945--953.

\bibitem{Kuiper}
N.~Kuiper, \textit{On {$C^1$} isometric imbeddings i, ii}, Proc. Kon. Acad.
  Wet. Amsterdam A \textbf{58} (1955), 545--556, 683--689.

\bibitem{kuksin}
S.~Kuksin, \textit{{The Eulerian Limit for 2D Statistical Hydrodynamics}}, J.
  Stat. Phys. \textbf{115} (2004), no. 1-2, 469--492.

\bibitem{Lasota1994}
A.~Lasota and M.~Mackey, \textit{Chaos, Fractals, and Noise: Stochastic Aspects
  of Dynamics}, \textit{Applied Mathematical Sciences}, vol.~97, 2nd edn.,
  Springer, 1994, \doi{10.1007/978-1-4612-4286-4}.

\bibitem{DeLellis2021}
C.~D. Lellis and V.~Giri, \textit{{Smoothing does not give a selection
  principle for transport equations with bounded autonomous fields}}, Annales
  math\'ematiques du Qu\'ebec \textbf{46} (2022), 27--39.

\bibitem{Leray}
J.~Leray, \textit{Sur le mouvement d'un liquide visqueux emplissant l'espace},
  Acta Math. \textbf{63} (1934), no.~1, 193--248.

\bibitem{Lorenz63}
E.~Lorenz, \textit{Deterministic nonperiodic flow}, J. Atmos. Sci. \textbf{20}
  (1963), no.~2, 130--141.

\bibitem{Lorenz69}
E.~Lorenz, \textit{The predictability of a flow which possesses many scales of
  motion}, Tellus \textbf{21} (1969), no.~3, 289--307.

\bibitem{Mailybaev2012}
A.~Mailybaev, \textit{Computation of anomalous scaling exponents of turbulence
  from self-similar instanton dynamics}, Physical Review E \textbf{86} (2012),
  no.~2, 025301.

\bibitem{Maily2012}
A.~Mailybaev, \textit{Renormalization and universality of blowup in
  hydrodynamic flows}, Physical Review E \textbf{85} (2012), no.~6, 066317.

\bibitem{Maily2016}
A.~Mailybaev, \textit{Spontaneous stochasticity of velocity in turbulence
  models}, Multiscale Modeling \& Simulation \textbf{14} (2016), no.~1,
  96--112.

\bibitem{AM_24end}
A.~Mailybaev, \textit{{RG} analysis of spontaneous stochasticity on a fractal
  lattice: linearization and bifurcations}, arXiv:2410.14903  (2024).

\bibitem{AM_24}
A.~Mailybaev, \textit{{RG} approach to the inviscid limit for shell models of
  turbulence}, arXiv:2408.04659  (2024).

\bibitem{AM_Ra23}
A.~Mailybaev and A.~Raibekas, \textit{Spontaneous stochasticity and
  renormalization group in discrete multi-scale dynamics}, Commun. Math. Phys.
  \textbf{401} (2023), 2643--2671.

\bibitem{AM_Rb23}
A.~Mailybaev and A.~Raibekas, \textit{Spontaneous stochasticity {A}rnold's
  {C}at}, Arnold Math J. \textbf{9} (2023), 339--357.

\bibitem{mailybaev2017toward}
A.~A. Mailybaev, \textit{Toward analytic theory of the rayleigh--taylor
  instability: lessons from a toy model}, Nonlinearity \textbf{30} (2017),
  no.~6, 2466.

\bibitem{Mengual}
F.~Mengual and L.~Székelyhidi~Jr, \textit{{Dissipative Euler Flows for Vortex
  Sheet Initial Data without Distinguished Sign}}, Communications on Pure and
  Applied Mathematics \textbf{76} (2023), no.~1, 163--221.

\bibitem{Nash}
J.~Nash, \textit{{$C^1$} isometric imbeddings}, Annals of Mathematics
  \textbf{60} (1954), 383--396.

\bibitem{Obukhov49}
A.~M. Obukhov, \textit{The structure of the temperature field in a turbulent
  flow}, Izvestiya Akad. Nauk SSSR. Ser. Geograf. Geofiz. \textbf{13} (1949),
  58--69.

\bibitem{Onsager49}
L.~Onsager, \textit{Statistical hydrodynamics}, Il Nuovo Cimento, Supplemento
  \textbf{6} (1949), 279.

\bibitem{Ortiz2025Spontaneous}
E.~Ortiz, C.~S. Campolina, and A.~A. Mailybaev, \textit{Spontaneous
  stochasticity in the fluctuating navier-stokes equations on a logarithmic
  lattice}, arXiv preprint arXiv:2507.03196  (2025).

\bibitem{Palmer14}
T.~Palmer, A.~Döring, and G.~Seregin, \textit{The real butterfly effect},
  Nonlinearity \textbf{27} (2014), no.~9, R123.

\bibitem{pomeau1984intrinsic}
Y.~Pomeau, A.~Pumir, and P.~Pelce, \textit{Intrinsic stochasticity with many
  degrees of freedom}, Journal of Statistical Physics \textbf{37} (1984),
  39--49.

\bibitem{Richardson1926}
L.~Richardson, \textit{Atmospheric diffusion shown on a distance-neighbour
  graph}, Proceedings of the Royal Society of London. Series A, Containing
  Papers of a Mathematical and Physical Character \textbf{110} (1926),
  709--737.

\bibitem{RSM}
R.~Robert and J.~Sommeria, \textit{Statistical equilibrium states for
  two-dimensional flows}, J. Fluid Mech. \textbf{229} (1991), 291–310.

\bibitem{Rotunno_Snyder08}
R.~Rotunno and C.~Snyder, \textit{{A Generalization of {L}orenz’s Model for
  the Predictability of Flows with Many Scales of Motion}}, J. of Atmos. Sci.
  \textbf{65} (2008), no.~3, 1063 -- 1076.

\bibitem{RoyKS}
D.~Roy and R.~Pandit, \textit{One-dimensional kardar-parisi-zhang and
  kuramoto-sivashinsky universality class: Limit distributions}, Phys. Rev. E
  \textbf{101} (2020), 030103.

\bibitem{salazar2009}
J.~Salazar and L.~Collins, \textit{Two-particle dispersion in isotropic
  turbulent flows}, Annual Review of Fluid Mechanics \textbf{41} (2009),
  405--432.

\bibitem{Sell1973}
G.~Sell, \textit{Differential equations without uniqueness}, Journal of
  Differential Equations \textbf{14} (1973), no.~1, 42--56.

\bibitem{Simsen2008}
J.~Simsen and C.~Gentile, \textit{On attractors for multivalued semigroups
  defined by generalized semiflows}, Set-Valued Analysis \textbf{16} (2008),
  105--124.

\bibitem{Taylor}
G.~Taylor, \textit{Observations and speculations on the nature of turbulent
  motion}, Reports and Memoranda of the Advisory Committee for Aeronautics
  \textbf{4} (1917), no. 345, 1--20.

\bibitem{Thalabard2020}
S.~Thalabard and J.~Bec, \textit{Turbulence of generalised flows in two
  dimensions}, J. Fluid Mech. \textbf{883} (2020), R1.

\bibitem{Simon_Jeremie_AM20}
S.~Thalabard, J.~Bec, and A.~Mailybaev, \textit{From the butterfly effect to
  spontaneous stochasticity in singular shear flows}, Commun. Phys. \textbf{3}
  (2020), no. 122.

\bibitem{Nicolasetal24}
N.~Valade, S.~Thalabard, and J.~Bec, \textit{Anomalous dissipation and
  spontaneous stochasticity in deterministic surface quasi-geostrophic flow},
  Ann. Henri Poincaré \textbf{25} (2024), 1261--1283.

\bibitem{Vishik1977}
M.~Vishik and A.~Fursikov, \textit{{L’équation de Hopf, les solutions
  statistiques, les moments correspondant aux systèmes des équations
  paraboliques quasi-linéaires}}, J. de Mathématiques Pures et Appliquées
  \textbf{59} (1977), no.~9, 85--122.

\bibitem{Walter1970}
W.~Walter, \textit{Differential and Integral Inequalities}, \textit{Ergebnisse
  der Mathematik und ihrer Grenzgebiete}, vol.~55, Springer, 1970,
  \doi{10.1007/978-3-642-86405-6}.

\bibitem{Yakhot1981Kuramoto}
V.~Yakhot, \textit{Large-scale properties of unstable systems governed by the
  kuramoto-sivashinsky equation}, Physical Review A \textbf{24} (1981), no.~1,
  642--644.

\end{thebibliography}
%%%%%%%%%%%%%%%%%%%%%%%%%%%%%%%%%%%%%%%%%%%%%%%%%%%%%%%%
\appendix 
\section{Basic Notions from Measure Theory and Dynamical Systems}\label{PFTP}
\begin{enumerate}
    \item{\bf Pushforward}.
    \begin{definition}[Pushforward]
       Let $(X,{\cal A})$ and $(Y,{\cal B})$ be two measurable spaces. Let $\mu$ be a measure on $X$ and a measurable map
       $$
       f : X \to Y,~f^{-1}(B) \in {\cal A},~\forall B \in {\cal B}.
       $$
    \end{definition}
    The pushforward measure $f_\# \mu $ is defined as
    $$
    (f_\# \mu)(B) = \mu(f^{-1}(B)),~\forall B \in {\cal B}.
    $$
    One can also write, for all (bounded continuous) test functions $F \in C_b(Y;\mathbb{R})$:
    $$
    \langle f_\# \mu,F \rangle = \int_Y F(y) (f_\# \mu)(dy) = \int_X (F\circ f)(x) \mu(dx).
    $$
    In words, it is merely a change of variables where the statistics encoded by $\mu$ on $X$ are mapped to statistics on $Y$ through the map $f:X \to Y$. It is of interest to notice that for a semigroup $S_t:X \to Y$, the pushforward $(S_t)_\#$ has another well-known name: it is the \emph{Perron-Frobenius/transfer operator}.
    Namely, one has 
    $$\langle ({S}_t)_\# \mu,F \rangle = \int_X F(S_t x) \mu(dx) = \int_X (U_t F)(x) \mu(dx) = \langle \mu, U_t F \rangle = \langle (U_t)^\star \mu, F \rangle. $$ Here $U_t$ is the \emph{Koopman operator} acting on observables:
    $$
    U_t: C_b(X;\mathbb{R}) \to C_b(Y;\mathbb{R}): (U_t F)(x) = F(S_t x),
    $$
    and $(U_t)^\star$ is the dual Perron-Frobenius operator acting on measures. Although, we do not use these concepts, it is useful to recall that the infinitesimal generator of the Perron-Frobenius operator for $S_t$ the solution of some deterministic autonomous ODE/PDE $\dot x = f(x)$  is precisely the \emph{Liouville} operator ${\cal L} \mu = -\nabla \cdot (f \mu)$ acting on densities $\mu$. It is often written as $(S_t)_\#  = e^{t {\cal L}} $.

 \item[]

\item{\bf Tightness and Prokhorov's Theorem}. 
     Let us now defined tightness of a family of probability measures and state Prokhorov's theorem linking tightness and relative compactness
\begin{definition}[Tightness]
        Let $(X,T)$ be a Hausdorff space and let $\Sigma$ be a $\sigma$-algebra on $X$ that contains the topology $T$. Let $\lbrace\mu_\kappa \rbrace$ be a family of probability measures defined on $\Sigma$. The collection $\lbrace\mu_\kappa \rbrace$ is called tight if for any $\varepsilon>0$ there exists a compact set $K_\varepsilon \subset X$ such that for all $\kappa$, 
        $$ \mu_\kappa(K_\varepsilon) \geq 1-\varepsilon $$
\end{definition}
Tightness of collections of probability measure ensures relative compactness and therefore subsequential weak convergence as stated by Prokhorov's theorem. For simplicity, we restrict to \emph{Polish spaces} which are separable completely metrizable topological spaces. In particular, Banach spaces are Polish.
%which are  our main concern in this article
        
\begin{theorem}[Prokhorov's Theorem]
         Let $(S,\rho)$ be a separable and complete metric space and $\mathcal{P}(S)$ be the set of probability measures on $S$. There is a metric $d_0$ on $\mathcal{P}(S)$ equivalent to the topology of weak convergence. Moreover, a family of probability measures $\lbrace\mu_\kappa \rbrace\subset \mathcal{P}(S)$ is tight if and only if the closure of $\lbrace\mu_\kappa \rbrace$ in $(\mathcal{P}(S),d_0)$ is compact.
\end{theorem}
\item[] 
\item{\bf Portmanteau Theorem}
There are many different formulations, we give only two of them:
Let $(\mu_n)$ and $\mu$ be probability measures on a metric space $X$.
The Portmanteau theorem: $\mu_n \rightharpoonup \mu$ \emph{if and only if} any one of the following equivalent conditions holds.
(Closed-set form) For every closed $F\subset X$,
$$
\limsup_{n\to\infty} \mu_n(F) \le \mu(F).
$$
(Equivalently, open-set form) For every open $G\subset X$,
$$
\liminf_{n\to\infty} \mu_n(G) \ge \mu(G).
$$
The rule of thumb is that, one uses the closed-set form to show $x \in \operatorname{supp}(\mu)$ and the open-set form to show that for $n$ large enough $x \in \operatorname{supp}(\mu_n)$.
\item[] 
\item{\bf Continuous mapping Theorem}. 
Let $X,Y$ be two Polish spaces and $\{\mu_\epsilon\}_{\epsilon > 0}$ a family of probability measures on $X$ such that $\mu_\epsilon \rightharpoonup \mu \in {\cal P}(X)$ as $\epsilon \to 0$. Let $h:X\to Y$ a continuous function, then
$$
h_\# \mu_\epsilon \rightharpoonup h_\# \mu \in {\cal P}(Y).
$$
\item[] 
\item {\bf Ergodic invariant measures}.
Let $(X, \mathcal{B}, \mu, (\mathcal{R}_t)_{t \in \mathbb{R}})$ be a measure-preserving flow, i.e., $(X, \mathcal{B}, \mu)$ is a probability space and each $\mathcal{R}_t : X \to X$ is measurable with $\mu(\mathcal{R}_t^{-1}B) = \mu(B)$ for all $B \in \mathcal{B}$ and $t \in \mathbb{R}$, or equivalently ${\cal R}_\# \mu = \mu$. A $\mathcal{R}_t$-invariant measure $\mu$ is \emph{ergodic} if every invariant set $A \in \mathcal{B}$ satisfies $\mu(A) \in \{0,1\}$. Equivalently, $\mu$ is ergodic if for all $F \in L^1(\mu)$, the time averages
$$
\frac{1}{T} \int_0^T F(\mathcal{R}_t x)\, dt \ \xrightarrow[T \to \infty]{} \ \int_X F(x)\, \mu(dx)~~\mu\text{-a.e.}
$$
Ergodicity means that the statistical behavior under the flow is indecomposable: $\mu$ cannot be expressed as a nontrivial convex combination of other invariant probability measures; see Proof in Section \ref{RG_auto}.
An interesting consequence is that is $F$ is ${\cal R}_t$-invariant:
$F({\cal R}_t x) = F(x)$ $\mu$-a.e. then from the above $F(x) = cst = \int F d\mu$ $\mu$-a.e.
\item[] 
\item {\bf Uniquely ergodic}.
A continuous flow $({\cal R}_t)_{t\in\mathbb{R}}$ on a compact metric space $X$ is \emph{uniquely ergodic} if there exists a single ${\cal R}_t$-invariant Borel probability measure on $X$.  
Equivalently, the space $\operatorname{Inv}({\cal R})$ of invariant probability measures is a singleton.  
In this case, time averages of every continuous function $F$ converge \emph{uniformly} to the same constant independently of $x \in X$.
\item[] 
\item {\bf Pointwise uniquely ergodic/generic}. A continuous flow $({\cal R}_t)_{t\in\mathbb{R}}$ on a compact metric space $X$ is \emph{pointwise uniquely ergodic} if there exists $\mu\in\operatorname{Inv}({\cal R})$ such that, for $\mu$-almost every $x\in X$ and every continuous $F$,
$
\frac{1}{T} \int_0^T F({\cal R}_t x)\,dt \ \xrightarrow[T\to\infty]{} \ \int_X F\,d\mu.
$
Such $x$ are called \emph{$\mu$-generic} or \emph{generic points} for $\mu$. Unlike unique ergodicity, the invariant measure need not be the only one; it is the unique limit for almost every starting point.
\item[] 
\item {\bf Axiom A}.
Let $({\cal R}_t)_{t\in\mathbb{R}}$ be a $C^1$ flow on a compact manifold $M$.  We need first to define few important concepts.

\emph{Nonwandering set:} $\Omega({\cal R})=\{x\in M:\ \forall\,$neighborhoods $U\ni x$ and $\forall T>0,\ \exists t\ge T$ with ${\cal R}_t(U)\cap U\neq\varnothing\}$.  
Intuitively, these are points whose trajectories return arbitrarily close after arbitrarily long times.  
\emph{Topologically transitive:} an invariant set $\Lambda\subset M$ is transitive if for all nonempty open $U,V\subset \Lambda$ (relative topology) there exists $t\in\mathbb{R}$ with ${\cal R}_t(U)\cap V\neq\varnothing$ (equivalently, $\Lambda$ has a dense orbit).  
It means the dynamics can move from any region of $\Lambda$ to any other, allowing orbits to explore the whole set.  
\emph{Locally maximal:} an invariant set $\Lambda$ is locally maximal if there is a neighborhood $U$ of $\Lambda$ with $\Lambda=\bigcap_{t\in\mathbb{R}}{\cal R}_t(U)$.  
Such a set is isolated from the rest of the dynamics: all nearby orbits remain in $\Lambda$ for all time.  
\emph{Basic set:} a compact invariant set $\Lambda$ is basic if it is hyperbolic (in the flow sense), locally maximal, and topologically transitive.  
A basic set is an indecomposable, self-contained chaotic component of the dynamics. It is not necessarily a (local) attractor.

\emph{Axiom A (flow):} ${\cal R}_t$ satisfies Axiom~A if $\Omega({\cal R})$ is hyperbolic and periodic orbits are dense in $\Omega({\cal R})$.  
Under Axiom~A, $\Omega({\cal R})$ is the disjoint union of finitely many basic sets.

\item[] 
\item{\bf Morse-Smale systems}. 
A $C^1$ flow $({\cal R}_t)_{t\in\mathbb{R}}$ on a compact manifold $M$ is \emph{Morse–Smale} if its nonwandering set consists of finitely many hyperbolic equilibria and periodic orbits, and the stable and unstable manifolds of these intersect transversely.  
Morse–Smale systems form a special case of Axiom~A systems, with $\Omega({\cal R})$ finite and each basic set reduced to a single periodic orbit or equilibrium.

\end{enumerate}

\section{Equivalence of Definitions: Proposition \ref{SPDEFS}}\label{SP_LSP}
We justify Proposition~\ref{SPDEFS}. Figure~\ref{tricho} illustrates the mutually exclusive scenarios.  
In the definition of weak $\SP$, it is not necessary to specify whether the subsequential limits are Dirac or non-Dirac.  
We begin by explaining how a Dirac weak limit can still be compatible with a lack of selection principle.

\subsection{$\delta\text{-}LSP$: Dirac weak limit with lack of selection principle}\label{DiracLSP}

We describe the mechanism responsible for
$\delta\text{-}LSP := LSP \setminus \{ LSP \setminus \delta \}$;  
see Definition~\ref{SPdef}.  
This situation corresponds to the absence of a selection principle,  
but only on a subset whose relative Lebesgue measure vanishes in the inviscid limit.  One can indeed provide a very simple example where
a regularization curve $\gamma$ can be discontinuous everywhere but still converges in measure to a Dirac mass. Take 
$\gamma(s) = \gamma_0 + \mathbf{1}_{\mathbb{Q}}(1/s)$, 
then $\gamma_\# \operatorname{Leb}_\epsilon \rightharpoonup \delta_{\gamma_0}$.
\\

One has indeed situation where $\gamma$ is
\emph{approximately continuous} at zero, i.e.
has an approximate inviscid limit, say $\gamma_0$. Equivalently,
$\gamma$ converges in density to $\gamma_0$
with an exceptional set having vanishing relative Lebesgue measure. There is therefore, strickly speaking, a lack of selection principle that we call $\delta$-$LSP$, but on a set of vanishing (relative) density. 
This is the following lemma:

\begin{lemma}\label{aaecontinuity}
Assume that, there is a $\gamma_0 \in H$ such that $\gamma_\# \operatorname{Leb}_\epsilon \rightharpoonup \delta_{\gamma_0}$, then there exists a Borel set $B$ such that for all $\delta>0$
\be 
%\lim_{s \to 0, s \in B} \gamma(s) = x_0~\text{with}~\lim_{\epsilon \to 0} \operatorname{Leb}_\epsilon(B)=1
\operatorname{Leb}_\epsilon\left( \|\gamma(s)-\gamma_0 \|> \delta 
\right) \to 0~\text{as}~\epsilon \to 0.
\de 
\end{lemma}
\begin{proof}
For all $F \in C_b(H;\mathbb{R})$, $\langle \gamma_\# \operatorname{Leb_\epsilon},F\rangle \to F(\gamma_0)$. We can therefore exhibit positive test functions $0 \leq F_n \leq 1 \in C_b(H;\mathbb{R}), n \in \mathbb{N}$ such that $F_n(\gamma_0)=0$, $F_n(\gamma(s)) = 1$ when $\| \gamma(s)-\gamma_0 \| > 1/n$,
 and
$\langle \gamma_\# \operatorname{Leb_\epsilon},F_n\rangle \to F_n(\gamma_0) = 0$. Define
$$
C_{n}:= \left\{ s ~:~\| \gamma(s)-\gamma_0 \| > \frac{1}{n} \right\}.
$$
One has, 
$$ 
\frac{\operatorname{Leb}_\epsilon(C_n)}{\operatorname{Leb}(C_n \cap [0,\epsilon])} \int_{C_n \cap [0,\epsilon]} 
\underbrace{F_n(\gamma(s))}_{=1} ds =
\langle \gamma_\# \operatorname{Leb}_\epsilon,F_n \rangle 
-  
\frac{\operatorname{Leb}_\epsilon(C_n^c)}{\operatorname{Leb}(C_n^c \cap [0,\epsilon])} \int_{C_n^c \cap [0,\epsilon]} 
\underbrace{F_n(\gamma(s))}_{\in [0,1]} ds \leq \langle \gamma_\# \operatorname{Leb}_\epsilon,F_n \rangle \to 0. 
$$
Therefore $\operatorname{Leb}_\epsilon(C_n) \to 0$ (and $\operatorname{Leb}_\epsilon(C_n^c) \to 1$) as $\epsilon \to 0$.
The technical difficulty is that this convergence does depend on $F_n$ above.
We need a diagonal argument. One can pick a decreasing sequence $\epsilon_n \to 0$ such that $\operatorname{Leb}_{\epsilon_n}(C_n) \leq e^{-n}$ for instance. Then we define
$$
B := \bigcup_{n \geq 1} [0,\epsilon_n] \setminus C_n.
$$
From this definition, for $\epsilon \in (0,\epsilon_n)$, one has $(0,\epsilon_n) \setminus C_n \subset [0,\epsilon] \cap B$, and
$$
1 \geq \operatorname{Leb}_\epsilon(B) \geq 1-\operatorname{Leb}_\epsilon(C_n) \geq
1-\operatorname{Leb}_{\epsilon_n}(C_n) \geq 1-e^{-n} \to 1.
$$
To conclude, for all $s \in (0,\epsilon_n)$, one has $\|\gamma(s)-\gamma_0\| \leq 1/n$ and taking $n=n(s):=\min \{ k~:~s \leq \epsilon_k \}$ and going to the limit, $\gamma(s) \to \gamma_0$ for $s \to 0 \in B$.
\end{proof}

For clarity, we now discuss in detail a simple example.  
Let $\gamma$ be a regularization curve (see~\eqref{gamma}), and assume --without making this link explicit -- that it corresponds to the regularization of an inviscid system~\eqref{P_0}.  
Let ${\cal K}_\epsilon \subset \mathbb{R}^+$ satisfy
$$
{\cal K}_\epsilon \cup {\cal K}_\epsilon^c = [0,\epsilon].
$$
As a concrete example, take
$$
K_\epsilon = \bigcup_{n \geq N(\epsilon)} 
\left[\frac1n-e^{-n},\frac1n+e^{-n}\right] \cap [0,\epsilon],  
\quad \lim_{\epsilon \to 0} N(\epsilon) = \infty.
$$
Let $\gamma_0, \gamma_1 \in {\cal S}_0 \subset H$ with $\gamma_0 \neq \gamma_1$,  
and define a smooth curve $\gamma: s \mapsto \gamma(s) \in H$ so that  
$\gamma(\frac1n) = \gamma_1$ and  
$\left. \gamma\right|_{{\cal K}_\epsilon^c} = \gamma_0$. 
\\

For $F \in C_b(H;\mathbb{R})$ we have
$$
\langle \gamma_\# \operatorname{Leb}_\epsilon, F \rangle 
= \frac{1}{\epsilon} \int_{{\cal K}_\epsilon^c} F(\gamma_0) \, ds 
+ \frac1\epsilon \int_{{\cal K}_\epsilon} F(\gamma(s)) \, ds
= \frac{|{\cal K}_\epsilon^c|}{\epsilon} F(\gamma_0)  
+ \frac{|{\cal K}_\epsilon|}{\epsilon} \frac{1}{|{\cal K}_\epsilon|} 
\int_{{\cal K}_\epsilon} F(\gamma(s)) \, ds.
$$
Since $F$ is bounded and $|{\cal K}_\epsilon| + |{\cal K}_\epsilon^c| = \epsilon$, it follows that
$$
\left| \langle \gamma_\# \operatorname{Leb}_\epsilon - \delta_{\gamma_0},F \rangle \right| 
\leq  C \left| \frac{|{\cal K}_\epsilon^c|}{\epsilon} -1 \right| 
+ C \frac{|{\cal K}_\epsilon|}{\epsilon} 
= 2 C \frac{|{\cal K}_\epsilon|}{\epsilon}.
$$
Moreover,
$$
|{\cal K}_\epsilon| = 2 \sum_{n \geq N(\epsilon)} e^{-n} 
= \frac{2}{e-1} e^{-(N(\epsilon)-1)}.
$$
Choosing, for instance,
$$
N(\epsilon) = \lfloor \epsilon^{-a} \rfloor, \quad a > 0,
$$
we deduce that
$$
\gamma_\# \operatorname{Leb}_\epsilon \rightharpoonup \delta_{\gamma_0}.
$$

In this construction, the regularization curve $\gamma$ oscillates rapidly as $\epsilon \to 0$.  
One can extract a subsequence $\epsilon_n = \frac1n$ with $\gamma(\epsilon_n) \to \gamma_1$,  
and another subsequence $\epsilon_n'$ with $\gamma(\epsilon_n') = \gamma_0$.  
Thus, although the regularization converges to the Dirac measure $\delta_{\gamma_0}$,  
there remains a lack of selection principle.

\subsection{Proof of Proposition \ref{SPDEFS}}
\paragraph{$LSP \Rightarrow \LSPO$}
\begin{proof}
It is simple: one has two subsequences $x^{\epsilon_n},x^{\epsilon_n'} \to x_1,x_2$ in $C([0,T];H)$, and since $x_1 \neq x_2$ there exists a $t \in (0,T)$ such that $x_1(t) \neq x_2(t)$. Defining ${\cal O}(x) = ||x(t)-x_1(t)||$, it is continuous so that ${\cal O}(x^{\epsilon_n}) \to 0$ and
${\cal O}(x^{\epsilon_n'}) \to {\cal O}(x_2(t)) > 0$. Therefore
$\LSPO$ holds.
\end{proof}
\bigskip
\paragraph{$\LSPO \Rightarrow LSP$}
The added difficulty is to show one can recover convergence in $C([0,T];H)$ from pointwise convergence. 
\begin{proof} Let us assume by contradiction $\LSPO$ holds and there is an unique solution of (\ref{P_0}). Let us denote
$\Psi_\epsilon(s) := \Phi_s[f(\cdot,\epsilon)] x_0$ and $\Psi_0(s) :=
\Phi_s[f_0(\cdot)] x_0$. We show in the following that $\forall s \in [0,t]$, $||\Psi_\epsilon(s)-\Psi_0(s)|| \to 0$ uniformly giving a contradiction.

We have $\Psi_\epsilon(s) = x_0 + \int_0^s f(\Psi_\epsilon(s'),\epsilon) ds'$ so that $||\Psi_\epsilon(s)|| \leq ||x_0|| + t \sup_{x} ||f(x,\epsilon)|| \leq C$ uniformly in $\epsilon$ and $s$. Similarly, since $\dot \Psi_\epsilon = f(\Psi_\epsilon,\epsilon)$, 
$||\dot \Psi_\epsilon|| \leq C$, the time derivatives being uniformly bounded, one obtains uniform equicontinuity (by the mean value theorem). One can therefore use the
Arzel\`a-Ascoli theorem which gives uniformly convergent subsequences 
$\{ \Psi_{\epsilon_j} \}$
converging to some $\Psi^\star$ (depending on the subsequence). We can write
$\Psi_{\epsilon_j}(s) = x_0 + \int_0^s f(\Psi_{\epsilon_j}(s'),\epsilon_j) ds'$ and by uniform convergence of the subsequence together with the uniform convergence $ ||f(\cdot,\epsilon_j)-f_0(\cdot)||_\infty \to 0$,
one obtains $\Psi^\star(s)  = x_0 + \int_0^s f_0(\Psi^\star(s')) ds'$.
Since by hypothesis, one has assumed the solution of 
(\ref{P_0}) to be unique, then $\Psi^\star = \Psi_0$ (for all subsequences). We therefore obtain the uniform convergence
$\Psi_\epsilon \to \Psi_0$. Then using the continuity of the observable, one 
also has ${\cal O}(\Phi_t[f(\cdot,\epsilon)] x_0) 
\to {\cal O}(\Phi_t[f_0(\cdot)] x_0)$, namely
$\limsup_{\epsilon \to 0} {\cal O}(\gamma(\epsilon)) = 
\liminf_{\epsilon \to 0} {\cal O}(\gamma(\epsilon))$ giving a contradiction to $\LSPO$.
\end{proof}

\paragraph{$\neg \SP \Longleftrightarrow \text{Selection principle} \bigcup \delta\text{-}LSP$}\label{bigissue}
\begin{proof}
We notice that the negation of $\SP$ is 
$\neg({\rm strong}~ \SP)$ and $\neg({\rm weak}~\SP)$.
%bien vu
Since the two notions are mutually exclusive. Then $\neg({\rm strong}~ \SP)$ is
either weak $\SP$ or $\gamma_\# {\rm Leb}_\epsilon \rightharpoonup \mu$ with $\mu$ a Dirac measure. Now $\neg({\rm weak}~\SP)$ is either strong $\SP$ or
$\gamma_\# {\rm Leb}_\epsilon \rightharpoonup \mu$ with $\mu$ a Dirac measure.
Therefore $\neg \SP$ is just weak convergence to a single Dirac mass, namely using Lemma \ref{aaecontinuity}, this is either a classical selection principle and existence of an unique solution of (\ref{P_0}) in the limit $\epsilon \to 0$, or a lack of selection principle but occurring on a set whose relative measure w.r.t. Lebesgue is going to zero in the limit $\epsilon \to 0$; see Lemma \ref{aaecontinuity}.

\end{proof}
\paragraph{Proof that ${\cal O}|_{{\cal S}_0}\not\equiv \text{const}$}
\begin{proof}
We show that $\LSPO \Rightarrow {\cal O}|_{{\cal S}_0}\not\equiv \text{const}$.
We notice that the connectedness of ${\cal S}_0$, from Kneser's theorem, can be used together with the continuity of ${\cal O} \in C_b(H;\mathbb{R})$ to infer that ${\cal O}({\cal S}_0)$ is necessarily a closed interval. Therefore if ${\cal O}$ is constant with value $c$, ${\cal O}(\gamma(s)) = c$ which contradicts $\LSPO$. 
\end{proof}

\begin{figure}[htbp]
\centerline{\includegraphics[width=0.5\columnwidth]{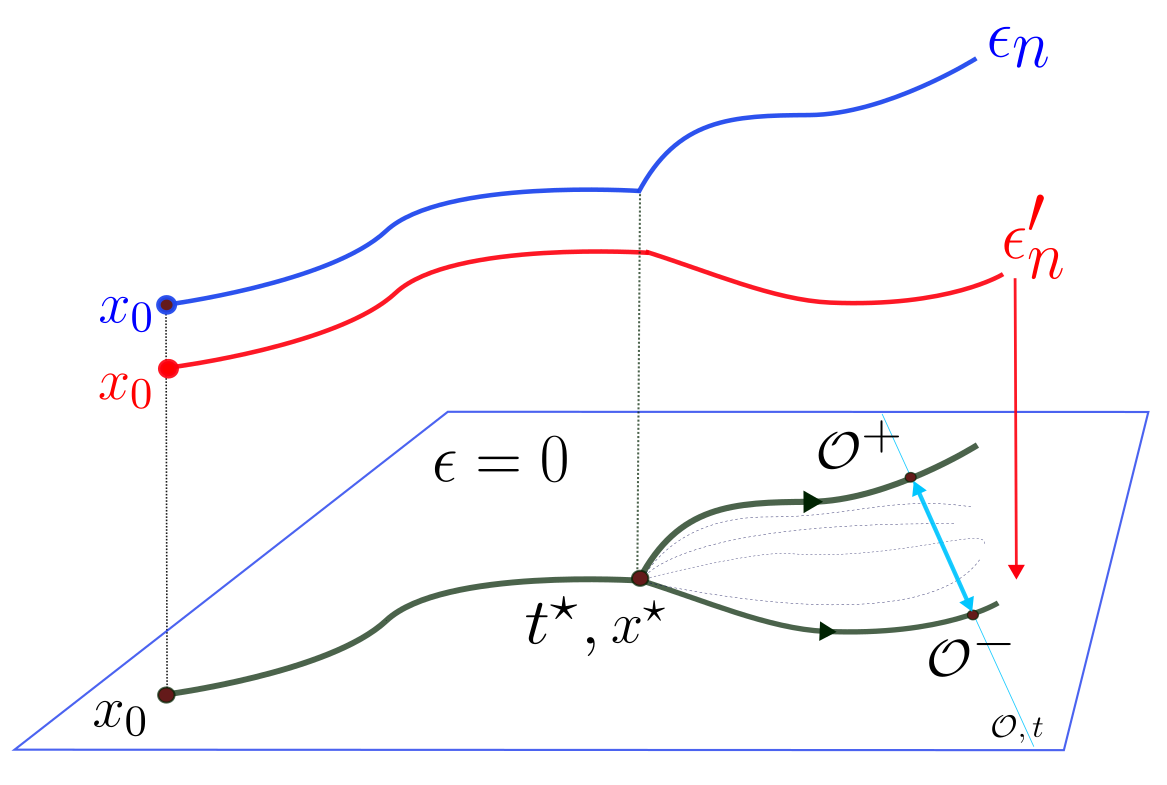}}
\caption{A sketch view of trajectory splitting in the singular limit $\epsilon \to 0$.
One can construct a multi-valued flow which gather all possible flows associated with convergent $\epsilon$ subsequences (at least two exist). These flows come 
from the well-posed regularized systems $({\cal P}_\epsilon)$ in the limit $\epsilon \to 0$. Here, there is a particular (finite) time $t^\star$ for which the system meets a non-Lipschitz singularity $x^\star$.
For $t > t^\star$, one has $ {\cal O}^- < {\cal O}^+$ for the observable ${\cal O}$ at time $t$ shown by a thin blue line.}
\label{split}
\end{figure}

The inviscid limit for $({\cal P}_\epsilon)$ thus selects at least two distinct limsup/liminf trajectories starting from the same $x_0$ among infinitely many trajectories in the inviscid system, which differ at time $t$. Due to that, one has automatically unbounded finite-time Lyapunov exponents, meaning that the
quantity (\ref{fts}) diverges in the inviscid limit:
\begin{equation} \label{fts}
\lim_{\epsilon \to 0} ~\Biggl| \langle \nabla{\cal O}(\Phi_t[f(\cdot,\epsilon)] x_0),  \Biggr. \left. {\cal T} {\rm exp} \left(\Int_0^t \frac{\partial f(\cdot,\epsilon)}{\partial x}(\Phi_s[f(\cdot,\epsilon)] x_0)~ds \right) v \rangle\right| = + \infty,
\end{equation}
where ${\cal T}$ is the time-ordering operator.
We give a simple sketch of this scenario in Fig.~\ref{split}.

\section{Choice of temporal cutoff functions}\label{APP:TimeCutoff}
Defining $\zeta_{m,k}(t)= \zeta(t/\tau_m-k)$, the time cutoff $\zeta$ must be such that \cite{AV24},
\begin{equation*}
    \begin{cases}
        &0\le \zeta \le \mathbbm{1}_{t\in[-2/3, 2/3]},\\
        & \sum \limits_{k\in \mathbb{Z} }\zeta(\cdot -k)=1,\\
        &\int_\mathbb{R} \zeta^2 =\dfrac{9}{10}.
    \end{cases}
\end{equation*}
In order to build such a function, take $\varphi$ a compact support $C^\infty$ function and consider,
$$\displaystyle \zeta(t)= \dfrac{\varphi(\alpha t)}{\sum \limits_{k\in \mathbb{Z} }\varphi(\alpha t -k)}.$$
The partition of unity property is verified and $\alpha$ is chosen such that the square $L^2$ norm of $\zeta$ is $9/10$. A simple change of variable in the integral yields
$$\alpha = \dfrac{10}{9} \int \left(\dfrac{\varphi(s)}{\sum \limits_{k\in \mathbb{Z} }\varphi( s -k)}   \right)^2 \, \mathrm{d}s.$$ 
With this construction, the partition of unity and $L^2$ norm property are verified but we don't necessarily have $\mathrm{Supp}( \zeta) \subset [-2/3, 2/3]$. Taking $\varphi(t)= \exp\left( 1/\left(\left(\dfrac{t}{2/3} \right)^2-1\right)\right)$ when $ |t|<2/3$ and zero elsewhere,  we have $\alpha \simeq 1.04$. The support of $\zeta$ is thus included in the right interval because $\alpha \ge 1$, the value for $\alpha$ is computed numerically once and stored.
This cutoff governs the smooth alternation between $x$ and $y$ shear flows, the $k-$shifted, $\tau_m$-dilated cutoffs are given by
\begin{equation}
    \zeta_{m,k}(t)= \zeta\left(t/\tau_m-k\right).
\end{equation}
Hence, $ \mathrm{Supp}(\zeta_{m,k}) \subset  \left[ (k-2/3)\tau_m, (k+2/3)\tau_m \right]$.

The cutoffs $\hat{\zeta}_{m,l}$ are not partitions of unity, they must satisfy

\begin{equation}
      \mathbbm{1}_{t\in[(l-1/2)\tau_m^{''} +2\tau_m^{'},(l+1/2)\tau_m^{''} -2\tau_m^{'}]} \le \hat{\zeta}_{m,l}\le \mathbbm{1}_{t\in [(l-1/2)\tau_m^{''} +\tau_m^{'},(l+1/2)\tau_m^{''} -\tau_m^{'}]}\label{eq:supportzetahat}
\end{equation}
In order to build the $\hat{\zeta}_{m,l}$, consider first
\begin{equation}
    f(t) = 
    \begin{cases}
        &0, \, t\le 0, \\
        &1, \, t\ge 1, \\
        & \dfrac{1}{\left(1+ \exp\left(\dfrac{1-2t}{t(1-t)} \right) \right)},\, 0<t<1.
    \end{cases}    
\end{equation}
and then define 
\begin{equation}
    \hat{\zeta}_{m,l}(t)= f\left( \dfrac{1}{2}\dfrac{\tau_m^{''}}{\tau_m^{'}}-1-\dfrac{t-l\tau_m^{''}}{\tau_m^{'}}\right)f\left( \dfrac{1}{2}\dfrac{\tau_m^{''}}{\tau_m^{'}}-1+\dfrac{t-l\tau_m^{''}}{\tau_m^{'}}\right).
\end{equation}
which indeed has the desired property.
\section{Algebraic reparameterizations: Proposition  \ref{ALGINV}}\label{alginv_proof}
\begin{proof}[Proof of Proposition \ref{ALGINV}]
We consider a $C^2$ diffeomorphism $g: \mathbb{R}^+ \to 
\mathbb{R}^+, \tau \mapsto g(\tau)$ such that
$\lim_{\tau \to \infty} g(\tau) = 0$ and is strictly decreasing $g' < 0$.
One assume without loss of generality that $\lim_{s \to \infty} \gamma(s)$ exists which correspond to the limit where $\epsilon$ for $({\cal P}_\epsilon)$ is large. We can thus extend $F\circ \gamma \circ g$ by continuity at  $\tau=0$.
Let us define $F_g(\tau) = F(\gamma \circ g(\tau))$ for all $\tau \geq 0$.
Before hand, we need the following easy result:
\be \label{tdg}
\lim_{\tau \to \infty} \tau g'(\tau) = 0.
\de 
\begin{proof}
  By using the mean value theorem,
one has for some $\theta > 1$, $g(\theta \tau) - g(\tau) = (\theta-1) \tau g'(\zeta),
\zeta \in [\tau,\theta \tau]$ so that $\tau g'(\zeta) \to 0$ and in particular $\tau g'(\tau) = o(1)$.  
\end{proof}
First notice that after the change of variable $s = g(\tau)$ with $\epsilon = g(R), R = g^{-1}(\epsilon)$, it gives 
$$
\frac{1}{\epsilon} \int_0^\epsilon F(\gamma(s))ds = J(R),
$$
where
\be 
J(R):=  {\displaystyle -\frac{1}{g(R)} \int_R^\infty F_g(\tau) g'(\tau) d\tau},~\text{and we write}~
B(R) :=  {\displaystyle \frac{1}{R} \int_0^R F_g(\tau) d\tau}.
\de 
\subsection{$B \to \rho \Rightarrow J \to \rho$}
We can express $J$ as a function of $B$ using (\ref{tdg}) like
$$
J(R) = -\frac{1}{g(R)}\int_R^\infty (\tau B(\tau))' g'(\tau) d\tau =
\Omega(R) B(R) + \frac{1}{g(R)} \int_R^\infty \tau g''(\tau) B(\tau) d\tau,~~\Omega(R):= \frac{R g'(R)}{g(R)}.
$$
By remarking that $\Omega = -\frac{1}{g(R)} \int_R^\infty (\tau g'(\tau))' d\tau$, $1 = -\frac{1}{g(R)} \int_R^\infty g'(\tau)d\tau$, and $\tau g''(\tau) = (\tau g'(\tau))' - g'(\tau)$, one has
$$
J(R)-\rho = \Omega(R)(B(R)-\rho) + \int_R^\infty \omega_R'(\tau) (B(\tau)-\rho) d\tau~\text{with}~\omega_R(\tau) = \frac{\tau g'(\tau)-g(\tau)}{g(R)}.
$$
It yields
$$
|J(R)-\rho| \leq |\Omega(R)| ~|B(R)-\rho| +
\left( \int_R^\infty |\omega_R'| ~d\tau \right) ~\sup_{s \in [R,\infty)} |B(s)-\rho|.
$$
It imposes that there exist constants $C$ independent of $R$ such that
\be 
|\Omega(R)| \leq C~\text{and}~\int_R^\infty \left|\frac{\tau g''(\tau)}{g(R)}
\right| d\tau \leq C.
\de 
\subsection{$J \to \rho \Rightarrow R \to \rho$}
Similarly, one can express $B$ as a function of $J$:
$$
B(R) = \frac{1}{R} \int_0^R \frac{(g(\tau)J(\tau))'}{g'(\tau)} d\tau =
\Omega^{-1}(R) J(R) - \frac{C_0}{R} + \frac{1}{R} \int_0^R \frac{gg''}{g'^2}(\tau) J(\tau) d\tau,~C_{0} = \frac{g(0)J(0)}{g'(0)}.
$$
Again, we use the fact that $\Omega^{-1}(R) = \frac{1}{R} \int_0^R \left(\frac{g}{g'} \right)' d\tau + \frac{1}{R} \frac{g(0)}{g'(0)}$ and $\frac{gg''}{g'^2}  = 1 - 
\left(\frac{g}{g'} \right)'$ giving
$$
B(R)-\rho = \Omega^{-1}(R)(J(R)-\rho) -\frac{D_0}{R}
+\frac{1}{R} \int_0^R \left( 1 - \left(\frac{g}{g'}\right)' \right) (J(\tau)-\rho)~d\tau,~D_0 = \frac{g(0)}{g'(0)} (J(0)-\rho).
$$
An upper bound is, for some constant $C$ independent of $T$:
\be \label{R_rho_upp}
|B(R)-\rho| \leq \frac{C}{R} + |\Omega^{-1}(R)|~|J(R)-\rho|
+ \left| \frac{1}{R} \int_0^R \omega'(\tau) (J(\tau)-\rho) d\tau \right|,~\omega'(\tau) = 1 - \left(\frac{g}{g'} \right)'.
\de 
Let us focus on the Cesaro term, 
Let $\epsilon  > 0$, $\exists R_0=R_0(\epsilon)$ such that
$\forall \tau \geq R_0, |J(\tau)-\rho| \leq \epsilon$. We have,
$$
 \left| \frac{1}{R} \int_0^R \omega'(\tau) (J(\tau)-\rho) d\tau \right| \leq 
 \frac{1}{R} \sup_{s \in [0,R_0]} |(J(s)-\rho) \omega'| +
\left( \frac{1}{R} \int_{0}^R 
 \left| 1-\left(\frac{g}{g'}\right)' \right| d\tau \right) \epsilon.
$$
Therefore, in order to have the right-hand side of (\ref{R_rho_upp}) going to zero, one needs that there exists some constants $C$ independent of $R$ such that
\be 
|\Omega^{-1}(R)| \leq C~\text{and}~\frac{1}{R}\int_0^R \left|\frac{gg''}{g'^2} \right|(\tau) d\tau \leq C.
\de 

Note that since $g$ is strictly decreasing, the boundary term $D_0 / R$ vanishes as $R \to \infty$. In conclusion, for the two limits to coincide, there must exist $R_0$ sufficiently large and constants $0 < c \leq C < \infty$ such that for all $R \geq R_0$, the following conditions hold:
\begin{equation} \label{asymptotic_conditions}
\begin{aligned}
\text{(1)}\quad & \frac{c}{R} \leq \frac{g'(R)}{g(R)} \leq \frac{C}{R}, \\
\text{(2)}\quad & \int_R^\infty \left| \frac{\tau g''(\tau)}{g(R)} \right| d\tau \leq C, \\
\text{(3)}\quad & \frac{1}{R} \int_0^R \left| \frac{g(\tau) g''(\tau)}{(g'(\tau))^2} \right| d\tau \leq C.
\end{aligned}
\end{equation}
It indicates that $g$ must have algebraic decay but with a prefactor which is slowly varying.
\end{proof}

\section{Proof of Theorem \ref{CN}} \label{DiniApp}
Although we do not use Dini derivatives explicitly, it is closely related to
Definition \ref{DiniME}. We just recall very few well-known properties.
\begin{definition}
Let $f(x)$ be a  real valued function defined in $(a,b)$, the four Dini derivatives at $x_0$ are defined as
\begin{align}
D^\pm f(x_0) = \limsup_{x \to x_0^\pm} \frac{f(x)-f(x_0)}{x-x_0}, \nonumber \\ 
D_\pm f(x_0) = \liminf_{x \to x_0^\pm} \frac{f(x)-f(x_0)}{x-x_0}. \nonumber
\end{align}
$\pm \infty$ is allowed.
\end{definition}
We state some Dini properties for general functions (not necessarily continuous)
\begin{itemize}
\item if $f$ is continuous in $(a,b)$, and $D(x) \in \{D_\pm^\pm f(x)\} > 0 (<0),~\forall x \in (a,b)$ then $f$ is
strictly increasing (decreasing). In other words, one needs only
one of the Dini derivative. 
\item Let $f$ be continuous on $(a,b)$ with at least one of the 
Dini derivatives bounded  (e.g. $|D^+ f(x)| \leq C, \forall x \in (a,b)$)
then f is  Lipschitz on $(a,b)$.
\end{itemize}
Deeper results can be found in \cite{giorgi1992dini,bruckner1978}.
\\\\
The proof of Theorem \ref{CN} is simple, since it is mostly a rephrasing of non-Lipschitz
property using Dini-like quantities but translated in the "Osgood framework".
\begin{proof}
We show the contrapositive. We restrict $x$ to a compact set
$M \subset \mathbb{R}^n$. Therefore, $\exists C, \forall (x,v) \in M \times \mathbb{S}^{n-1}, 
\Lambda^+_\Omega(x,v) \leq C < + \infty$. We need to show that in this case
$\dot x =f(x), x(0)=x_0 \in M$ has a unique solution. Note that $f$ might not be necessarily Lipschitz. Consider $v = \frac{y-x}{||y-x||}$. Using the uniform bound above, for small enough $t>0$, one has
$\langle f(x+tv)-f(x),v \rangle \leq C \Omega(t)$, and taking $t=||y-x||$  gives the one-sided inequality
$$
\langle f(y)-f(x),y-x \rangle \leq C ||y-x|| \Omega(||y-x||).
$$
Denote $g(||y-x||^2) = ||y-x|| \Omega(||y-x||)$. Then by the one-sided 
Osgood Lemma, uniqueness
occurs if $\Int_{0^+} \frac{dz}{g(z)} = \Int_{0^+} \frac{dz}{\Omega(z)} = +\infty$ which is just the hypothesis of Theorem \ref{CN}. 
\end{proof}
Note that the nonautonomous case holds
as well: this is Giuliano's uniqueness Theorem with a time-dependent measurable constant (see Theorem 3.5.1 in 
\cite{Agarwal_Lak}).
\section{Regularization encoding all statistical behaviors: Proposition \ref{CHOCBAR2}}\label{Chocbar2proof}
\begin{proof}[Proof of Proposition \ref{CHOCBAR2}]
Let $b_n := (x_n,y_n,\theta_n) \in B:={\cal S}_0 \times {\cal S}_0 \times [0,1]$ a sequence which is sequentially dense, meaning that for all $b = (x,y,\theta) \in B$ one has a convergent subsequence $b_{n_k} \to b$. Let $T_n \to \infty$ a given sequence to choose. Call $\chi_n:\mathbb{R}^+ \to [0,1]$ a partition of unity such that $\sum_n \chi_n(s) = 1$ and
$\chi_n(s)=1, s \in [T_n+\delta,T_{n+1}-\delta]$. We define $\gamma_{\rm un}(\tau) = \sum_n \chi_n(\tau) \gamma_{x_n,y_n,\theta_n}(\tau)$ where $\gamma_{x,y,\theta}$ is the regularization curve constructed in the proof of Theorem \ref{M0M}, Section \ref{Proof_M0M}.
We must choose $T_n$ so that $T_{n+1}-T_n$ is large enough to distinguish at least one period of $a_\theta$ in (\ref{controlswitch}).
In the interval $(T_n,T_{n+1})$, the curve is approximating $\theta_n x_n + (1-\theta_n) y_n$. Due to the sequential density, for all $(x,y,\theta) \in B$, one can exhibit subsequences $n_k \to \infty$ such that $x_{n_k},y_{n_k},\theta_{n_k} \to x,y,\theta$, namely $\mu_{T_{n_k}} := \frac{1}{T_{n_k}} \int_0^{T_{n_k}} \delta_{\gamma_{\rm un}(\tau)} d\tau \rightharpoonup  \theta \delta_{x} + (1-\theta) \delta_{y}$. The set of accumulation points is therefore
$\overline{\operatorname{co}}({\cal E}) = {\cal M}_0$.
\end{proof}

\section{Discussion on the admissible set ${\cal K}$}\label{Kgen}

Let $(g_m)_{m \in \mathbb{N}}$ be a sequence with $g_m > 0$ and $\sum_{k} g_k < +\infty$.  
Fix $M \in \mathbb{N}$ and suppose $a_m > 0$ with $a_m \to 0$ as $m \to \infty$.  
We wish to study the behavior of the backward sequence \eqref{avseq} for $m \leq M$  
and to determine $a_M$  such that the following sequence remains bounded for all $M$ and $m$:
\be 
\kappa_{m-1} = \kappa_m + \frac{g_m}{\kappa_m},  
\quad \frac{1}{A} a_M \leq \kappa_M \leq A a_M, \quad A > 1.
\de 
Introduce the prefactor sequence $(s_m)_{m \leq M}$ via
\be 
\kappa_m = \gamma_m s_m.
\de
Then $s_m$ satisfies
\be 
s_{m-1} = \frac{\gamma_m}{\gamma_{m-1}} s_m + \frac{g_m}{\gamma_m \gamma_{m-1}} \frac{1}{s_m},  
\quad \frac{1}{A} \frac{a_M}{\gamma_M} \leq s_M \leq A \frac{a_M}{\gamma_M}.
\de 
We assume $\gamma_m$ can be chosen to satisfy
\be \label{ggg}
\gamma_m \gamma_{m-1} = g_m.
\de 
In this setting,
$$
s_{m-1} = F_m(s_m),  
\quad F_m(s) := \sigma_m s + \frac{1}{s},  
\quad \sigma_m := \frac{\gamma_m}{\gamma_{m-1}},
$$
and we define
\be \label{defRbM}
{\cal R}(x) := \max\left\{ x,\frac{1}{x} \right\},  
\quad b_M := \frac{a_M}{\gamma_M}.
\de 
A straightforward inspection yields
$$
{\cal R}(s_M) \leq A {\cal R}(b_M),\quad  
{\cal R}(s_{M-1}) \leq A {\cal R}(F_M(b_M)), \quad  
{\cal R}(s_{M-2}) \leq A {\cal R}(F_{M-1} \circ F_{M}(b_M)), \ \ldots
$$
and, recursively, for all $k \geq 1$:
\be \label{Ybound_1}
{\cal R}(s_{M-k}) \leq A{\cal R}\bigl( F_{M-k+1} \circ F_{M-k+2} \cdots F_{M}(b_{M}) \bigr).
\de 
A second bound can be obtained as follows:  
if $s_m \geq 1$, then $s_{m-1} \leq (1+\sigma_m) s_m$;  
if $s_m \leq 1$, then $\frac{1}{s_{m-1}} \leq (1+\sigma_m) s_m$,  
hence ${\cal R}(s_{m-1}) \leq (1+\sigma_m) {\cal R}(s_m)$.  
Applying this recursively for $1 \leq m \leq M-k-1$ yields
\be \label{Ybound_2}
{\cal R}(s_m) \leq (1+\sigma_{m+1}) \cdots (1+\sigma_{M-k}) \,{\cal R}(s_{M-k}).
\de 
Combining \eqref{Ybound_1} and \eqref{Ybound_2} gives, for $m \leq M-k-1$, $k \geq 1$,
\be 
{\cal R}(s_m) \leq A \,\Pi_{m,M-k} \,{\cal R}_{k,M},
\de 
where
\be 
{\cal R}_{k,M} := {\cal R}\bigl( F_{M-k+1} \circ \cdots \circ F_{M}(b_{M}) \bigr) \geq 1,  
\quad \Pi_{m,M-k} := \prod_{j=m+1}^{M-k} (1+\sigma_j) > 1.
\de 
Since $A > 1$ is fixed and the series $\sum_k g_k$ converges,  
the product $\Pi_{m,M-k}$ is uniformly bounded in $m$.  
We thus obtain the uniform bound
\be 
{\cal R}(s_m) \leq C \,{\cal R}_{k,M},
\de 
with $C > 1$. Consequently, for $m \leq M-k-1$, the sequence obeys
\be  \label{Xbounded}
\frac{\gamma_m}{C {\cal R}_{k,M}} \leq \kappa_m \leq C \gamma_m {\cal R}_{k,M}.
\de

We now apply this estimate to the hypergeometric sequence  
$g_m = c_0 \epsilon_m^{2\beta}$.  
From the constraint~\eqref{ggg}, we obtain  
\be 
\gamma_m = \sqrt{c_0} \,\epsilon_m^{ \frac{2 q \beta}{q+1}},  
\quad \sigma_m = \epsilon_m^{2 \beta \frac{q-1}{q+1}}.
\de 
Since $k$ can be chosen freely, we take $b_M^{(k)}$ such that  
${\cal R}_{k,M} = 1$. The condition ${\cal R}(x)=1$ holds only for $x=1$, namely  
\be \label{Feq1}
F_{M-k+1} \circ F_{M-k+2} \cdots F_{M}(b_{M}^{(k)}) = 1.
\de 
Choosing $k=1$, the equation $\sigma_M b_M^{(1)} + \frac{1}{b_M^{(1)}} = 1$ has two roots, and we select the larger one,  
$b_M^{(1)} \approx \sigma_M^{-1}$. In this case, the definition~\eqref{defRbM} yields  
$
a_M \approx \sqrt{c_0} \,\epsilon_M^{\frac{2 \beta}{q+1}}.
$
Using~\eqref{Xbounded} and~\eqref{Feq1} with $\kappa_M = \kappa$, we obtain~\eqref{avseqstaybounded} with $A$ instead:
$$
\kappa \in \left[ \frac{1}{A} \sqrt{c_0} \,\epsilon_M^{\frac{2\beta}{q+1}},  
\; A \sqrt{c_0} \,\epsilon_M^{\frac{2\beta}{q+1}}\right]  
\;\Rightarrow\; 
\kappa_m \in \left[ c \,\epsilon_m^{\frac{2q \beta}{q+1}}, \; C \,\epsilon_m^{\frac{2 q\beta}{q+1}} \right].
$$
\\
One may alternatively take the smallest root, namely $b_M^{(1)} \approx 1$.  
In this case, $a_M \approx \sqrt{c_0} \,\epsilon_M^{\frac{2 \beta q}{q+1}}$,  
which results in a smaller set ${\cal K}$ since $q > 1$.  
\\
It is also possible to proceed to the next order in $k$.  
For example, consider $k=2$. One must solve  
$\sigma_{M-1} F_M(b_M^{(2)}) + \frac{1}{F_M(b_M^{(2)})} = 1$.  
If the smallest root $\approx 1$ is chosen, we recover $b_M^{(2)} \approx b_M^{(1)}$.  
Instead, taking the largest root yields  
$b_M^{(2)} \approx (\sigma_M \sigma_{M-1})^{-1}$,  
which in turn gives  
$a_M \approx \sqrt{c_0} \,\epsilon_M^{\frac{2\beta}{q(q+1)}}$.  
Proceeding recursively in this manner, we arrive at the following result:
\begin{lemma}
 Let $A > 1$,  let $1 \leq k \leq M-m-1$, then 
 \be
 \kappa \in \left[ \frac1A \sqrt{c_0} \epsilon_M^\frac{2\beta}{q^{k-1}(q+1)},A \sqrt{c_0} \epsilon_M^\frac{2\beta}{q^{k-1}(q+1)}\right] \Rightarrow 
\kappa_m \in \left[ c \epsilon_m^\frac{2q \beta}{q+1}, C \epsilon_m^\frac{2 q\beta}{q+1} \right].
 \de 
\end{lemma}
Although $\mathcal{K}$ is significantly larger, it remains disconnected.  
Intriguingly, numerical evidence suggests that it becomes connected once $\beta$ exceeds a critical value of approximately $1.1428$. We have, however, been unable to establish a rigorous proof of this fact.

%%%%%%%%%%%%%%%%%%%%%%%%%%%%%%%%%%%%%%%%%%%%%%%%%%%%%%%%%%%%%%%%%
%%%%%%%%%%%%%%%%%%%%%%%%%%%%%%%%%%%%%%%%%%%%%%%%%%%%%%%%%%%%%%%%%
%%%%%%%%%%%%%%%%%%%%%%%%%%%%%%%%%%%%%%%%%%%%%%%%%%%%%%%%%%%%%%%%%
\end{document}